\documentclass{article}

\usepackage{epsfig} 
\usepackage{amsmath}
\usepackage{amsfonts}
\usepackage{graphicx}
\usepackage{cite}

 \usepackage{amssymb}

\date{\today}

\newcommand{\insertplot}[5]{\begin{figure}
 \hfill\hbox to 0.05in{\vbox to #5in{\vfill
 \inputplot{#1}{#4}{#5}}\hfill}
 \hfill\vspace{-.1in}
 \caption{#2}\label{#3}
 \end{figure}}
 \newcommand{\inputplot}[3]{
 \special{ps: plotfile #1}
}
\newcounter{fig}   

\numberwithin{equation}{section}

\newcommand{\ee}{\end{equation}}
\newcommand{\eea}{\end{eqnarray}}
\newcommand{\be}{\begin{equation}}
\newcommand{\bea}{\begin{eqnarray}}

\newcommand{\re}[1]{(\ref{#1})}

\begin{document}

\title{\bf \Large Construction and physical properties of \\ Kerr black holes with scalar hair}

 \author{
{\large Carlos Herdeiro}\footnote{herdeiro@ua.pt} \
and
{\large Eugen Radu}\footnote{eugen.radu@ua.pt}
\\ 
\\
{\small Departamento de F\'\i sica da Universidade de Aveiro and CIDMA} \\ 
{\small   Campus de Santiago, 3810-183 Aveiro, Portugal}
}
\date{January 2015}
\maketitle

\begin{abstract}
Kerr black holes with scalar hair are solutions of the Einstein-Klein-Gordon field equations describing regular (on and outside an event horizon), asymptotically flat black holes with scalar hair~\cite{Herdeiro:2014goa}. These black holes interpolate continuously between the Kerr solution and rotating boson stars in $D=4$ spacetime dimensions. Here we provide details on their construction, discussing properties of the ansatz, the field equations, the boundary conditions and the numerical strategy. Then, we present an overview of the parameter space of the solutions, and describe in detail the space-time structure of the black holes exterior  geometry and of the scalar field for a sample of reference solutions. Phenomenological properties of potential astrophysical interest are also discussed, and the stability properties and possible generalizations are commented on. 
As supplementary material to this paper we make available numerical data files for the sample of reference solutions discussed, for public use.
\end{abstract}

\newpage

\tableofcontents

\newpage

\section{Introduction}

There is strong observational evidence that extremely compact and massive objects populate the Universe. One piece of evidence comes from our own galactic centre, from the bright radio source Sagittarius A$^\star$. From the Keplerian orbits of stars in its vicinity, its mass has been estimated as $4.1\times 10^6$ M$_\odot$ (where M$_\odot$= 1 solar mass)~\cite{Ghez:2008ms} and a size constraint of $6$ light hours has been placed on the object~\cite{2005ApJ}. The best theoretical candidate from well established physical models which fits these observational data is a black hole (BH). Thus Sagittarius A$^\star$, as well as other similar compact objects at the center of other spiral and elliptic galaxies, is commonly referred to as a \emph{supermassive black hole}.
BH candidates of this type have been found within a mass range between $10^6$ and $10^{10}$ M$_\odot$~\cite{Narayan:2013gca} and they are thought to play a central role in both the formation and growth of their host galaxies \cite{Kormendy:2013pja}; thus understanding them is of vital importance for the models of structure formation in the Universe.

Another case of extremely compact and massive objects is found in binary systems in our galaxy, where strong X-ray sources exist.
One example of such an object is Cygnus X-1, known since the 1960s~\cite{1965Sci...147..394B}.
The estimated masses of these $24$ binaries were found to range between $5$ and $30$ M$_\odot$~\cite{Narayan:2013gca}.
For instance, Cygnus X-1 has an estimated mass of 14.8 M$_\odot$~\cite{Reid:2011nn}.
\textit{Neutron stars}, as the most compact directly observable objects currently known, have masses which are below $3$ M$_\odot$ \cite{Kalogera:1996,Rhoades:1974}.
Thus, these $24$ X-ray sources are thought to be BHs, which, within this mass range, are dubbed \textit{stellar mass black holes}.

\bigskip

At present, it is unknown if the BH candidates discussed in the previous two paragraphs are the paradigmatic  BHs of general relativity, BHs as described by some alternative model, or even other types of compact objects without an event horizon. 
The next decade promises to shed light on both of these issues: we are on the verge of gathering observational evidence that will map the spacetime geometry close to these BH candidates.
This evidence will be obtained by both gravitational wave astronomy~\cite{Hild:2011np,Hobbs:2009yy,Seoane:2013qna} and by large baseline interferometry measurements of the galactic center, using the Event Horizon Telescope (EHT).
The latter, promises to resolve angular scales of the order of the horizon scale for the Sagittarius A$^\star$ BH candidate~\cite{Loeb:2013lfa}.
The EHT will study the so-called BH \textit{shadows}~\cite{Falcke:1999pj}: the gravitational lensing and redshift effect due to the BH on the radiation from background sources, with respect to the observer.
These forthcoming experiments make it particularly timely to explore alternative models to the general relativity BH paradigm, and their associated phenomenology~\cite{Berti:2015itd}.

\bigskip

According to the general relativity BH paradigm the myriads of BHs that populate the Universe should, when near equilibrium, be well described by the Kerr metric~\cite{Kerr:1963ud}. The paradigm is supported on both a set of mathematical theorems -- the uniqueness theorems~\cite{Robinsoon:2004zz,Chrusciel:2012jk} -- but also on a conjecture -- the no-hair conjecture~\cite{Ruffini:1971}. The former established that, for vacuum Einstein's equations, the only regular (on and outside an horizon) BH solution is Kerr. The latter extrapolates that, even for more generic forms of matter, the end-point of gravitational collapse should still be an exterior Kerr solution. Progress in this context was obtained for particular types of matter. One of the simplest types of ``matter'' often considered by physicists is provided by \textit{scalar fields}.
Since 2012, there is observational evidence that scalar fields exist in nature, by virtue of the discovery of a scalar particle at the Large Hadron Collider, at CERN, identified with the standard model Higgs boson~\cite{Aad:2012tfa,Chatrchyan:2012ufa}.
But for decades, scalar fields have been considered in phenomenological models, in particular within gravitational physics.
A notable example is cosmology, where various types of scalar fields have been used to model dark energy and dark matter.
One reason is that scalar fields are well motivated by some high energy physics models such as string theory or scalar extensions of the standard model of particle physics.
Yet another reason is that scalar fields may be considered as a proxy to realistic matter, such as some perfect fluids.

It is therefore quite natural that in testing the no-hair idea, scalar fields were one of the first types of ``matter'' considered.
This program was initiated by Chase~\cite{Chase:1970} who established that ``every zero-mass scalar field which is gravitationally coupled, static and asymptotically flat, becomes singular at a simply-connected event horizon''.
In other words, a BH spacetime cannot support a regular massless scalar field in equilibrium with it; $i.e.$ no \textit{BH (massless) scalar hair}.
Further ``no-scalar-hair'' theorems were developed by Bekenstein~\cite{Bekenstein:1971hc,Bekenstein:1972ky} who also considered massive scalar, vector and spin 2 fields (see also the review~\cite{Bekenstein:1996pn}) and Hawking~\cite{Hawking:1972qk}, who showed that in the Brans-Dicke theory of gravity~\cite{Brans:1961sx}, in which there is a scalar field non-minimally coupled to the geometry, the regular BH solutions are the same as in general relativity.
Hawking's theorem has actually been recently generalized~\cite{Sotiriou:2011dz} to more general \textit{scalar-tensor theories of gravity}, of which the Brans-Dicke model is an early example.\footnote{Scalar-tensor theories with higher derivatives (albeit with second order equations) can, however, accommodate spherically symmetric hairy BHs~\cite{Sotiriou:2013qea,Babichev:2013cya,Sotiriou:2014pfa,Charmousis:2014zaa}.} 

A few remarks are in order concerning ``no-scalar-hair'' results for BHs.
First of all, we are focusing on regular configurations satisfying the weak energy condition\footnote{
Hairy BH solutions can be constructed by allowing a scalar potential which is not strictly positive, 
 see $e.g.$ the recent work \cite{Kleihaus:2013tba}
and the references therein.}.
Solutions where the scalar field diverges at the horizon are known (see e.g.~\cite{Bocharova:1970}), but they appear to have no physical relevance.
We are also considering independent scalar fields.
Considering \textit{simultaneously} gauge fields to which the scalar fields are non-minimally coupled leads to solutions~\cite{Gibbons:1982ih,Gibbons:1987ps} - notably $p$-brane type solutions in supergravity~\cite{Horowitz:1991cd}; but these have no independent scalar charge and the scalar field vanishes when the electromagnetic field vanishes.
Finally, we are focusing on four dimensional, asymptotically flat BHs.
Considering, for instance, Anti-de-Sitter asymptotics can allow for hairy BHs, since the asymptotic nature of the spacetime yields a trapping mechanism from which the scalar field cannot escape. A thorough review of asymptotically flat BHs with scalar hair is given in~\cite{Herdeiro:2015waa}.

Considering more general kinds  of matter, many counter-examples to the no-hair conjecture -- at least in its weakest version, stating that no BH solutions with other forms of matter should exist, regardless of their stability -- have actually been shown to exist. 
An influential pioneering example was built upon the Bartnik-Mckinnon ``particle-like'' solution of the Einstein-Yang-Mills equations~\cite{Bartnik:1988am}.
Soon after that, it has been found 
that a BH could be added at the center of these solitons 
\cite{Volkov:1989fi,Bizon:1990sr}, 
yielding \textit{BHs with Yang-Mills hair} also dubbed \textit{colored BHs}.
Other examples with a similar spirit, were obtained with other non-Abelian gauge fields, $e.g.$ \cite{Droz:1991cx,Lavrelashvili:1992cp}; see also the reviews~\cite{Bizon:1994dh,Volkov:1998cc}. 
Such counter-examples however, invariably use non-linear matter fields, 
and the resistance of the matter field against collapse into the BH 
is anchored to these non-linearities. 
Furthermore, many of these counter-examples use matter fields which, probably, 
have  little astrophysical relevance, at macroscopic scales. 
As such, even though these examples show clearly the mathematical limitations of the no-hair idea, 
 the question remains if there are astrophysically more realizable models of hairy BHs.

\bigskip

Recently a new family of BHs with scalar hair was found~\cite{Herdeiro:2014goa} that is continuously connected to the Kerr family and yields a qualitatively new example of hairy BHs dubbed \textit{Kerr BHs with scalar hair} (KBHsSH). Firstly because the scalar hair is not anchored on non-linear effects. The scalar field possesses a mass term but no self-interactions. Consequently, the hair can be seen in linearized theory, by considering the massive Klein-Gordon equation around the Kerr BH, and it was interpreted in~\cite{Herdeiro:2014goa} as a zero mode of the superradiant instability. We recall that the superradiant instability of a Kerr BH, in the presence of a massive scalar field, is a mechanism via which a field mode with frequency $w$ and azimuthal harmonic index $m$ is amplified when $w<m\Omega_H$, where $\Omega_H$ is the Kerr BH horizon angular velocity~\cite{Press:1972zz}. By solving the Klein-Gordon equation on the Kerr background, real frequency bound states can be obtained when $w=m\Omega_H$, corresponding to linearized (hence non-backreacting) hair, called \textit{scalar clouds}~\cite{Hod:2012px,Hod:2013zza,Herdeiro:2014goa,Hod:2014baa,Benone:2014ssa}. These are therefore, precisely between the decaying ($w>m\Omega_H$) and superradiant regimes, and thus can be faced as zero modes of the latter. Secondly, because they provide an example of scalar hair around rotating BHs, which, as discussed above has more potential to be astrophysically relevant.

In this paper we will provide details of the construction and of the physical properties of this new type of hairy BHs, complementing the construction in~\cite{Herdeiro:2014goa} and the discussions in~\cite{Herdeiro:2014ima,Herdeiro:2014jaa}. The paper is organized as follows. In Section~\ref{sec_eq_motion} we describe the Einstein-Klein-Gordon theory, the ansatz taken (complemented in Appendix~\ref{appendixa}) and exhibit the corresponding system of equations obtained. In Section~\ref{sec_bc_num} we discuss the boundary conditions to be imposed on the solutions (complemented in Appendix~\ref{coef_asym_exp}), introduce physical quantities and physical relations of interest, such as the first law of BH mechanics and Smarr relations and we discuss the numerical procedure that is used to obtain the solutions. In Section \ref{sec_solutions} we present the solutions, starting by describing relevant properties of the two limiting cases (boson stars (BSs) and a test scalar field around Kerr BHs). In particular we shall present a set of five reference configurations that illustrate qualitatively distinct regions of the solution space and provide illustrative plots of these reference solutions (in Appendix~\ref{sec_plots}). In Section~\ref{sec_pheno} we discuss physical properties of KBHsSH with astrophysical phenomenological interest, as the quadrupole moment and the orbital frequency at the innermost stable circular orbit. In Section~\ref{sec_stability} we address the issue of stability, exhibiting some of the known facts and the open questions.  Finally, in Section~\ref{sec_conclusions} we enumerate a set of research directions and various generalizations of the solutions that can be addressed.

\section{Action, ansatz and equations of motion}
\label{sec_eq_motion}

\subsection{Action}
\label{sec_eq_motion_action}

We shall be working with the Einstein-Klein-Gordon (EKG) field theory, describing a massive complex scalar field 
$\Psi$ minimally coupled to Einstein gravity. The model has the following action and corresponding EKG  field equations, obtained from the variation of the action with respect to the metric and scalar field, respectively: 
\begin{equation}
\label{action}
S=\int  d^4x \sqrt{-g}\left[ \frac{1}{16\pi G}R
   -\frac{1}{2} g^{ab}\left( \Psi_{, \, a}^* \Psi_{, \, b} + \Psi _
{, \, b}^* \Psi _{, \, a} \right) - \mu^2 \Psi^*\Psi
 \right] ,
\end{equation}
\begin{eqnarray}
\label{E-eq}
&&
E_{ab}\equiv R_{ab}-\frac{1}{2}g_{ab}R-8 \pi G~T_{ab}=0 \ ,
\ \
T_{ab}\equiv  
 \Psi_{ , a}^*\Psi_{,b}
+\Psi_{,b}^*\Psi_{,a} 
-g_{ab}  \left[ \frac{1}{2} g^{cd} 
 ( \Psi_{,c}^*\Psi_{,d}+
\Psi_{,d}^*\Psi_{,c} )+\mu^2 \Psi^*\Psi\right]  , \nonumber
 \\
&&
\label{KG-eq}
\Box \Psi=\mu^2\Psi \ ,
\end{eqnarray}  
where $G$ is Newton's constant, $\mu$ the scalar field mass and 
$T_{ab}$ is the energy-momentum tensor  of the scalar field. Observe that this model containing one complex scalar field is equivalent to a model with two real scalar fields. Indeed, writing $\Psi=\Psi^R+i\Psi^I$, where $\Psi^R,\Psi^I$ are two real scalar fields, the action becomes
 \begin{equation}
\label{action_real}
S=\int  d^4x \sqrt{-g}\left[ \frac{1}{16\pi G}R
   - g^{ab}\left( \Psi_{, \, a}^R \Psi_{, \, b}^R + \Psi _
{, \, a}^I \Psi _{, \, b}^I \right) - \mu^2\left[ (\Psi^R)^2+(\Psi^I)^2\right]
 \right] ,
\end{equation}
 which describes Einstein's gravity minimally coupled to two real massive scalar fields, with the same mass. The fact that one has two real scalar degrees of freedom is the reason why the solutions we describe circumvent the theorem in~\cite{Graham:2014ina}.

The action (\ref{action}) is invariant under the global $U(1)$ transformation $\Psi\rightarrow e^{i\alpha}\Psi$, where $\alpha$ is constant. Thus, the scalar 4-current, $j^a=-i (\Psi^* \partial^a \Psi-\Psi \partial^a \Psi^*)$, is conserved:  $j^a_{\ ;a}=0$. It follows that integrating the timelike component of this 4-current in a spacelike slice $\Sigma$ yields a conserved quantity -- the \textit{Noether charge}:
\begin{eqnarray}
\label{Q}
Q=\int_{\Sigma}~j^t \ .
\end{eqnarray}
At a microscopic level, this Noether charge counts the number of scalar particles. In the case an event horizon is present, Noether charge conservation does not prevent the scalar field from falling into the BH; rather, there is a continuity equation relating the decrease of the Noether charge with the scalar flux through the horizon. As such the scalar field may completely disappear through the horizon; moreover, since there is no Gauss law associated to the scalar field, it would leave no signature in the exterior spacetime.

\subsection{Ansatz}
\label{sec_eq_motion_ansatz}
 
 Kerr BHs with scalar hair  (KBHsSH) solutions are obtained by using the following ansatz for the metric and scalar field
\begin{eqnarray}
\label{ansatz}
ds^2=e^{2F_1}\left(\frac{dr^2}{N }+r^2 d\theta^2\right)+e^{2F_2}r^2 \sin^2\theta (d\varphi-W dt)^2-e^{2F_0} N dt^2, \qquad 
~~{\rm with}~~N=1-\frac{r_H}{r}\ ,
\end{eqnarray} 
\begin{eqnarray}
\Psi=\phi(r,\theta)e^{i(m\varphi-w t)}~,
\label{scalar_ansatz}
\end{eqnarray} 
 where $w$ is the scalar field frequency and $m=\pm 1,\pm 2$\dots is the azimuthal harmonic index;
 without loss of generality, we take $w>0$.
The full configuration is therefore described by five functions of $(r,\theta)$: $F_0,F_1,F_2,W,\phi$. We observe that for $r_H=0$ this is basically 
the ansatz used for obtaining rotating BSs~\cite{Schunck:1996he,Yoshida:1997qf}. We further remark that, in the Kerr limit, this ansatz will yield the Kerr solution in a coordinate system which does not coincide with standard textbook coordinates for Kerr. The coordinate transformation from the coordinate system in \eqref{ansatz} to Boyer-Lindquist coordinates is, however, quite simple. It is provided in Appendix \ref{appendixa}. 

Observe that KBHsSH are described by a metric ansatz with two Killing vector fields
\begin{equation}
  \xi=\partial_t, \qquad  {\rm and} \qquad \eta=\partial_\varphi.  
  \end{equation}
  $\xi$ and $\eta$ do not, however, generate symmetries of the full solution, since they do not preserve the expression of the scalar field. The only symmetry of the full solution is generated by the helicoidal vector field
  \begin{equation} 
  \chi =\xi+\frac{w}{m} \eta,
  \label{helicoidal}
  \end{equation}
since $\chi \Psi=0$. This combination is reminiscent of the null horizon generator for rotating BHs. KBHsSH will be obtained by choosing $\chi$ to precisely coincide with such generator. 

Expanding a bit more on the two real fields picture, we remark that for $\Psi$ given by (\ref{scalar_ansatz}), it holds independently that $\chi \Psi^R=0$ and $\chi \Psi^I=0$, where
\begin{equation}
\Psi^ R=\phi(r,\theta)\cos{(m\varphi-w t)} \ , \qquad \Psi^ I=\phi(r,\theta)\sin{(m\varphi-w t)} \ .
\end{equation}
That the real and imaginary part of the complex scalar field are independently preserved by $\chi$ implies that \textit{one} real scalar field coinciding with either $\Psi^R$ or $\Psi^I$ can form stationary waves around a Kerr BH at linear level. These \textit{scalar clouds} will be further discussed in Section \ref{sec_scalar_clouds}.
 The existence of a fully non-linear solution with a stationary metric, however, requires the existence of \textit{two} real scalar fields with opposite phases and the same mass, 
corresponding to one complex scalar field. 
The phase difference guarantee the cancellation of the $t$ and $\varphi$ dependence in the total scalar energy-momentum tensor.

\subsection{Equations of motion}
\label{sec_eq_motion_eom}

Let us now address in detail the system of coupled partial differential equations (PDEs) obtained from this ansatz. Firstly, the explicit form of the Klein-Gordon (KG) equation (\ref{KG-eq}) reads

\begin{eqnarray}
\label{eq-phi}
&&
\phi_{,rr}+\frac{1}{r^2 N}  \phi_{,\theta\theta} 
+\phi_{,r}(F_{0,r}+F_{2,r})+\frac{1}{r^2 N}  \phi_{, \theta} (F_{0,\theta}+F_{2,\theta})
+
\left(1+\frac{rN'}{2N}\right)\frac{2 }{r}\phi_{,r}+\frac{\cot \theta}{r^2 N} \phi_{, \theta}
\\
\nonumber
&&
{~~~~~~~~~~}
-
\left(
\frac{e^{-2F_2} m^2}{r^2\sin^2\theta}
-\frac{e^{-2F_0}(w-m W)^2}{N}+\mu^2
\right)
\frac{e^{2F_1}}{N}\phi=0\ .
\end{eqnarray}

This is a second order PDE for the function $\phi$. Secondly, concerning the Einstein equations \eqref{E-eq}, the non-trivial components are $E_t^t, E_r^r,E_\theta^\theta,E_\varphi^\varphi,E_\varphi^t, E_r^\theta$. These six equations are divided into two groups: four of these equations are solved together with the KG equation \eqref{eq-phi}, yielding a coupled system of five PDEs on the five unknown functions. The remaining two Einstein equations are treated as constraints and used to check the numerical accuracy of the method. 

Each of the four Einstein equations we shall solve simultaneously with \eqref{eq-phi}, should, as \eqref{eq-phi}, have second derivatives of a single function. This is achieved by using the following combinations of the Einstein equations:
 \begin{eqnarray}
 \nonumber
&&
E_r^r+E_\theta^\theta-E_\varphi^\varphi-E_t^t=0\ ,
\\
%
%
\label{EKG-eqs}
&&
E_r^r+E_\theta^\theta-E_\varphi^\varphi+E_t^t+2W E_\varphi^t=0\ ,
%
%
\\
\nonumber
&&
E_r^r+E_\theta^\theta+E_\varphi^\varphi-E_t^t-2W E_\varphi^t=0\ ,
%
%
\\
\nonumber
&&
 E_\varphi^t=0\ .
\end{eqnarray}
These four equations, multiplied by suitable factors, yield, respectively, second order equations for $F_1,F_2,F_0$ and $W$:
\begin{eqnarray}
\nonumber
&&
F_{1,rr}+\frac{1}{r^2 N}  F_{1,\theta\theta} 
-\left(F_{0,r}F_{2,r}+\frac{1}{r^2 N}F_{0, \theta}  F_{2,\theta}\right)
-\frac{e^{-2F_0+2F_2}r^2\sin^2\theta}{4N}\left(W_{,r}^2+\frac{1}{r^2 N}W_{, \theta}^2\right)
-\frac{F_{0,r}}{r}
-\frac{N'F_{2,r}}{2N}
\\
\label{eq-F1}
\nonumber
&&
{~}+\left(1+\frac{r N'}{2N}\right)\frac{F_{1,r}}{r}
-\frac{\cot \theta F_{0,\theta}}{r^2 N}
+8\pi G
\left(
 \phi_{,r}^2+\frac{1}{r^2 N}\phi_{, \theta}^2 
 +\frac{e^{2F_1}}{N^2}
 \left[
 e^{-2F_0}(w-m W)^2-\frac{e^{-2F_2}m^2N}{r^2\sin^2\theta }
 \right]
 \phi^2
\right)
=0\ ,
\end{eqnarray}
\begin{eqnarray}
\nonumber
&&
 F_{2,rr}+\frac{1}{r^2 N}  F_{2,\theta\theta} 
 +F_{2,r}^2+\frac{1}{r^2 N}F_{2, \theta}^2 
+F_{0,r}F_{2,r}+\frac{1}{r^2 N}F_{0, \theta}  F_{2,\theta} 
+\frac{e^{-2F_0+2F_2}r^2\sin^2\theta}{2N}\left(W_{,r}^2+\frac{1}{r^2 N}W_{, \theta}^2 \right)
\\
&&
{~}+\frac{1}{r}
\left(
F_{0,r}+\frac{\cot \theta F_{0,\theta}}{rN}
\right)
+
\left(1+\frac{rN'}{3N}\right)\frac{3F_{2,r}}{r}
+\frac{2\cot\theta F_{2,\theta}}{r^2N}
+8\pi G\frac{e^{2F_1}}{N}
\left(
\mu^2+\frac{2e^{-2F_2}m^2}{r^2\sin^2\theta}
\right)
\phi^2
=0\ ,
\nonumber
\end{eqnarray}
 \begin{eqnarray}
\nonumber
&&
 F_{0,rr}+\frac{1}{r^2 N}  F_{0,\theta\theta} 
 +F_{0,r}^2+\frac{1}{r^2 N}F_{0, \theta}^2 
+F_{0,r}F_{2,r}+\frac{1}{r^2 N}F_{0, \theta}  F_{2,\theta} 
-\frac{e^{-2F_0+2F_2}r^2\sin^2\theta}{2N}\left(W_{,r}^2+\frac{1}{r^2 N}W_{, \theta}^2 \right)
\\
\nonumber
&&
{~}+\left(1+\frac{3rN'}{4N}\right)\frac{2F_{0,r}}{r}
+\frac{\cot\theta F_{0,\theta}}{r^2N}
+\frac{N'F_{2,r}}{2N}
-8\pi G\frac{e^{2F_1}}{N}
\left(
\frac{2e^{-2F_0}(w-mW)^2}{N}-\mu^2
\right)
\phi^2
=0\ ,
\end{eqnarray}
 \begin{eqnarray}
 &&
W_{,rr}+\frac{1}{r^2 N} W_{,\theta\theta}
+(3 F_{2,r}- F_{0,r})W_{,r}
+\frac{1}{r^2 N} (3 F_{2,\theta}- F_{0,\theta})W_{,\theta}
\nonumber
\\
\nonumber
&&
{~}+\frac{4}{r}\left(W_{,r}+\frac{3\cot\theta W_{,\theta}}{4rN}\right)
+32\pi G \frac{e^{2F_1-2F_2}m (w-mW)}{r^2\sin^2\theta N}\phi^2=0\ .
\end{eqnarray}

On the other hand, the two constraint equations are chosen to be
 \begin{eqnarray}
 E_r^r-E_\theta^\theta=0\ ,
\end{eqnarray}
and
 \begin{eqnarray}
 E_r^\theta =0,
\end{eqnarray}
which yield, respectively,
\begin{eqnarray}
\nonumber
&&
F_{0,rr}-\frac{1}{r^2 N}  F_{0,\theta\theta} 
+F_{2,rr}-\frac{1}{r^2 N}  F_{2,\theta\theta} 
+F_{0,r}^2-\frac{1}{r^2 N}  F_{0,\theta}^2
-2\left(
F_{0,r}F_{1,r}-\frac{1}{r^2 N}F_{0, \theta}  F_{1,\theta} 
\right) 
\\
\nonumber
&&
{~}-2\left(
F_{1,r}F_{2,r}-\frac{1}{r^2 N}F_{1, \theta}  F_{2,\theta} 
\right)
-\frac{e^{-2F_0+2F_2}r^2\sin^2\theta}{2N}
\left(
W_{,r}^2-\frac{1}{r^2 N}W_{, \theta}^2
\right)
+
F_{2,r}^2-\frac{1}{r^2 N}F_{2, \theta}^2
\\
\nonumber
&&
{~}+\left(\frac{3rN'}{2N}-1\right)\frac{F_{0,r}}{r}
+\frac{1}{r}\left(1+\frac{rN'}{2N}\right)
(F_{2,r}-2F_{1,r})
+\frac{2\cot\theta}{r^2N}
(F_{1,\theta}-F_{2,\theta})
+16\pi G 
\left(\phi_{,r}^2-\frac{1}{r^2 N}\phi_{, \theta}^2\right)
=0\ ,
\end{eqnarray}
and
\begin{eqnarray}
\nonumber
&&
 F_{0,r\theta}+
 F_{2,r\theta}+
 F_{0,r} F_{0,\theta}+
 F_{2,r} F_{2,\theta}-
 ( F_{0,r} F_{1,\theta}+
 F_{1,r} F_{0,\theta})
 - ( F_{1,r} F_{2,\theta}+
 F_{2,r} F_{1,\theta})
 \\
 \nonumber
 &&
{~} +\left(\frac{rN'}{2N}-1\right)\frac{F_{0,\theta}}{r}
 -\left(1+\frac{rN'}{2N}\right)\frac{F_{1,\theta}}{r}
 -\cot\theta(F_{1,r}-F_{2,r})
 -\frac{e^{-2F_0+2F_2}r^2\sin^2\theta}{2N} W_{,r} W_{,\theta}
 +16\pi G  \phi_{,r} \phi_{,\theta}=0\ .
\end{eqnarray}

In the next section we will address the boundary conditions and the numerical methods used to solve these equations.

\section{Boundary conditions, quantities of interest and numerics}
\label{sec_bc_num}
KBHsSH are asymptotically flat solutions, which are regular on and outside an event horizon.
In order to perform the numerical integration of the system of equations described 
in Section~\ref{sec_eq_motion}, appropriate boundary conditions must be imposed. In our study, we shall not consider the behaviour of the solutions 
inside the event horizon. The boundary conditions we have chosen implement asymptotic flatness and regularity at the horizon and at the symmetry axis. 
Let us describe these boundary conditions in detail.

\subsection{Boundary conditions}
\label{sec_abc}

{\bf~~ Asymptotic boundary conditions.} 
For the solutions to approach,  at spatial infinity described by $r\rightarrow \infty$, a Minkowski spacetime background we require
\begin{equation}
\lim_{r\rightarrow \infty}{F_i}=\lim_{r\rightarrow \infty}{W}=\lim_{r\rightarrow \infty}{\phi}=0.
\end{equation}
For any input parameters,
one can obtain an asymptotic expression of the solution, compatible with these boundary conditions.
 This expression is given in 
given in Appendix~\ref{coef_asym_exp}.

\bigskip


{\bf Axis boundary conditions.} Axial symmetry and regularity impose
the following boundary conditions on the symmetry axis, $i.e.$ at $\theta=0,\pi$:
\begin{equation}
\partial_\theta F_i = \partial_\theta W = \phi = 0.
\end{equation}
Moreover, the absence of conical singularities implies also that 
\begin{equation}
F_1=F_2,
\end{equation}
on the symmetry axis.

Also, all solutions discussed in this work are symmetric $w.r.t.$
a reflection on the equatorial plane\footnote{
We have also found solutions with an anti-symmetric scalar field $w.r.t.$
reflections along the equatorial plane, while the metric functions are still even parity.
Such configurations, however, are hard to study systematically and are likely to be more unstable.
}.
As a result, it is enough to consider 
the range $0\leq \theta \leq \pi/2$
for the angular variable, the functions
$F_i,~W$ and $\phi$; these
satisfy the following boundary conditions on the equatorial plane
\begin{equation}
\partial_\theta F_i\big|_{\theta=\pi/2} = \partial_\theta W\big|_{\theta=\pi/2} =\partial_\theta \phi\big|_{\theta=\pi/2} = 0.
\end{equation}



\bigskip

{\bf Event horizon boundary conditions.} The event horizon is located at a surface with constant 
radial variable $r=r_H>0$.
The boundary conditions there and also the numerical treatment of the problem 
are simplified by introducing a new
radial coordinate 
\begin{equation}
x=\sqrt{r^2-r_H^2}.
\end{equation}
Then a power series expansion near the horizon 
yields  
\begin{eqnarray}
&&
\label{c1}
F_i=F_i^{(0)}(\theta)+x^2 F_i^{(2)}(\theta)+{\cal O}(x^4),
\\
\label{c2}
&&
W=\Omega_H +{\cal O}(x^2),
\end{eqnarray}
and 
\begin{equation}
\label{c3}
\phi=\phi_0(\theta)+{\cal O}(x^2),
\end{equation}
where the constant $\Omega_H>0$ is shown to be the horizon angular velocity, see (\ref{OmegaH}). 
The field equations together with (\ref{c1})-(\ref{c3}) imply that this quantity obeys the condition 
\begin{eqnarray}
\label{cond}
w=m\Omega_H.
\end{eqnarray}
This guarantees that the null geodesic generators of the horizon are tangent to the Killing vector field $\chi$, defined in \eqref{helicoidal}. The physical significance of such identification is that there is no flux of scalar field into the BH, 
\begin{equation}
\chi^{\mu}\partial_\mu \Psi=0,
\end{equation}
which is central to the existence of regular BHs with a stationary geometry and scalar hair. As discussed in the Introduction, the condition \eqref{cond} is also related to the superradiance phenomenon. Furthermore, we note that  the Einstein equation $E_r^\theta=0$ implies that
 the difference $F_1-F_0$ is constant on the horizon.
 In our scheme, however, we do not impose this condition,
 but rather use it as another test of the numerical accuracy of the solutions.

To summarize, the boundary conditions at the horizon are  
\begin{equation}
\partial_x F_i \big|_{r=r_H}= \partial_x \phi  \big|_{r=r_H} =  0, \qquad W \big|_{r=r_H}=\frac{w}{m}.
\end{equation}

\subsection{Quantities of interest and a Smarr relation}
\label{sec_q}
Most of the quantities of interest are 
encoded in the expression for the metric functions at the horizon or at infinity.
Considering first horizon quantities, we introduce the 
Hawking temperature $T_H={\kappa}/({2\pi})$, where $\kappa$ is the surface gravity
defined as $\kappa^2=-\frac{1}{2}(\nabla_a \chi_b)(\nabla^a \chi^b)|_{r_H}$, and the event horizon area $A_H$ of KBHsSH. These are computed as
\begin{eqnarray}
\label{THAH}
&&
T_H=\frac{1}{4\pi r_H}e^{F_0^{(2)}(\theta)-F_1^{(2)}(\theta)} \ ,
\\
&&
A_H=2\pi r_H^2 \int_0^\pi d\theta \sin \theta~e^{F_1^{(2)}(\theta)+F_2^{(2)}(\theta)} \ .
\end{eqnarray}
Moreover, $S={A_H}/({4G})$ is, as usual, the BH entropy.
Also, 
the event horizon velocity $\Omega_H$ is fixed by the horizon value of the metric function $W$,
\begin{eqnarray}
\label{OmegaH}
\Omega_H=-\frac{\xi^2}{\xi\cdot \eta}=-\frac{g_{\varphi t}}{g_{tt}}\bigg|_{r_H}=W \bigg|_{r_H}.
\end{eqnarray}

The ADM mass $M$ and the angular momentum $J$ are read from 
the asymptotic sub-leading behaviour of the metric functions:
\begin{eqnarray}
\label{asym}
g_{tt} =-e^{2F_0}N+e^{2F_2}W^2r^2 \sin^2 \theta
=-1+\frac{2GM}{r}+\dots,~~g_{\varphi t}=-e^{2F_2}W r^2 \sin^2 \theta=-\frac{2GJ}{r}\sin^2\theta+\nonumber \dots. \\
\end{eqnarray}

As usual in (asymptotically flat) BH mechanics, 
the temperature, entropy and the global charges
are related through a Smarr mass formula \cite{Townsend:1997ku},
which for the KBHsSH reads
\begin{eqnarray}
\label{smarr}
M=2 T_H S +2\Omega_H (J-m Q)+ M^\Psi,
\end{eqnarray}
where $M^\Psi$, given by
\begin{eqnarray}
\label{Mpsi}
-M^\Psi\equiv  \int_{\Sigma} dS_a (2T_{b}^a \xi^b-T\xi^a)=
 4\pi \int_{r_H}^\infty dr \int_0^\pi d\theta~r^2\sin \theta ~e^{F_0+2F_1+F_2}
 \left(
 \mu^2-2 e^{-2F_2}\frac{w(w-mW)}{N}
 \right)\phi^2  ~~
\end{eqnarray}
is the scalar field energy $outside$ the BH, and 
\begin{eqnarray}
\label{Q-int}
Q=4\pi \int_{r_H}^\infty dr \int_0^\pi d\theta 
~r^2\sin \theta ~e^{-F_0+2F_1+F_2}  \frac{m(w-mW)}{N}\phi^2 \ ,
\end{eqnarray}
is the conserved  Noether charge  (\ref{Q}).

Some of these physical quantities are also connected via the first law
\begin{eqnarray}
\label{first-law}
dM=T_H dS +\Omega_H dJ .
\end{eqnarray}

A natural question in the context of the hairy BHs we are discussing is how much ADM energy and angular momentum is \textit{in} the BH and how much is in the scalar hair \textit{outside} the BH. This question can be addressed by noting that the ADM quantities $M$ and $J$ can be expressed as
\begin{eqnarray}
\label{MH-hor}
 M=M^\Psi+M_H,~~ J=mQ+J_H,
\end{eqnarray}
where $M_H$ and $J_H$ are the horizon mass and angular momentum, computed as Komar integrals, which, from~\re{smarr} and \re{MH-hor}, satisfy the relation
\begin{eqnarray}
\label{rel-hor}
M_H=2T_H S+2 \Omega_H J_H.
\end{eqnarray}

\subsection{Numerical implementation}
\label{sec_numerics}

As usual when dealing with gravitating massive scalar fields, 
the numerical integration is performed 
with dimensionless variables
introduced
by using
natural units set by $\mu$ and $G$,
\begin{eqnarray}
r\to r \mu,~~
\phi \to \phi M_{Pl}/\sqrt{4\pi},~~
w \to w/\mu,
 \end{eqnarray}
 where $M_{Pl}^2=G^{-1}$ is the Planck mass.
 As a result, the dependence on both $G$ and $\mu$
 disappears from the equations.
 Also, the global charges and all other quantities of interest are 
 expressed in units set by $\mu$ and $G$ (note that,
 in order to simplify the output,
  we set $G=1$ in what follows).

In our approach, the EKG equations reduce to a set of five 
coupled non-linear elliptic partial differential equations for the functions 
${\cal F} =(F_0, F_1, F_2, W; \phi)$,
which are displayed in Section~\ref{sec_eq_motion_eom}. 
These equations 
 have been solved numerically subject to the boundary conditions 
 introduced above.
 An important issue here concerns the status of the two constraint equations $E_\theta^r =0,~E_r^r-E_\theta^\theta  =0$, 
 also presented in Section~\ref{sec_eq_motion_eom}. This is addressed following Ref. \cite{Wiseman:2002zc}.
 One notes that
the Bianchi identities $\nabla_\mu E^{\mu r} =0$ and $\nabla_\mu E^{\mu \theta}=0$, 
imply the Cauchy-Riemann relations
$
\partial_{\bar r} {\cal R}_2  +
\partial_\theta {\cal R}_1  
= 0 ,~~
 \partial_{\bar r} {\cal R}_1  
-\partial_{\theta} {\cal R}_2
~= 0 ,
$
where we have defined ${\cal R}_1\equiv \sqrt{-g} E^r_\theta$, ${\cal R}_2\equiv \sqrt{-g}r \sqrt{N}(E^r_r-E^\theta_\theta)/2$
and the new variable $d\bar r\equiv \frac{dr}{r \sqrt{N}}$.
Therefore the weighted constraints still satisfy Laplace-type equations.
Then they are obeyed  when one of them is satisfied on the boundary and the other 
at a single point
\cite{Wiseman:2002zc}. 
From the boundary  conditions we are imposing,
it turns out that this is indeed the case,
 $i.e.$ the numerical scheme is consistent.

Our numerical treatment can be summarized as follows.
The first step is to introduce a new radial variable  
$\bar x=x/(1+x)$ 
which maps the semi--infinite region $[0,\infty)$ to the finite region $[0,1]$
[we recall $x=\sqrt{r^2-r_H^2}$, with $r$ the radial variable in 
the line element 
(\ref{ansatz})].
This involves the following substitutions in the differential equations
\begin{eqnarray}
x {\cal F}_{,x}   \longrightarrow   
 (1-\bar x) {\cal F}_{,\bar x},
~~~
x^2 {\cal F}_{,x x}   \longrightarrow
(1-\bar x)^2    {\cal F}_{,\bar x \bar x}
  - 2 (1-\bar x) {\cal F}_{,\bar x }.
\end{eqnarray}
 
Next, the equations for ${\cal F}$
are discretized on a grid in $\bar x$ and $\theta$. 
Various grid choices have been considered,  
the number of grid points ranging between $280 \times 20$ and $90 \times 70$.
 Most of the results in this work have actually been found for  
 an equidistant grid with $250 \times 30$ points. 
The grid covers the integration region
$0\leq \bar x \leq 1$ and $0\leq \theta \leq \pi/2$.  

All numerical calculations have been 
performed by using the professional package  \textsc{fidisol/cadsol} \cite{schoen},
 which uses a finite difference method with 
an arbitrary grid and arbitrary consistency order.
This package has been
extensively tested in the past by recovering numerous exact solutions in general relativity and field theory. Furthermore, some of the new solutions derived by using \textsc{fidisol/cadsol} were rederived
subsequently by other groups with different numerical methods.
  
This code requests the system of nonlinear partial differential equations 
to be written 
in the form
$
P(\bar x,\theta;{\cal F};{\cal F}_{\bar x},{\cal F}_{\theta};
{\cal F}_{\bar x \theta},{\cal F}_{\bar x \bar x},{\cal F}_{\theta\theta})=0,
$
subject to a set of boundary conditions on a rectangular domain. 
Besides  that, \textsc{fidisol/cadsol} requires
 the Jacobian matrices for the equations $w.r.t.$ the
 functions ${\cal F}$ and their first and second derivatives,
the boundary conditions, as well as some initial guess for the functions ${\cal F}$. 
Indeed, this solver uses a Newton-Raphson method,
which requires a good first guess in order
to start a successful iteration procedure.
Also, this software package provides error estimates for each unknown function,
which allows judging the quality of the computed solution.
The  numerical error
for the solutions reported in this work is estimated to be typically  $<10^{-3}$. 
However, errors increase dramatically when studying
solutions close to central inspiralling region of the BSs curve, see Figure~\ref{kbhsh_fig} below (left panel).
A detailed description of  the numerical method and explicit examples are provided in
\cite{schoen}. 

As a further check of numerics, we have verified that the families of solutions 
with a varying frequency satisfy  with a very good accuracy
the first law of
thermodynamics (\ref{first-law})
and also the Smarr relation (\ref{smarr}).

In the scheme we have used, there are three input parameters: 
${\bf i)}$ the 
frequency $w$ and
 ${\bf ii)}$ the winding number $m$ in the ansatz (\ref{scalar_ansatz}) for the scalar field $\Psi$,
together with  ${\bf iii)}$ the event horizon radius $r_H$ in the metric ansatz (\ref{ansatz}).
The number of nodes $n$ of $|\phi|$
on the equatorial plane, as well as all other quantities of interest ($e.g.$ mass, angular momentum, Noether charge, Hawking temperature and horizon area) are computed from the numerical
solution. Both here and in~\cite{Herdeiro:2014goa}, for simplicity, we have restricted our study to fundamental configurations, $i.e.$
with a nodeless scalar field, $n=0$. 
Excited states are more difficult
to investigate systematically and, in any case, are expected to be more unstable. 
Also, we have studied in a systematic way the BH solutions with $m=1$;
sets of solutions with $m=2,3,4$
have also been constructed for several fixed values of $w$. 

In some of the calculations, we interpolate the resulting
configurations on points between the chosen grid points,
and then use these for a new guess on a finer grid.
Finally, the compilation of the numerical output is done by using the software \textsc{mathematica}.
 
Let us close this section with some technical details 
on the systematic procedure we have used to scan the parameter space of the KBHsSH solutions.
We start with
a spinning BS solution -- $i.e.$ having $r_H=0$ -- with given $m,w$, as initial guess
for a KBHSH with a small event horizon radius.
Then we  slowly  increase
the value of $r_H$, keeping $m,w$ fixed (thus also the event horizon velocity $\Omega_H$).
The iterations converge, and, in principle, repeating the procedure we obtain in this
way solutions for higher and higher values of $r_H$.
Then, for the coordinate system we are using,
 a maximal value of $r_H$ is a approached  and
a second branch of solutions emerges, extending backwards in $r_H$.
The limiting behaviour of this secondary branch of  KBHsSH
depends on the value of the input parameters $w,m$.
We find numerical evidence for the existence of  
 three different possible limiting configurations on this secondary branch; they can be: 
{\bf i)} another 
BS solution  with $r_H =0$; 
{\bf ii)}  an extremal  KBHSH (which is also approached as $r_H\to 0$), and, 
{\bf iii)}  a special set of vacuum Kerr solutions with $r_H>0$. 
Thus, in principle, the full set of  KBHsSH with a given $m$
can be constructed in this way, by repeating this procedure for different
values of  the scalar field frequency $w$.
The results reported in this work are obtained from around three thousand
solution points.
For all these solutions we have monitored the Ricci and the Kretschmann scalars,
and, at the level of the numerical accuracy,
we have not observed any sign of a singular behaviour.

\section{KBHsSH: exterior space-time and scalar field structure}
\label{sec_solutions}
The solutions obtained by solving the equations exhibited in Section~\ref{sec_eq_motion} with the boundary conditions and numerical method discussed in Section~\ref{sec_bc_num} form a ${\bf 5}$-parameter family. Three of these parameters are continuous: 
{\bf i)} the ADM mass $M$, 
{\bf ii)} the ADM angular momentum $J$ and {\bf iii)} the Noether charge $Q$. 
The remaining two parameters are discrete: 
{\bf iv)} the azimuthal hamonic index $m\in \mathbb{Z}$, appearing in (\ref{scalar_ansatz}) and 
{\bf v)} the number of nodes (along the equatorial plane) of the scalar function $\phi(r,\theta)$, $n\in \mathbb{N}_0$. 
Observe that the latter does not appear explicitly in the ansatz. Observe also that this 5-parameter family contains both horizonless solutions -- BSs -- and hairless solutions -- Kerr BHs. To describe these two limiting solutions it is convenient to introduce the normalized Noether charge
\begin{equation}
q\equiv \frac{Q}{mJ} \ .
\end{equation}
Then, the domain of solutions consists of $q\in \, [0,1]$, with Kerr BHs corresponding to $q=0$ and BSs to $q=1$. This latter statement will be expanded below. To contextualize KBHsSH let us start by
 introducing two ingredients:
   {\bf i)}  the BSs that arise in the $q=1$ limit of this family; and {\bf ii)} a test scalar field in the Kerr geometry, which arises in the $q=0$ limit.

\subsection{Solitonic limit: boson stars ($q=1$)}
In this case $r_H=0$ and the horizon is replaced with a regular origin.
The corresponding solutions are well known in the literature -- 
 BSs --, arguably the physically most interesting gravitating solitons.
BSs have found a variety of applications, being considered
as possible BH mimickers and dark matter candidates -- see $e.g.$, for a recent review  \cite{Liebling:2012fv}.

The study of BSs was initiated by the work of Kaup \cite{Kaup:1968zz} and 
Ruffini and Bonazzala \cite{Ruffini:1969qy} more than 40 years ago.
They found globally regular, asymptotically flat, equilibrium solutions
of the Einstein equations coupled with a
massive complex scalar field,
 providing an explicit realization of Wheeler's geons 
 \cite{Wheeler:1955zz}
 in EKG theory.
The BSs were regarded as {\it `macroscopic quantum states'},    
which are prevented 
from gravitationally collapsing by Heisenberg's uncertainty principle.

The BSs studied in the pioneering works
\cite{Kaup:1968zz,Ruffini:1969qy}  
are  spherically symmetric.
Rotating generalizations,
were first studied in the work of 
 Schunck and  Mielke
\cite{Schunck:1996he}
where such configurations were constructed
in the weakly relativistic regime for a large range of
winding numbers $m$.
Highly relativistic spinning BSs have been constructed
for the first time by 
Yoshida and  Eriguchi
\cite{Yoshida:1997qf} for winding numbers $m=1,2$.
These results have been extended recently~\cite{Grandclement:2014msa},
in particular by constructing
higher winding number solutions.
For completeness, let us mention that 
spinning BSs with a self--interacting potential of the  $Q$-ball type (see the discussion of Section~\ref{sec_conclusions})
have been studied in~\cite{Kleihaus:2005me,Kleihaus:2007vk}, where odd-parity solutions were also first addressed.

In the present context, rotating BSs are a particular limit of KBHsSH; as such they form part of the boundary of their domain of existence. 
Therefore we have performed an independent study of their
 properties for $m=1,2,3$,
by using the same methods described above,
$i.e.$
by solving a boundary value problem for ${\cal F}$, with two
input parameters: $w$ and $m$.
The  ansatz is again
(\ref{ansatz}) (with $r_H=0$) together with (\ref{scalar_ansatz}).
The boundary
conditions at the origin, $r=0$ read\footnote{Note that, however,
most of the BS solutions have been computed with $W=\bar W/r$.
 }
\begin{eqnarray}
\label{r=0}
\partial_r F_1=\partial_r F_2=\partial_r F_0=\partial_r W =\phi=0.
\end{eqnarray} 
The boundary conditions as $r\to \infty$
and at $\theta=0,\pi$
are similar to those that apply to the KBHsSH case and that were described in Section~\ref{sec_bc_num}.

\begin{figure}[h!]
\begin{center}
\includegraphics[height=.25\textheight, angle =0]{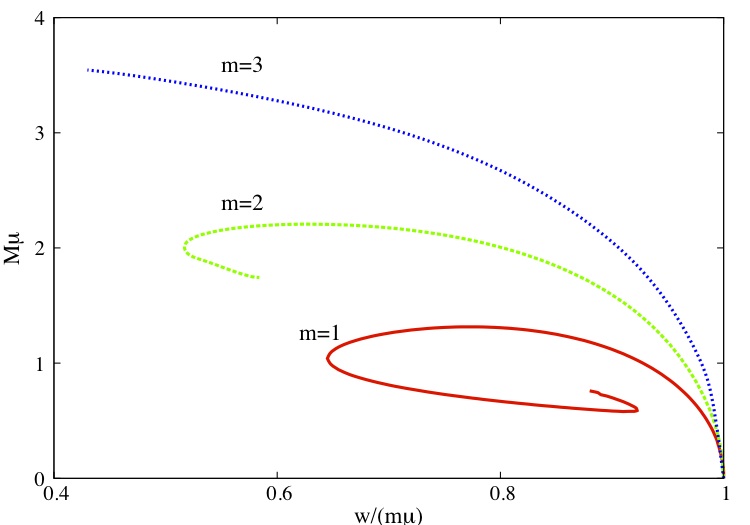} \ \ \ 
\includegraphics[height=.25\textheight, angle =0]{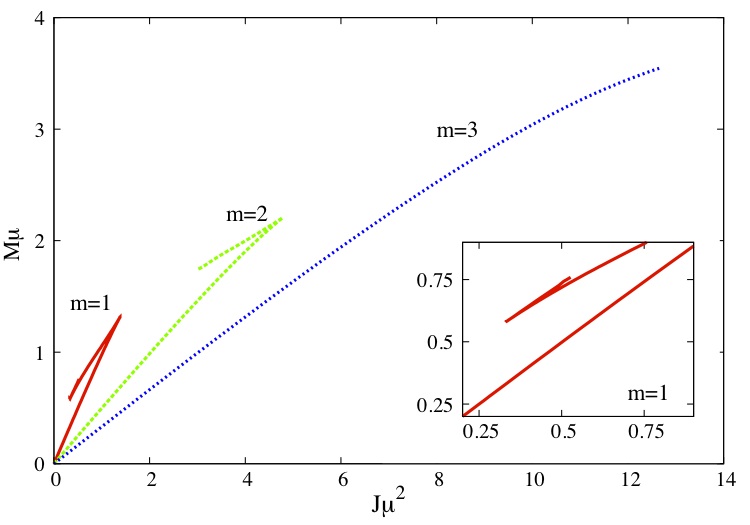}
\end{center}
\caption{Rotating BSs with $m=1~,2,~3$ and $n=0$ in an ADM mass $vs.$ scalar field frequency diagram (left panel) 
and an ADM mass versus ADM angular momentum diagram (right panel).}
\label{f-1}
\end{figure}

In Figure \ref{f-1} we illustrate the BSs solutions. 
In the left panel we can see the ADM mass $M$ $vs.$ the frequency $w$ distribution for BSs with $m=1,2,3$, represented by the solid red, dashed green and dotted blue lines, respectively. 
We focus on nodeless solutions, $n=0$, since these are typical the most stable ones, and we use natural units set by $\mu$. 
On this plot we see that BSs exist for $w<\mu$; this is a bound state condition. As we decrease the frequency the mass increases until a maximum value. This value is of the order of $1/\mu$ (or $M_{Pl}^2/\mu$, reintroducing the Planck mass $M_{Pl}$). Thus in order to have such a BS with the mass of the Sun we would need extremely light scalar particles, with masses around $10^{-11}$ eV. While such light scalars have been suggested in string compactifications - the Axiverse~\cite{Arvanitaki:2009fg} -, more reasonable masses lead to so called mini-BSs~\cite{Schunck:2003kk}. Further decreasing $w$ one finds a minimal frequency  (seen for both $m=1$ and $2$; for $m=3$ the data collected did not reach the minimal value of $w$ but we expect a similar pattern) below which no BS solutions are found. The BS curve then seems to spiral towards a central region of the diagram where numerics become increasingly challenging. Qualitatively, this is also the behaviour
found for spherically symmetric BSs, $m=0$, in which case, a detailed investigation of the inspiraling behaviour was possible.

In the right panel of Figure \ref{f-1}, we exhibit the ADM mass $M$ $vs.$ the ADM angular momentum $J$ distribution for the same BSs.  This curve zig-zags, as can be seen for $m=1$ and partly for $m=2$. Each branch of this zig-zag pattern corresponds to a branch of the curve on the left panel where the mass  increases or decreases. In Section~\ref{kbhsh_sec} we will see how KBHsSH fit in these two diagrams. 

Let us remark on the conserved Noether charge carried by the BSs. As observed above, this cannot be transformed into a flux at infinity; it is given by an integral over a space-like slice of the time component of the 4-current. A simple calculation shows that the angular momentum carried by rotating BSs relates to the Noether charge as $J=mQ$~\cite{Schunck:1996he,Yoshida:1997qf,Kleihaus:2005me}. Thus $q=1$ for BSs as advertised above. This will be useful in parameterizing KBHsSH.

\begin{figure}[h!]
\begin{center}
\includegraphics[height=.28\textheight, angle =0]{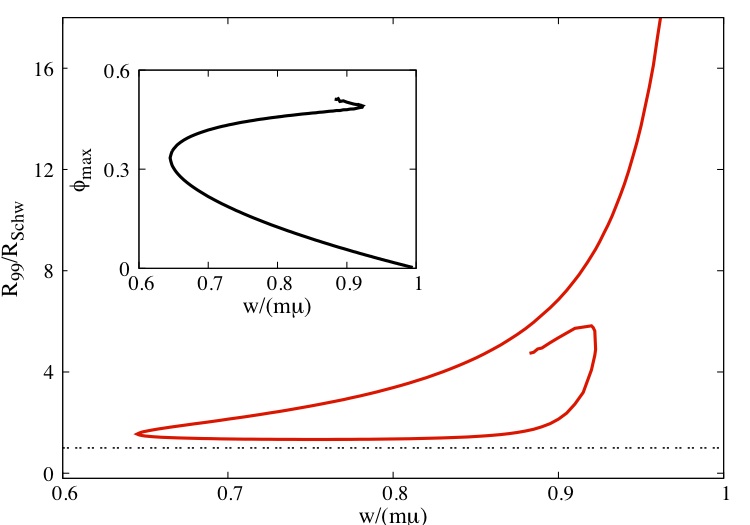} \ \ \ 
\end{center}
\caption{Inverse compactness of BSs with $m=1$. The inset shows the maximal value of the scalar field along the BS line.}
\label{effective_radius}
\end{figure}

Finally, one may wonder how `compact' these BSs are. Since BSs have no surface, $i.e.$ the scalar field decays exponentially towards infinity, $cf.$~(\ref{scalar_asy}), there is no unique definition of the BS's `radius'. One estimate is provided in the following way. Firstly, we note that the `perimeteral' radius, $i.e.$, a radial coordinate $R$ such that a circumference along the equatorial plane has perimeter $\simeq 2\pi R$, is related to the radial coordinate used in~\re{ansatz} as $R=e^{F_2}r$. Secondly, we compute $R_{99}$, the perimeteral radius containing 99\% of the BS mass, $M_{99}$. Finally, we define the inverse compactness by comparing $R_{99}$ with the Schwarzschild radius associated to 99\% of the BS's mass, $R_{Schw}=2M_{99}$
\cite{AmaroSeoane:2010qx}:
\begin{equation}
{\rm Compactness}^{-1}\equiv  \frac{R_{99}}{2M_{99}} \ .
\label{compactness}
\end{equation}
The result for the inverse compactness of BSs with $m=1$ is exhibited in Figure~\ref{effective_radius}. With this measure, the inverse compactness is always greater than unity; in other words, BSs are less compact than BHs, as one would expect.

As can be seen from Figure~\ref{effective_radius}, the least compact BSs are the ones closer to vacuum, which can be orders of magnitude less compact than a BH with the same total mass. At maximal mass, BSs are already only a few times less compact than a BH and during the whole spiral remain very close to the compactness of a BH. In the inset of the figure one can also see the maximal value of the scalar field along the BS line. For each solution this value occurs at a different radial coordinate. Observe that $\phi_{\rm max}$ increases monotonically along the BS curve; thus it is non-degenerate and could be used to label uniquely BS solutions. Indeed, historically, spherically symmetric BS solutions were labelled by the value of $\phi$ at the origin. For rotating BS solutions this value is zero, but $\phi_{\rm max}$ could be used as a label.

\subsection{Kerr limit: scalar clouds ($q=0$)}
\label{sec_scalar_clouds}
In this case $\Psi$ is `small' and we can linearize the coupled EKG equations (\ref{E-eq}) in $\Psi$; these become simply $R_{\mu\nu}=0$ and $\Box \Psi=\mu^2\Psi$. Thus we are led to study the KG equation for a complex massive scalar field on a vacuum solution of the Einstein equations. We consider the Kerr solution. As we shall see, a particular subset of Kerr solutions defines another limit of KBHsSH and form another part of the boundary of their domain of existence. 

Consider the KG equation on the Kerr background in Boyer-Lindquist coordinates and separate variables in the standard fashion: $\Psi=e^{-i(wt-m\varphi)}S_{lm}(\theta)f(r)$, where $S_{lm}$ are spheroidal harmonics. One obtains a linear, second order, ordinary differential equation for $f(r)$~\cite{Brill:1972xj,Teukolsky:1972my,Teukolsky:1973ha,Press:1973zz,Herdeiro:2014goa}. One may then search for bound-state type solutions of this equation.
Requiring that the radial function decays exponentially with $r$, however, yields generically \textit{quasi-bound} states which have a complex, rather than real, frequency. The imaginary part of the frequency manifests that, in general, the scalar field cannot be in equilibrium with the BH; in fact one would expect it to fall into the BH, and this is precisely what one finds if the background is a Schwarzschild BH. In this case the imaginary part of the frequency is always negative corresponding to the scalar field decaying into the BH. 
In the Kerr case, however, it turns out that there is a critical frequency
given by the product of the azimuthal harmonic index, $m$, and the horizon angular velocity,
$\Omega_H$: $w_c=m\Omega_H$. It defines 3 qualitatively different cases.
If the real part of the frequency is larger than the critical frequency, $\mathcal{R}(w)>w_c$, then the quasi-bound state decays
with time. This is the behaviour described above for Schwarzschild and it is the typical behavior expected due to the purely ingoing boundary
condition at the horizon.
If $\mathcal{R}(w)<w_c$, however, the quasi-bound state grows in
time, signaling an instability. This is the superradiant instability of Kerr BHs in the presence
of a massive scalar field~\cite{Press:1972zz,Damour:1976kh,Zouros:1979iw,Detweiler:1980uk,Cardoso:2013krh,Dolan:2012yt}.
Precisely when the frequency equals the critical frequency there are true bound states,
with real frequency and a time independent energy-momentum tensor. These are referred to as \textit{scalar
clouds} around Kerr BHs~\cite{Hod:2012px,Hod:2013zza,Herdeiro:2014goa,Hod:2014baa,Benone:2014ssa}.

\begin{figure}[h!]
\begin{center}
\includegraphics[height=.28\textheight, angle =0]{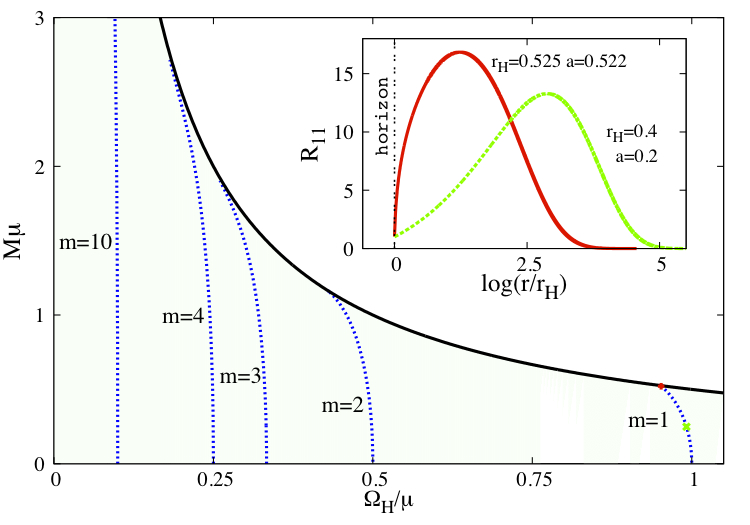} \ \ \ 
\end{center}
\caption{Existence lines for scalar clouds around Kerr BHs. The inset shows the radial profile along the equator for two clouds with $n=0$, $m=l=1$. The colors correspond to the points with the same color in the existence line. The clouds are regular at the horizon, they attain a maximum and decay exponentially towards infinity. From~\cite{Herdeiro:2014goa}.}
\label{clouds}
\end{figure}

Scalar clouds around Kerr BHs form a discrete set labelled by three quantum
numbers $(n,l,m)$, where $n$ is the number of nodes of the scalar field. An analytic treatment can be made for extremal Kerr BHs~\cite{Hod:2012px}. In this case the existence of scalar clouds yields 
a quantization condition determining one physical possible (physical) value of the BH mass, which  determines the angular momentum. For the non-extremal case, the clouds with a fixed $(n,l,m)$ exist for a 1-parameter subspace of the 2-dimensional Kerr parameter space. This is shown in Figure~\ref{clouds}, where we exhibit a mass   \textit{vs.} horizon angular velocity diagram for Kerr BHs. The latter exist below the black line which corresponds to extremal Kerr. The blue dotted lines correspond to the backgrounds that support scalar clouds with $n=0$ and $m=l$, for $m=1,2,3,4$ and $10$.

It is important to observe that, fixing $m$ and $n$, the scalar cloud whose existence line occurs for smaller angular velocity -- for fixed M -- is the line with $m=l$~\cite{Herdeiro:2014goa,Benone:2014ssa}. This line divides the Kerr parameter space into Kerr BHs which are unstable against some scalar modes with that value of $m$ and Kerr BHs that are stable against all scalar modes with that value of $m$. Fixing $m$, therefore, this line, plays a special role and we shall call it the \textit{fundamental existence line} for the azimuthal harmonic index $m$.

\subsection{KBHsSH solutions ($0<q<1$)}
\label{kbhsh_sec}

\subsubsection{Overview}
\label{sec_over}
\begin{figure}[h!]
\begin{center}
\includegraphics[height=.25\textheight, angle =0]{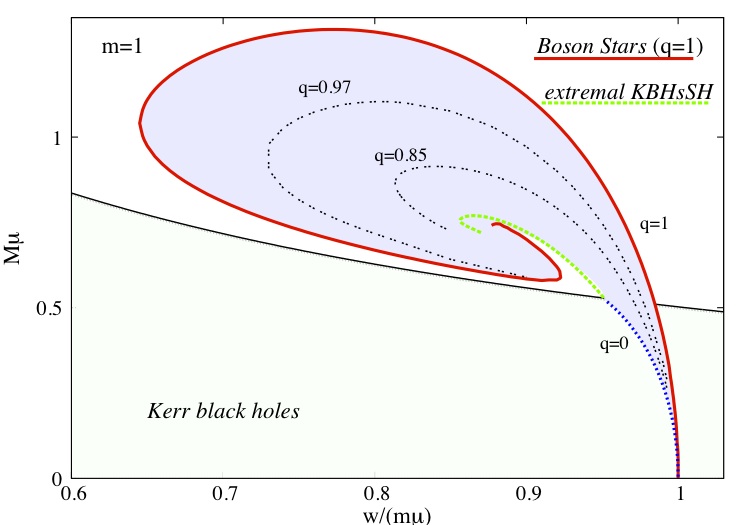} \ \ \ 
\includegraphics[height=.25\textheight, angle =0]{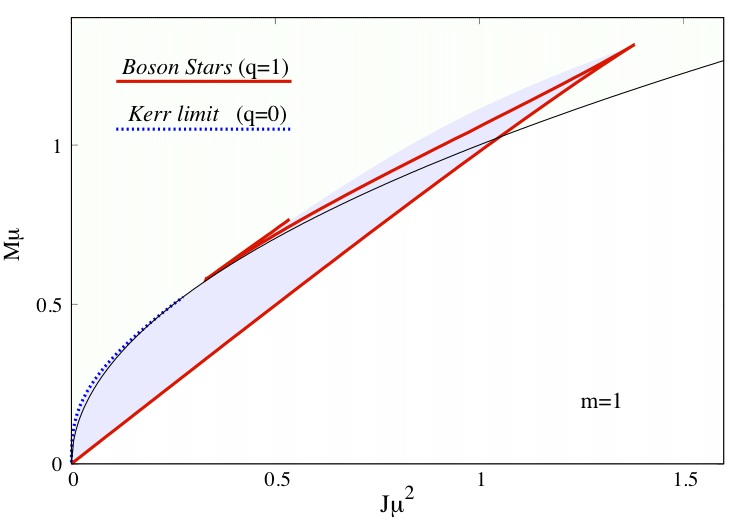}
\end{center}
\caption{Domain of existence for KBHsSH with $m=1$ and $n=0$ (shaded blue region) in an ADM mass $vs.$ scalar field frequency diagram (left panel) and an ADM mass $vs.$ ADM angular momentum diagram (right panel). Improved from~\cite{Herdeiro:2014goa}; note, in particular that the extremal KBHsSH line contains more data then in~\cite{Herdeiro:2014goa}. }
\label{kbhsh_fig}
\end{figure}

Let us start with an overview of the full space of KBHsSH solutions. 
In Figure~\ref{kbhsh_fig} we fill in the plots shown in Figure~\ref{f-1} 
with the domain of existence of KBHsSH -- the shaded blue region, for solutions with $n=0$, $m=1$. 
Let us mention that
here and in figures below, the domain was obtained by extrapolating to the continuum the results from discrete sets of (thousands of) numerical solutions, $cf.$ Section~\ref{sec_numerics}.
This can safely be done for most of the parameter space;
however, we do not exclude a more complicated picture for a small region around the center of the BS spiral,
which is more difficult to explore numerically.  

In the left panel of Figure~\ref{kbhsh_fig} we see the domain of existence in the ADM mass $vs.$ scalar field frequency diagram. The region where KBHsSH exist is delimited by:
\begin{itemize}
\item[i)]  the BS curve already discussed in Figure~\ref{f-1} -- where $q=1$;
\item[ii)] by the subset of Kerr solutions that support the fundamental existence line of scalar clouds with $n=0$, $m=1$ -- where $q=0$. 
In particular, this demonstrates that these hairy BHs are the non-linear realization of scalar clouds;
\item[iii)] and by an yet unseen (green dashed) curve, corresponding to \textit{extremal (i.e. zero temperature) KBHsSH}. We have evidence, but not a complete analysis that along this line $q$ varies from zero
to one and that it is the end point of all constant $q$ lines, which inspiral in a similar way to
the BS line. We shall further comment on these extremal solutions in Section~\ref{sec_extremal}.
\end{itemize}

Based on the existing numerical data, we are confident that the same pattern for the domain of existence of KBHsSH occurs for other values of $m$. In Figure~\ref{f-1} (see also~\cite{Herdeiro:2014goa}) the BS lines for $m=1,2,3$ are also exhibited. 

In the right panel of Figure~\ref{kbhsh_fig} we return to the ADM mass $vs.$ angular momentum plot. In Figure~\ref{f-1} only the red solid 
curve had been shown, corresponding to BSs with $m=1$. Kerr BHs exist in the upper part of
the diagram, above the black solid line, which corresponds to extremal Kerr and KBHsSH exist in the blue shaded area. As before, the dotted blue line is the Kerr limit,
corresponding to $q=0$. Three general observations can be made:
\begin{itemize}
\item[a)] The first observation is that KBHsSH can violate the Kerr bound, since there are
solutions below the black solid line. This is not surprising, since it is known that BSs can violate this bound~\cite{Ryan:1996nk}. Since KBHsSH are continuously connected to BSs one
would expect the same to occur -- at least some solutions with $q$ close to one -- and that is
precisely what we see in the plot. But one may wonder if the Kerr bound is still violated in terms of the \textit{horizon} angular velocity and mass. This question turns out to have interesting implications and will be discussed in detail elsewhere~\cite{HR_overspinning}.
\item[b)] The second observation is that there are KBHsSH with the same mass and
angular momentum as Kerr BHs. In this sense, and because $M,J$ are the only
asymptotic charges, there is non-uniqueness. Further specifying $q$, however, seems to
completely raise the degeneracy. At least we found no evidence that there are two distinct
solutions with the same $(M,J,q)$.
\item[c)] The third observation is that, in contrast to the vacuum case, KBHsSH do not possess a static limit.\footnote{Which is in agreement with the theorem in~\cite{Pena:1997cy}.}
A lower bound for the horizon angular velocity is set by the minimal
value of the scalar field frequency of the corresponding BSs.
Also, the maximal value of the BSs mass and angular momentum set an upper bound for the
global charges of KBHsSH.
\end{itemize}

\begin{figure}[h!]
\begin{center}
\includegraphics[height=.25\textheight, angle =0]{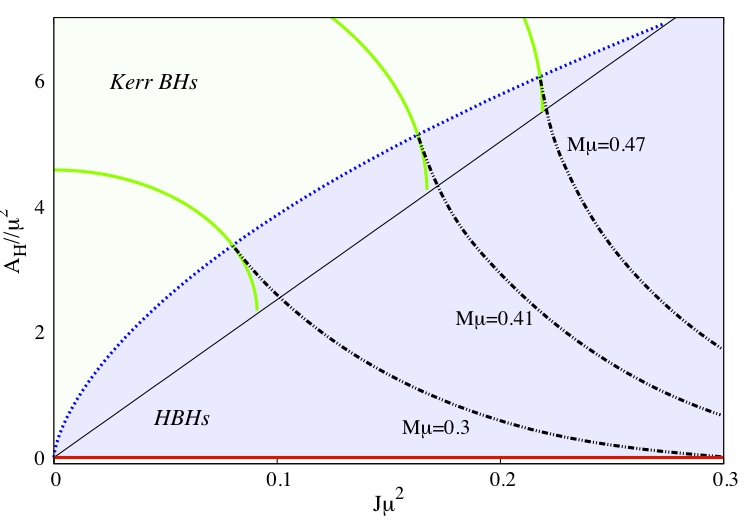} \ \ \ 
\includegraphics[height=.25\textheight, angle =0]{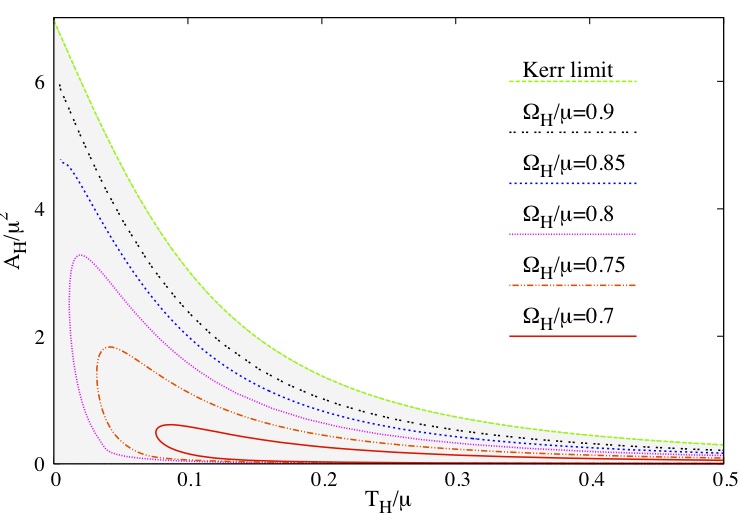}
\end{center}
\caption{(Left panel) Domain of existence of KBHsSH (shaded blue region) in a BH area $A_H$ $vs.$ an ADM angular momentum $J$ diagram. Kerr BHs exist above the black solid line; the blue dotted line corresponds to the fundamental existence line for scalar clouds and the red solid line to the BS limit. We display  curves with constant ADM mass $M$ connecting Kerr BHs (solid green) and KBHsSH (dashed black). In the region of non-uniqueness -- above the black solid line and below the blue dotted line -- the KBHsSH always have larger area for the same $M$ and $J$. (Right panel) Domain of existence of KBHsSH (shaded blue region) in a BH area $A_H$ $vs.$ BH temperature $T_H$ diagram. The solid green line corresponds to the Kerr limit, being delimited by extremal BHs ($T_H=0$) and arbitrarily small BHs ($T_H\rightarrow \infty$). Along constant $\Omega_H$ lines (or equivalently, constant $w$ lines), KBHsSH can either get to extremal -- as for $\Omega_H/\mu=0.9$ or become arbitrarily small and hence high temperature at both limits of the line, $cf.$ Figure~\ref{kbhsh_fig} (left panel).}
\label{AH-qJ}
\end{figure}

In the region of non-uniqueness one can compare the area or entropy of the Kerr BHs and KBHsSH with the same $M,J$~\cite{Herdeiro:2014goa}.  One
observes that in the common region KBHsSH are entropically favoured which
means that they cannot decay adiabatically to Kerr BHs - Figure~\ref{AH-qJ} (left panel) .

\begin{figure}[h!]
\begin{center}
\includegraphics[height=.25\textheight, angle =0]{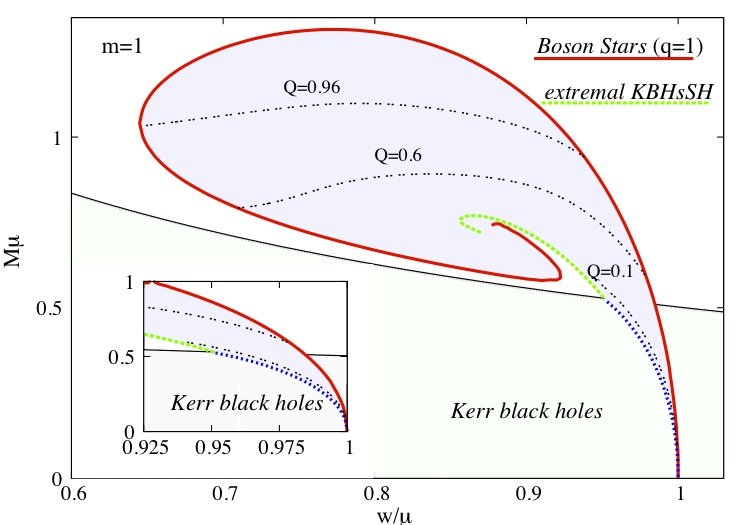} \ \ \ 
\includegraphics[height=.25\textheight, angle =0]{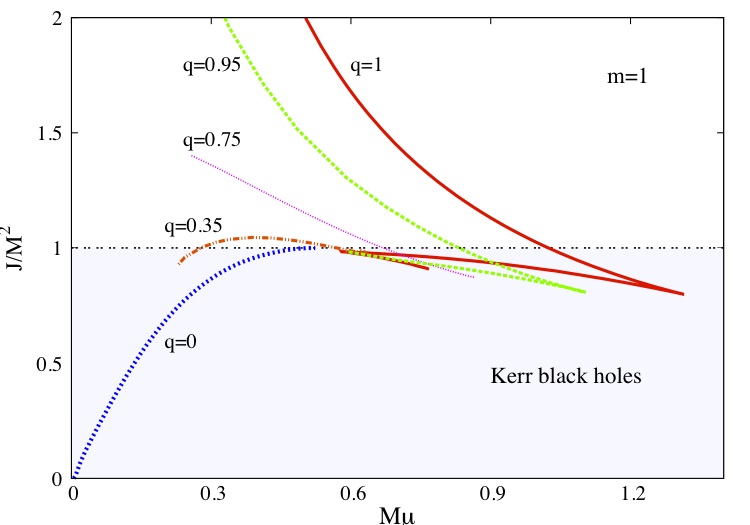}
\end{center}
\caption{(Left panel) Constant $Q$ KBHsSH in a ADM mass $vs.$ scalar field frequency diagram. (Right panel) Constant $q$ KBHsSH in a $J/M^2$ $vs.$ ADM mass diagram.}
\label{BH1}
\end{figure}

In Figure~\ref{BH1} (left panel) we exhibit again the ADM mass $vs.$ scalar field diagram, but now showing  constant $Q$ lines. We recall that $Q$ is a measure of the amount of scalar field outside the BH. Comparing with Figure~\ref{kbhsh_fig} (left panel) we observe that constant $Q$ lines span the domain of existence of KBHsSH in a very different way to constant $q$ lines, which inspiral in a similar way to the BS curve; the former interpolate between two BS solutions and roughly, more massive KBHsSH have also larger $Q$. Note that here the mass is the ADM mass, as such taking into account both the BH (horizon) mass and the energy of the scalar field outside the horizon, $cf.$~\re{MH-hor}. In Figure~\ref{BH1} (right panel) we exhibit lines of constant $q$ KBHsSH in a $J/M^2$ $vs.$ ADM mass diagram. In this `phase space' KBHsSH can coexist with Kerr BHs (which exist in the pale blue shaded region) on either side of the fundamental existence line ($q=0$), but lower (larger) mass solutions than the Kerr limiting solution occur for small (large) $q$. 

\subsubsection{A sample of reference configurations}
\label{sec_sample}
To illustrate the solutions of KBHsSH, as well as the limiting cases, we present five examples of qualitatively different cases, all with $m=1$, $n=0$, namely:
\begin{itemize}
\item[{\bf I}]:  a typical BS, with frequency $w=0.85$;
\item[{\bf II}]: a Kerr BH in the region of non-uniqueness. We have chosen its mass and angular momentum to be $M\simeq 0.415$, $J\simeq 0.172$; this corresponds to $r_H\simeq 0.066$;
\item[{\bf III}]: a KBHSH in the region of non-uniqueness with the same $M,J$ as the Kerr BH in example {\bf II}. This KBHSH is Kerr-like and has $ w= 0.975$ and $r_H=0.2$;
\item[{\bf IV}]: a KBHSH with $ w=0.82$ and $r_H=0.1$ (this solution is close to the main branch of BSs which is the most stable branch);
\item[{\bf V}]: a KBHSH with $ w= 0.68$ and $r_H= 0.04$ (the nearby BSs are in the secondary branch, being unstable).
\end{itemize}

In Appendix~\ref{sec_plots} we provide a set of plots that describe properties of these example solutions.  Each figure has ten (or eight) panels divided into two colums and five (or four) rows. The figures exhibit the following quantities: 
\begin{itemize}
\item[$\bullet$] the metric functions $F_0,F_1,F_2,W$, $cf.$ equation~(\ref{ansatz}), in Figures \ref{F0}, \ref{F1}, \ref{F2}, \ref{W}, respectively;
\item[$\bullet$] the metric coefficient $g_{tt}$, in Figure~\ref{gtt};
\item[$\bullet$] the scalar field amplitude $\phi$, $cf.$ equation~(\ref{scalar_ansatz}), in Figure~\ref{Z};
\item[$\bullet$] the scalar field energy-momentum tensor component  $E=-T_t^t$, in Figure~\ref{E};
\item[$\bullet$] the scalar field energy-momentum tensor component $J=T_{\varphi}^t$, in Figure~\ref{J};
\item[$\bullet$] the Ricci scalar $R$, in Figure~\ref{R};
\item[$\bullet$] The Kretschmann scalar $K=R_{abcd}R^{abcd}$, in Figure~\ref{K};
 \end{itemize}

In all of the figures described above, Figures~\ref{F0}--\ref{K}, the left column displays 3D plots, whereas the right column shows 2D plots of the corresponding function in terms of the radial variable for three different angular coordinates.
Concerning the left column, the axes for the 3D plots are $\rho=r\sin \theta$ (with $\rho\geq r_H$) and $z=r\cos \theta$ (with $-z_{max}\leq z\leq z_{max}$,
where the value of the $z_{max}$ is chosen for convenience for each case). In these 3D plots, only the near horizon region is shown.
Concerning the 2D plots, we show the full radial dependence of the functions for three different angles:
$\theta=0$ (red line),
$\theta=\pi/4$ (blue line)
and
$\theta=\pi/2$ (green line). We recall that the solutions are invariant under a 
reflection on the equatorial plane. The radial coordinate there is $x=1-r_H/r$ (for BHs),
and $x=r/(1+r)$ (for the BS example), such that the asymptotic values are approached 
at $x=1$. 

The numerical data for these reference configurations can be found in~\cite{gravwebsite}.


\subsubsection{Horizon and ergo-regions}
\label{sec_ergo}
\begin{figure}[h!]
\begin{center}
\includegraphics[height=.25\textheight, angle =0]{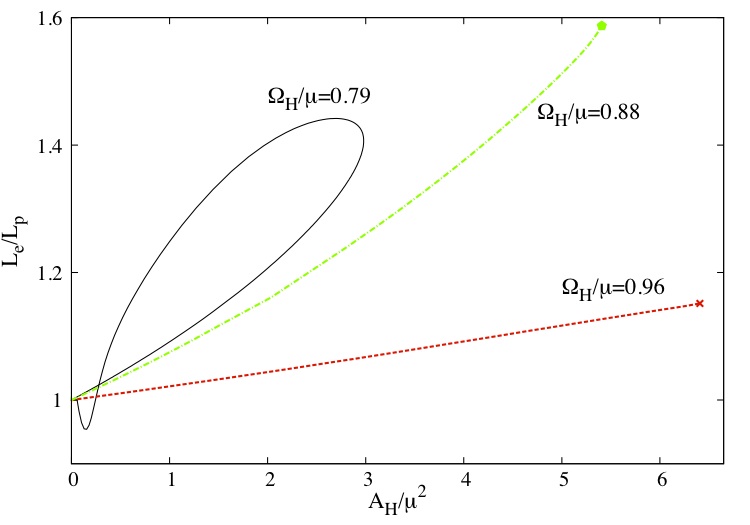} \ \ \ 
\includegraphics[height=.25\textheight, angle =0]{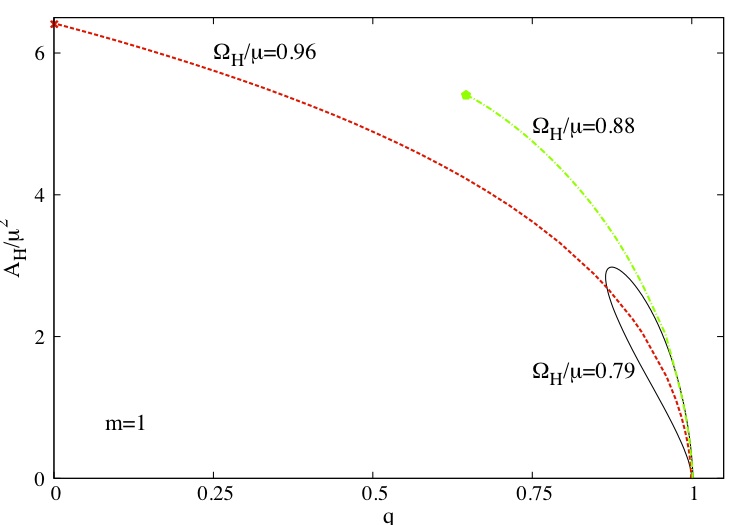} \ \ \ 
\end{center}
\caption{(Left panel) The ratio $L_e/L_p$ for three sets of KBHsSH solutions with fixed values of $\Omega_H$ (or, equivalently, $w$). In the three cases, the KBHsSH interpolate between a BS and i) another BS ($\Omega_H/\mu=0.79$), ii) an extremal KBHSH ($\Omega_H/\mu=0.88$) and iii) a Kerr BH ($\Omega_H/\mu=0.96$). (Right panel) The horizon area $vs.$ $q$ for the same three sets of solutions.}
\label{BH2}
\end{figure}

As for Kerr BHs, KBHSH have a topologically spherical horizon at $r=r_H$. Geometrically, however, the horizon is a squashed sphere. This can be seen by evaluating the circumference of the horizon along
the equator, 
\begin{equation}
L_e=2 \pi r_H e^{F_2(r_H,\pi/2)} \ ,
\end{equation}
 and the circumference of the horizon along the poles, 
\begin{equation}
L_p=2 r_H \int_0^\pi d\theta e^{F_2(r_H,\theta)} \ .
\end{equation} 
In Figure \ref{BH2} (left panel) we show the ratio of the equatorial circumference to the polar circumference for some KBHsSH. As expected the squashing of the horizon produced by the rotation is such that  $L_e/L_p$ is typically larger than one; but close to the secondary branch of BSs, one finds $L_e/L_p$ slightly smaller than one.

\begin{figure}[h!]
\begin{center}
\includegraphics[height=.18\textheight, angle =0]{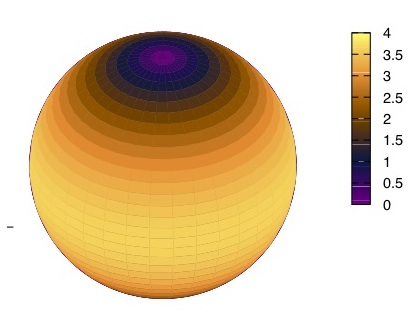} \ \ \ \ 
\includegraphics[height=.18\textheight, angle =0]{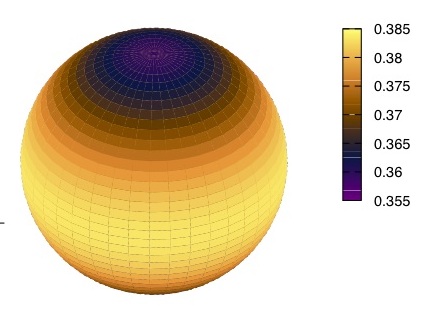} 
\includegraphics[height=.18\textheight, angle =0]{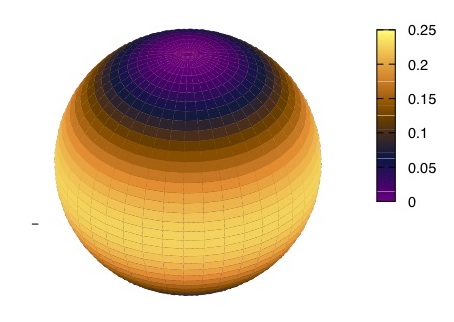} 
\end{center}
\caption{ The scalar field (left panel), its ``energy density" (middle panel) and ``angular momentum density" (right panel) on the horizon, for an example of a KBHSH (all panels have been multiplied by a factor of $10^3$).}
\label{BH3}
\end{figure}

On the horizon, the scalar field profile function $\phi(r_H,\theta)$, ``energy density" $-T^t_{\ t}$ and ``angular momentum density" $T_\varphi^{\ t}$ vary with the angular coordinate, as it is manifest in Figures~\ref{Z},~\ref{E} and \ref{J} for the sample solutions discussed in Section~\ref{sec_sample}. In Figure~\ref{BH3} we show the scalar field value, its ``energy density" $-T^t_{\ t}$ and its ``angular momentum density" $T_\varphi^{\ t}$ for a KBHSH solution with $r_H=0.2$, $m=1$ and $w=0.9$. Note in particular that both the scalar field value and the ``angular momentum density" vanish on the horizon poles -- a property common to all solutions -- but not the ``energy density".

%
It would be interesting to perform a study of the Gaussian curvature of the horizon spatial sections. It is well known that for Kerr, beyond a certain rotation parameter, this curvature becomes negative on the polar caps~\cite{Smarr:1973zz}. This, in turn, prevents a global isometric embedding in Euclidean 3-space, albeit such embedding is possible in curved embedding 3-spaces, as hyperbolic 3-space~\cite{Gibbons:2009qe}. Since an isometric embedding is a tool to gain insight into the geometry of BHs (see e.g.~\cite{Andrianopoli:1999kx,Gibbons:2009qe} and references therein) a study of the Gaussian curvature and global isometric embeddings for KBHsSH would be useful.

\begin{figure}[h!]
\begin{center}
\includegraphics[height=.5\textheight, angle =0]{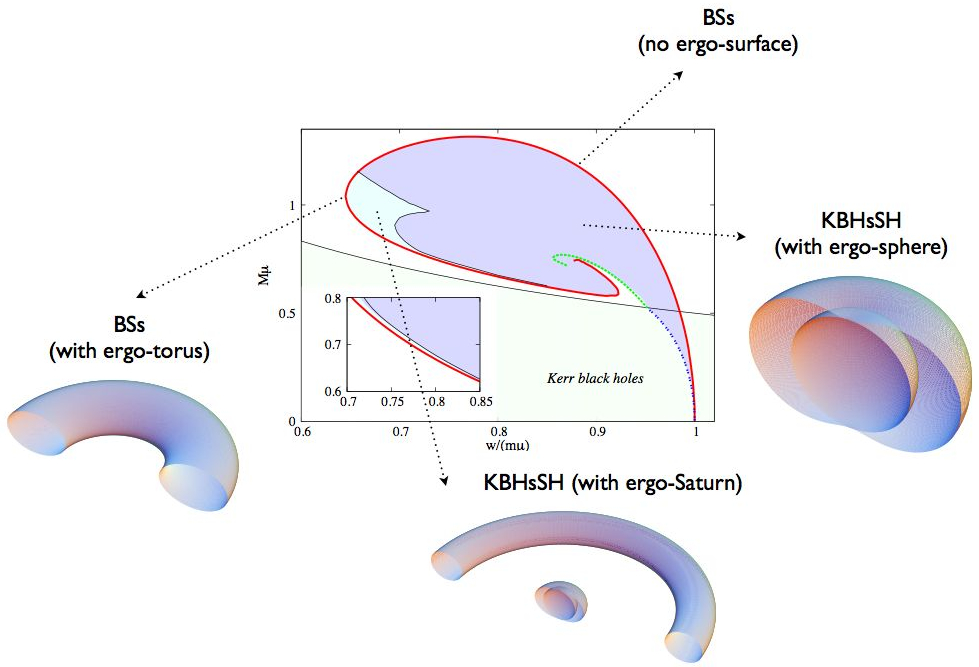} \ \ \ 
\end{center}
\caption{Ergo-surface distribution in the domain of existence for KBHsSH with $m=1$ and $n=0$ (shaded blue region) in an ADM mass $vs.$ scalar field frequency diagram. Adapted from~\cite{Herdeiro:2014jaa}. The 3D plots illustrate the ergo-surfaces and, for KBHsSH, also the horizon, which is the innermost half-sphere. These are obtained from the numerical data for specific solutions.}
\label{ergo_fig}
\end{figure}

We now turn our attention to the ergo-surfaces and ergo-regions of KBHsSH. These have been studied in detail in~\cite{Herdeiro:2014jaa}. Three different qualitative cases can be actually seen in Figure~\ref{gtt}, concerning the example solutions of the previous subsection. For example {\bf I}, $g_{tt}$ is always negative and there is no ergo-region. Indeed this is the case for some BSs. Some other BSs have a toroidal ergo-surface: an \textit{ergo-torus}. Example {\bf II}, the Kerr BH example, exhibits the usual ergo-sphere. An ergo-sphere also occurs for examples {\bf III} and {\bf IV}. Example {\bf V}, exhibits a more complex structure: there is both an ergo-sphere and an ergo-torus, $i.e.$ it exhibits an \textit{ergo-Saturn}. In Figure~\ref{ergo_fig} we exhibit the distribution of the different structure of ergo-regions in the $M$-$w$ diagram for KBHsSH.

\subsubsection{Extremal KBHsSH}
\label{sec_extremal}

With the ansatz (\ref{ansatz}),  extremal  KBHsSH
are found as a different $r_H\to 0$  limit of the hairy solutions from that of BSs (see Appendix~\ref{appendixa} for an explicit proof in the Kerr limit). In the former case, however, a number of metric functions diverge;
as a result, only near-extremal solutions can be constructed within the scheme described above.
Therefore, extremal  KBHsSH solutions are studied 
by constructing them directly, within the same numerical scheme, although for a different metric ansatz.
A systematic study of these configurations
will be presented elsewhere;
here we review their basic properties only. 

Our approach to study directly extremal KBHsSH is to use  
 a slightly different
version of  (\ref{ansatz}):
again with four unkwnown functions $(F_i,W)$ and the same scalar field ansatz,
but now we take
$N=(1-{r_H}/{r})^2$, where $r_H>0$ is a constant. Note that the extremal Kerr solution can be written in this form.
The numerical treatment of the problem is similar to the non-extremal case;
in particular, a new radial coordinate $x=\sqrt{r^2-r_H^2}$
is again introduced.
The boundary conditions are similar to those used in the non-extremal case, 
except for the scalar field $\phi$,
which now vanishes on the horizon. 

The extremal  KBHsSH  have finite horizon size and global charges and possess a regular horizon in terms of the Kretschmann invariant; they may, however, exhibit other more subtle pathologies,
see $e.g.$ \cite{Dias:2011at}.
Fixing $(m,n)$, these solutions form a line in the pararameter space of KBHsSH,
delimiting a part of the boundary of the domain of existence, as described in Section~\ref{sec_over}.
This line appears to have the same qualitative behaviour of the BS line;
however, instead of the zero mass, flat space limit,
the extremal KBHsSH line starts at a point corresponding to a scalar cloud line around an extremal Kerr background. Incidentally, we remark that the analytical estimates in \cite{Hod:2012px}
provide a very good approximation for the position of this point.
Then, in a mass-frequency diagram, this line increases up to a point corresponding 
to a maximal value of mass; 
then it decreases until a minimal value of the frequency is approached,
where is
backbends and keeps decreasing, until a minimal value of the mass is reached.
We expect this curve to inspiral towards a central
value where, we conjecture, it meets the endpoint of the BS spiral in a singular solution for
a critical configuration.
This is the behaviour noticed 
for five dimensional extremal BHs with scalar hair in
Anti-de Sitter spacetime \cite{Dias:2011at}, and also (up to some details)
in a Minkowski spacetime background \cite{Brihaye:2014nba}.


\section{Phenomenological properties}
\label{sec_pheno}
If fundamental, sufficiently stable, scalar fields exist in Nature, of the type that can source KBHsSH, the solutions exhibited herein could play a role in astrophysical systems. Moreover, if such fields only interact gravitationally, as some dark matter candidates, strong gravity systems may be the only arenas where they leave observational signatures. As such it is of interest to understand physical properties of KBHsSH with  phenomenological relevance. This discussion becomes even more interesting in view of the observation in~\cite{Herdeiro:2014goa} that some of these properties deviate considerably from those of the Kerr solution, an uncommon feature for BHs in alternative theories of gravity (see $e.g.$ the discussion in~\cite{Barausse:2014oca}). In the following we shall expand on two properties of phenomenological interest, already briefly discussed in~\cite{Herdeiro:2014goa}: the quadrupole moment and the orbital frequency at the innermost stable circular orbit (ISCO), and mention some other possible phenomenological directions of research.

\subsection{Quadrupole Moment}
In Newtonian gravitational systems, the multipolar expansion provides a complete description of the gravitational field for a distribution of (static) masses.  In relativistic gravity, Geroch~\cite{Geroch:1970cd} and Hansen~\cite{Hansen:1974zz} have devised strategies to define a physical significant multipolar expansion. In particular, the quadrupole moment is of great interest; because in principle it can be measured, say, by gravitational wave signals of a star or a smaller BH orbiting a central (larger) BH (see $e.g.$~\cite{Ryan:1995wh}) and it can be used to test the no-hair idea~\cite{Loeb:2013lfa}, since for the Kerr solution the quadrupole moment is completely determined in terms of the ADM mass $M$ and angular momentum $J$ as $-J^2/M$.

In computing the quadrupole moment of KBHsSH,
we shall follow the
general procedure described in
\cite{Ryan:1995wh,Berti:2003nb,Pappas:2012ns}
for extracting it from the asymptotics of
a stationary and axially symmetric
spacetime.
The above references use a line element
written in quasi-istropic coordinates, with
\begin{eqnarray}
\label{isotropic}
ds^2=e^{2(\zeta-\nu )}\left(dR^2+R^2 d\theta^2 \right)
+R^2\sin^2\theta B^2  e^{-2\nu }\left(d\varphi-W  dt \right)^2
-e^{2\nu }dt^2\ ,
\end{eqnarray}
where $\zeta,~\nu,~B$ and $W$ 
are functions of $R,\theta$.
The quadrupole moment of the spacetime
is encoded in the large-$R$ asymptotics
of 
the metric functions 
$ \nu,~B $;
to leading order, one finds
\begin{eqnarray}
\label{as1}
\nu=-\frac{M}{R}+\left(\frac{B_0M}{3}+\nu_2 P_2(\cos\theta) \right)\frac{1}{R^3}+\dots,~~
B=1+\frac{B_0}{R^3}+\dots,
\end{eqnarray}
(with $P_n$ the Legendre polynomials)
such that the quadrupole moment $Q$ is given by 
\begin{eqnarray}
\label{quadrupole}
Q=-\nu_2-\frac{4}{3}
\left(
\frac{1}{4}+\frac{B_0}{M^2} 
\right)M^3.
\end{eqnarray}
In evaluating (\ref{quadrupole})
for KBHsSH solutions,
we need first to bring the 
line element (\ref{ansatz})
into the form (\ref{isotropic}).
This is done by using the coordinate transformation
$r=R(1+\frac{r_H}{4R})^2$.
 Then the coefficients $B_0$ and $\nu_2$
which enter the asymptotics (\ref{as1}) can be read from 
the far field form (\ref{inf-1}), (\ref{inf-2}) of the solution 
(expressed in terms of the quasi-isotropic $R$),
\begin{eqnarray}
\label{quadrupole1}
B_0=a_5-\frac{1}{16}(2c_t-r_H)^2,~~
\nu_2=\frac{1}{12}
\left (
-24 b_1-8 a_5c_t+2c_t^3+4 a_5 r_H-3c_t^2r_H-3c_t r_H^2
\right).
\end{eqnarray}

\begin{figure}[h!]
\begin{center}
\includegraphics[height=.26\textheight, angle =0]{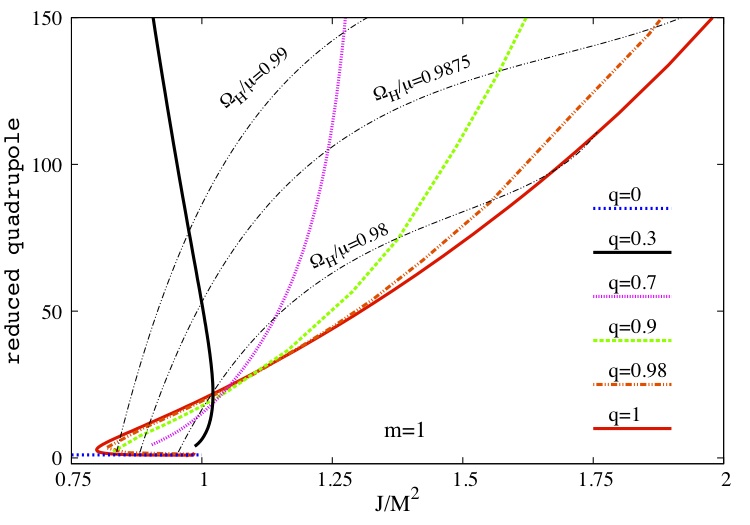} 
\includegraphics[height=.26\textheight, angle =0]{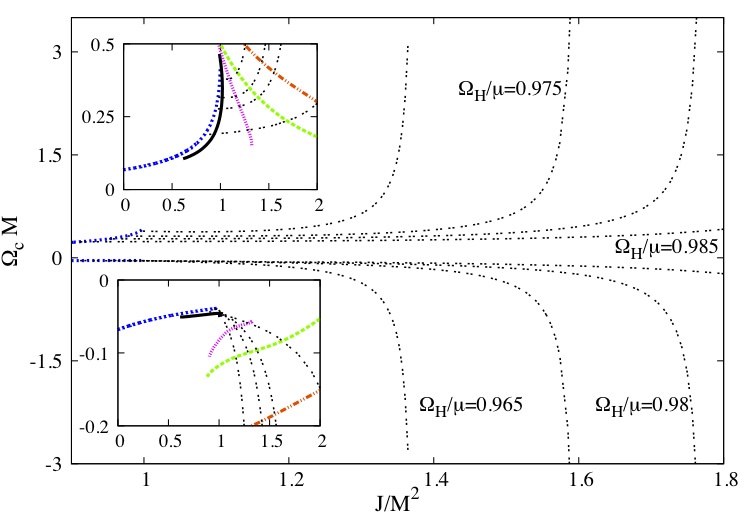} \ \ 
\end{center}
\caption{(Left panel) reduced quadrupole. (Right panel) angular frequency at the ISCO. From~\cite{Herdeiro:2014goa}.}
\label{BH5}
\end{figure}

In Figure~\ref{BH5} (left panel) we exhibit the reduced quadrupole ($i.e$ the ratio of the KBHSH quadrupole to the Kerr quadrupole) for a set of $q=$constant as well as $\Omega_H=$constant KBHsSH (and BSs) in terms of the dimensionless parameter $J/M^2$. One observes that KBHsSH can have a quadrupole moment one order of magnitude larger than Kerr BHs within the Kerr bound and up to two orders larger than Kerr solutions beyond the Kerr bound.

\subsection{Orbital frequency at the ISCO}

The study of geodesics represent a standard way
to analyze a given spacetime geometry. Moreover, in astrophysical environments it is thought that the edge of accretion disks is determined by the ISCOs around a given BH. The accelerated charges that orbit the BH will emit synchroton radiation which, in the simplest model, will have a cut-off at the orbital frequency of geodesics at the ISCO. Thus, measurements of the ISCO via observations of accretion disks can, in principle, be used to evaluate the properties of an astrophysical BH.
 As such we have studied the angular frequency at the ISCO for a large set of KBHsSH.

The geodesic  motion is studied along the equatorial plane, $\theta=\pi/2$; then
the Lagrangian of a timelike test particle (the only case studied here) is
\begin{eqnarray}
2{\cal L}=e^{2F_1 } \frac{\dot r^2}{N }+e^{2F_2 } r^2(\dot \varphi-W \dot t)^2
-e^{2F_0} N \dot t^2=-1\ .
\end{eqnarray}
Note that $F_i,W$ depend only on $r$;
also a dot denotes a derivative $w.r.t.$ the proper time.
The stationarity and axisymmetry of the KBHsSH metric implies the existence of the
 first integrals
\begin{eqnarray}
&&
e^{2F_2}r^2(\dot \varphi-W \dot t)=L\ ,
\\
\nonumber
&&
(e^{2F_0} N-e^{2F_2} r^2 W^2)\dot t+e^{2F_2} r^2 W \dot \varphi=E\ ,
\end{eqnarray}
with $E$ and  $L$
the specific energy and angular momentum of the test particle.
Then the orbital angular velocity is expressed as
\begin{eqnarray}
\Omega_c=\frac{\dot \varphi}{\dot t}=W-\frac{e^{2F_0-2F_2}LN}{r^2(LW-E)}\ .
\end{eqnarray}

The equation governing the variation of the radial coordinate $r$
for an orbit on the equatorial plane is
\begin{eqnarray}
\label{V}
\dot r^2=V(r)=e^{-2F_1}N
\left(
-1-e^{-2F_2}\frac{L^2}{r^2}
+\frac{e^{-2F_0}(E-LW)^2}{N}
\right) \ .
\end{eqnarray}
To determine circular orbits we need the first derivative of $V$, which is
\begin{eqnarray}
\label{derV}
&&
V'(r)=e^{-2F_1}
\bigg (
2N F_1'-2e^{-2F_0}(E-L W)^2(F_0'+F_1')-N'
\\
&&
\nonumber
{~~~~~~~~~}+\frac{e^{2F_2}L^2}{r^3}
\left(
2N(1+r(F_1'+F_2'))
-2N'
\right)
+2 e^{-2F_0}L(LW-E)W' 
\bigg)\ .
\end{eqnarray}
The second derivative of $V$ is also of
interest; however, its expression is very long and
not particularly enlightening; so we shall not exhibit it here.

The requirement for a circular orbit at $r=r_c$ is 
$V(r_c)=V'(r_c)=0$.
From (\ref{V}), (\ref{derV}), 
this results in two algebraic equations
for $E,L$ which are solved analytically, possessing
two distinct pairs of solutions $(E_{+},L_{+})$ and $(E_{-},L_{-})$,
corresponding to co-rotating and counter-rotating trajectories.

The solutions for $E,L$
are then replaced in the expression of $V''(r_c)$, asking for 
$V''(r_c)\leq 0$.
For the configurations studied so far we have noticed a 
(qualitative) analogy with the Kerr BH.
First, the circular geodesic motion is possible for 
$r_c>r_{min}$ only, a  constraint imposed by the
 requirement for the energy $E$
to be real.
Then for $r_{min}<r_c<r_{ISCO}$
only unstable circular orbit can exist, $i.e.$ with
$V''(r_c)< 0$.
For $ r_c>r_{ISCO}$
the circular orbits are stable.

In Figure~\ref{BH5} (right panel) we exhibit the angular frequency at the ISCO for co-rotating and counter-rotating geodesics and one can observe the deviations relative to the Kerr solution~\cite{Herdeiro:2014goa}. A similar study, but describing the angular frequency at the ISCO in terms of the horizon mass and angular momentum will be presented elsewhere~\cite{HR_overspinning}.


\subsection{Other phenomenological studies}
As discussed in the Introduction, the next decade may bring the first detection of gravitational waves, and with it a tool to test the general relativity BH paradigm. It is therefore important to understand how alternative models to this paradigm, for instance models with scalar fields, can have different phenomenological consequences. 

The strongest gravitational wave signal is expected to occur in the merger of two compact objects (such as neutron stars or BHs). To study such mergers, the only available tool is to resort to fully non-linear numerical relativity simulations. For the case of scalar-tensor theories of gravity, numerical relativity simulations are still in their infancy, $ e.g$.~\cite{Berti:2013gfa,Healy:2011ef,Barausse:2012da}, but qualitatively different features have already been reported -- see~\cite{Cardoso:2014uka} for a review. For the case of minimally coupled scalar fields, numerical simulations, and in particular collisions, of BSs have been performed (see~\cite{Liebling:2012fv} for a review), but not yet of KBHsSH. This will certainly be of interest, also to test the stability of these solutions -- see Section~\ref{sec_stability}. 

The interaction of minimally coupled scalar fields with BHs has, nevertheless, been considered using various approximation schemes, with the goal of extracting gravitational wave emission, $e.g.$~\cite{Nunez:2011ej,Witek:2012tr,Okawa:2014nda,Degollado:2014vsa,Brito:2014wla}. In particular, in~\cite{Degollado:2014vsa} it was shown that the gravitational wave response of a perturbed BH surrounded by a `dirty' environment composed of a quasi-stationary scalar cloud has an imprint in its late time tail. Even though such tail is unlikely to be measured in the near future, this example suggests how the gravitational wave signal from a KBHSH can be different from that of the conventional BHs in general relativity.

Another potentially observable feature of BHs also discussed in the Introduction is the BH shadow. Firstly, the presence of a shadow is regarded as a smoking gun for the presence of an event horizon (more precisely, of an apparent horizon). By contrast, for BSs there are unbound geodesics that can approach arbitrarily close to the centre of the BS~\cite{Grandclement:2014msa} and there is no shadow (see also~\cite{Eilers:2013lla}). As KBHsSH interpolate continuously between Kerr BHs and BSs, their shadows should vary continuously between the Kerr shadow and no shadow at all. It is of interest to understand quantitatively how this variation occurs and how distinctive can these shadows be.

\section{Stability?}
\label{sec_stability}

\subsection{General comments on stability of BHs}
\label{sec_stability_1}
The stability properties of any exact (analytical or numerical) solution of the Einstein equations are central to its physical relevance. It is important to remark, however, that absolute stability is not mandatory for physical relevance. Metastable states ($e.g.$ Uranium or Plutonium) can play a physical role and, ultimately, essentially all systems in the Universe are only metastable. Thus, if  instabilities are present, the discussion should focus on how large/small the timescales of the instabilities are, as compared to other typical timescales in the physical processes where the solution may play a role. 

Establishing the stability of BH solutions is highly non-trivial. One may address, by increasing order of complexity, the i) mode stability, ii) linear stability or iii) fully non-linear stability. Mode stability, in \textit{vacuum}, was established for Schwarzschild BHs in the classical works of Regge-Wheeler~\cite{Regge:1957td} and Zerilli~\cite{Zerilli:1970se}, and for Kerr in the work of Whiting~\cite{Whiting:1988vc}. It relies on an analysis of quasi-normal modes of the BHs. In the Kerr case, the analysis is made possible by the algebraic specialty of the solution (Petrov type D), which allows the decoupling of gravitational perturbations, using the Newman-Penrose formalism. We have verified that KBHsSH are (generically) algebraically general (Petrov type I). Thus, no decoupling of the coupled gravitational-scalar system is expected to occur and thus a similar mode analysis to that performed for Kerr cannot be carried through.

Mode stability does not, however, guarantee linear stability (quasi-normal modes do not form a basis of the space of functions). As an illustration of this difference, recently, it has been argued that extremal Kerr BHs are unstable against a special class of linear perturbations~\cite{Aretakis:2012ei}. Even for non-extremal Kerr, the linear stability has not yet been established (see $e.g.$~\cite{Dafermos:2010hd}). Finally, linear stability does not guarantee non-linear stability. An example which has arisen great interest over the last few years is Anti-de Sitter space which has been shown, by numerical simulations, that it is non-linearly stable against BH formation~\cite{Bizon:2011gg}. Establishing non-linear stability is a mathematical \textit{tour de force}; for Minkowski spacetime this was proven by Christodoulou~\cite{Christodoulou:1993uv}. No rigorous statement can be made for BHs, but the many fully-non linear numerical simulations performed with Kerr and Schwarzschild BHs have piled up evidence that, at least in the time and spatial scales probed by the simulations, no non-linear instabilities exist~\cite{Cardoso:2014uka}. 

\subsection{Stability of BSs}
\label{sec_stability_2}

Spinning BSs form a part of the
boundary of the domain of existence of KBHsSH, $cf.$ Figure~\ref{kbhsh_fig}. The stability of
these objects is even more complicated to establish than 
for the (vacuum) Kerr BH case, even at the level of mode stability, as we shall discuss below. But let us start by overviewing \textit{non-spinning} BSs ($m=0$), for which, as for the case of BHs, stability has been analysed from various viewpoints:
by considering either $i)$ mode stability for linear perturbations,  $ii)$ fully numerical non-linear evolutions or $iii)$ catastrophe theory.

Historically, the stability of spherically symmetric BSs was  first addressed in the late 80s, by means of a mode analysis in linear perturbation theory
\cite{Lee:1988av,Gleiser:1988rq,Gleiser:1988ih}.
 This approach leads to an eingenvalue problem, which is of Sturm-Liouville type
and which determines the normal modes of the radial oscillations and their eingenvalues $\sigma^2$. 
The sign of
the lowest eingenvalue $\sigma_0^2$
is crucial; if $\sigma^2_0>0$
  then the BS is stable, otherwise it is unstable. 
As already noticed above, the spherically symmetric BSs
 can be parametrized in terms of the central value of
the scalar field $\phi_0$.
Then one can show that 
the transition
from stability to instability always  occurs at critical points of the ADM mass $M$ (or Noether charge, $Q$) against the value of the
scalar field at the origin, where $dM(\phi_0)/d \phi_0=dQ (\phi_0)/d\phi_0=0$. The pattern of $\phi_0$ in a non-spinning BS is similar to that of $\phi_{\rm max}$ seen in the inset of Figure~\ref{effective_radius}. Observing also Figure~\ref{kbhsh_fig} or Figure~\ref{f-1}, one observes that critical points occur at the cusps of the BS line in the $M-J$ diagram (Figure~\ref{kbhsh_fig} or Figure~\ref{f-1} -- right panel). A physical understanding of this fact is related to the
binding energy of the BSs, $E_b = \mu Q-M$: it becomes negative after the first critical point, indicating instability.

Numerical non-linear evolutions of BSs started relatively recently -- 
in the last 15 years --  and allow 
the study of possible end-states of some initial data.
These 
include dispersion to infinity of the scalar field, 
transition from an unstable to a 
stable configuration,
or collapse to a BH.
A recent review of these aspects 
is given in Ref. 
\cite{Liebling:2012fv},
together with a large set of reference.
Here we only mention that the conclusions 
reached in this way on the stability question
are in agreement with those within the linear perturbation
theory. Note that, 
as shown by the 'no hair'
 theorem in  \cite{Pena:1997cy}
all spherically symmetric 
 BHs which occur as endstates of some initial data
are described by the  Schwarzschild vacuum metric.
 
Catastrophe theory \cite{Thom:1975,Zeeman:1977,Poston:1978}
provides yet another approach to 
investigate the stability of BSs
(see $e.g.$
 \cite{Kusmartsev:1990cr,Kusmartsev:1991pm}).
 In this approach, an appropriate set of behavior variable(s) and control
parameter(s) is chosen, and the series of solutions is constructed in terms of these.
Under certain conditions, such a series generates a curve smooth everywhere except
for certain points.
According to catastrophe theory, passing through one of these points means changing the stability of the
BS configurations. 
Between these points, the solutions form branches,
sharing the same stability properties.
The results obtained in this way confirm
 the previous conclusions.
 
Concerning the stability of $spinning$ BSs considerably less is known.
Firstly, such solutions have not yet been analysed 
even within linear perturbation theory. This study appears challenging,
since no decoupling of the coupled gravitational-scalar
system is expected to occur.
Additionally, no numerical evolutions of rotating
BS initial data has yet been performed~\cite{Liebling:2012fv}, although simulations of BS binaries \cite{Palenzuela:2007dm}
have found rotating BSs as a result of merger.
To our knowledge, the only results on the 
stability of rotating BSs are those in \cite{Kleihaus:2011sx}
obtained by using catastrophe theory.
As shown therein, the  rotating BSs share a similar stability
picture as the non-rotating solutions.
In particular, 
the branch of rotating  solutions between the vacuum 
$M=J=0$ 
and the point where the global charges approach the absolute maximum
is predicted to be stable.
A more complicated picture occurs in the inspiraling region (see Figure~\ref{f-1}),
with both stable and unstable branches.
The BSs in that region, however, possess an ergo-region, $cf.$ Figure~\ref{ergo_fig}. The existence of this region is expected to originate $superradiant$ instabilities, as we address in the next subsection.

\subsection{Ergo-regions and superradiant instabilities}
Kerr BHs in the presence of massive bosonic fields are afflicted by superradiant instabilities that amplify sufficiently low frequency modes, extracting rotational energy and angular momentum from the BH~\cite{Press:1972zz,Damour:1976kh,Zouros:1979iw,Detweiler:1980uk,Cardoso:2004nk,Dolan:2007mj,Hod:2009cp,Rosa:2009ei,Dolan:2012yt,Witek:2012tr,Cardoso:2013krh}.\footnote{Superradiant instabilities should be distinguished from superradiant \textit{scattering}, which can occur for bosonic massive or massless fields around Kerr BHs~\cite{Starobinsky:1973a,Starobinsky:1973b}. Superradiant scattering of charged bosonic fields can also occur around spherically symmetric charged BHs, leading to the extraction of Coulomb energy and charge~\cite{Bekenstein:1973mi}. But superradiant instabilities only occur around spherically symmetric charged BHs if these are enclosed in a cavity, in Anti-de-Sitter spacetime~\cite{Hod:2012zz,Herdeiro:2013pia,Hod:2013eea,Hod:2013nn,Wang:2014eha,Hod:2013fvl,Degollado:2013bha} or in higher dimensional brane-world scenarios~\cite{Zhang:2013haa}.} This instability is often regarded as a field analogue of the particle Penrose process~\cite{Penrose:1969pc}, which occurs due to the existence of an ergo-region. For a scalar field of mass $\mu$ in the background of a Kerr BH with ADM mass $M$ and angular momentum $J$, the time scale for the fastest growing unstable mode is~\cite{Detweiler:1980uk,Herdeiro:2014jaa}, for $\mu M\ll 1$,
\begin{equation}
\tau=\frac{1}{(\mu M)^9}\frac{M^2}{J}\frac{M}{M_\odot}\ 0.12 \ {ms} \ . 
\label{sits}
\end{equation}
As an estimate, for $M\mu\sim 0.1$ and an extremal Kerr BH, $J=M^2$ (to minimize the time scale of the instability), one gets $\tau\sim 1$ day to $\tau\sim 10^6$ years, for a BHs of $M=1M_\odot$ to $M=10^9 M_\odot$, respectively. As a comparison -- for astrophysical relevance -- the latter timescale is approximately that of the lifetime of blue supergiant stars. An estimate from the results of~\cite{Dolan:2012yt}, is that for $\mu M\sim 1$ these timescales may decrease by two orders of magnitude.

KBHsSH always have an ergo-region~\cite{Herdeiro:2014jaa} and thus should be afflicted by superradiant instabilities in the presence of massive bosonic fields. The same applies to BS that have an ergo-torus~\cite{Cardoso:2007az,Kleihaus:2007vk}. The novelty, with respect to Kerr, is that a massive scalar field is already present in the background solution, and thus one needs not consider extra fields in order to obtain superradiant instabilities, as it must be done for Kerr. Thus, the superradiant instabilities of KBHsSH should be seen within linear perturbations of the solutions in the theory~\re{action}. As we have argued in Section~\ref{sec_stability_1} the relevant question is the time scale of these instabilities, which, at the moment is not known, and a linear perturbation analysis faces the same challenges as for the case of BSs, discussed in Section~\ref{sec_stability_2}. Let us therefore remark on some expectations, mainly comparing to the timescale of the instabilities observed above for the Kerr case:
\begin{description}
\item[1)] Superradiant instabilities of KBHsSH only occur for sufficiently high $m$ modes; these are expected to be longer-lived than the fastest growing mode in the Kerr case. Indeed, as pointed out in~ \cite{Herdeiro:2014goa} (see also~\cite{Dias:2011at}) a KBHSH solution with a given $m$ azimuthal harmonic index should only be unstable against perturbations with higher azimuthal harmonic index. In the Kerr case, it is known that the timescale of the instabilities decays with increasing angular harmonic index $\ell$ of the spheroidal harmonic perturbation $S_{\ell m}$. Since a higher $m$ requires a higher $\ell$ we expect that the unstable modes for KBHsSH have a larger time scale than the fastest growing mode ($\ell=1$) in the Kerr case.
\item[2)] Ergo-regions of Kerr BHs are `larger' than those of KBHsSH for comparable solutions, which suggests longer timescales for the instabilities in the latter case. In~\cite{Herdeiro:2014jaa} it was observed that a measure of the size of the ergo-region (therein dubbed \textit{ergo-size}) can be introduced, which is positively correlated to the strength of superradiant instabilities, at least in the regime of validity of formula \re{sits}. Then, comparing the ergo-size of Kerr BHs and KBHsSH in the region of non-uniqueness, for the same global charges, it was observed that the ergo-size is smaller for the latter, thus suggesting instabilities are weaker. The region where this reasoning applies is close to BSs that have no ergo-region -- see Figure~\ref{ergo_fig}. If one regards KBHsSH as bound states of Kerr BHs with BSs, it is therefore natural that these bounds states have a smaller ergo-region.
\item[3)] In the $q\rightarrow 0$ limit, scalar clouds should be dynamical attractors, rather than unstable solutions. It was argued in~\cite{Benone:2014ssa} that scalar clouds around Kerr BHs ($i.e.$ the $q\rightarrow 0$ limit of KBHsSH) are dynamically stable configurations; they are in synchronous rotation with the BH in a tidal locking configuration with some analogy to the Earth-Moon orbital tidal locking. Slightly superradiant modes are under-spinning whereas slightly decaying modes are over-spinning and, in both cases, dynamics tries to synchronize the horizon angular velocity with that of the scalar perturbation.\footnote{Bear in mind this reasoning is simplistic as the role of accretion is not properly considered; see~\cite{Brito:2014wla}.} This motivates from a different direction why a superradiant perturbation in the background of a KBHSH grows more slowly than in the background of Kerr: as it grows, the perturbation extracts energy and angular momentum from the background BH which decreases its angular velocity. Then, the background scalar mode becomes over-spinning compared to the horizon and competes with the under-spinning perturbation, slowing down (stopping?)  its growth.
\end{description}

To summarize, whereas spinning BSs seem to possess a branch of stable solutions, all KBHsSH should be afflicted by instabilities of superradiant type. It is reasonable to expect, however, that the timescales for (at least some) KBHsSH can be larger than those observed for Kerr solutions, which, in some cases can already be interesting astrophysically. Some light into this question may be shed by numerical simulations, for which the technology exists for evolving spinning BSs and KBHsSH~\cite{Cardoso:2014uka}. We anticipate that, at least for a subset of KBHsSH solutions, these will appear stable even in long term numerical time evolutions, as the  timescale for the potential instabilities will be too small. 

As a final note, let us recall that, in the region of non-uniqueness, the entropy of KBHsSH is larger than that of a Kerr solution with the same global charges. This prevents the former from decaying into the latter, adiabatically.

 \section{Conclusions and Generalizations}
 \label{sec_conclusions}
  The main purpose of this work was to provide a detailed description
of the construction and of the (basic) physical properties of a new type
of hairy BH
reported in \cite{Herdeiro:2014goa}.
Our results show that stationary
scalar field matter distributions surrounding rotating BHs can exist even in the absence 
of scalar self--interactions or non-minimal couplings between the scalar field and the geometry. 

The existence of KBHsSH provides an unexpected connection between the scalar superradiant instability of Kerr BHs and the BSs of soliton physics. Moreover, they clarify a long standing puzzle: all field theory horizonless, particle-like solutions  were known to possess BH generalizations \textit{except} BSs. It is now clear that BSs also admit a BH generalization -- but only if they are $spinning$. 

There are many possible continuations
of this work -- a few already mentioned along the paper. The most pressing one is to deepen the issue of stability, for which a first doable step is to perform a fully dynamical evolution of KBHsSH. Equally relevant, especially in view of the experiments discussed in the Introduction, is to further explore astrophysical signatures: $e.g.$  to compute shadows of KBHsSH, contrasting them with those for Kerr BHs, and analysing possible gravitational wave signatures~\cite{Degollado:2014vsa}. A related research line is to understand the dependence of phenomenological aspects on the horizon (Komar) mass and angular momentum rather than on the ADM global quantities. On a more theoretical side, discussing the intrinsic BH horizon geometry, the extremal limit and its properties and eventually considering the BH interior, seem worthwhile problems.

In parallel to the aspects discussed in the previous paragraph, one may search for new solutions of BHs with scalar hair in more general theories, which belong to the same family as the KBHsSH considered here, in the sense of relying on the \textit{syncrhonization condition} $\omega=m\Omega_H$ (or appropriate extensions). Indeed, the study in this paper, as well as those in Refs.
\cite{Herdeiro:2014goa,Herdeiro:2014ima,Herdeiro:2014jaa}, 
 has been restricted to the simplest case of
a non-self interacting scalar field.
It should be possible, however, to adapt the approach described in Sections~\ref{sec_eq_motion} and \ref{sec_bc_num} to include a scalar field potential.
Apart from the theoretical and technical relevance, such solutions could be interesting 
from yet another point of view. 
As noticed above, the  maximal values of the mass of the BSs  
set an upper bound for the mass of KBHsSH,
which is of the order of $M_{Pl}^2/\mu$.
Thus these KBHsSH can only be relevant in an astrophysical context if extremely light bosons exist 
 (although they might
affect the dynamics of small primordial BHs).
The situation is very different for a self-interacting
scalar field.
As shown in \cite{Colpi:1986ye},
when switching on a quartic self--interaction term for $\Psi$,
the maximal mass of stable BSs is of the order of
$\lambda M_{Pl}^3/\mu^2$, with $\lambda$ the scalar field self--coupling.
Therefore we expect 
the typical masses for the corresponding hairy BHs to be
much larger 
than in the $\lambda=0$ case.

A more involved picture is expected to 
exist for a complex massive scalar field with 
a non-renormalizable self-interaction potential
allowing for finite mass 
solitons even in the absence of gravity -- $Q$-balls.
The existence of BH generalizations of the spinning gravitating Q-balls  has been 
discussed in \cite{Herdeiro:2014pka}
at the probe level, $i.e.$
for a fixed Kerr BH background.
The corresponding solutions have been dubbed therein $Q-clouds$. 
Such configurations are also in synchronous rotation with the BH horizon,
satisfying the condition $w=m\Omega_H$.
In constrast to the non-selfinteracting case, however,
Q-clouds exist on a 2-dimensional subspace of the Kerr parameter space, 
delimited by a minimal horizon angular velocity
and by the corresponding $m=l$ existence line, wherein the nonlinear terms become irrelevant and the
Q-clouds reduce  to the linear clouds discussed in Section~\ref{sec_scalar_clouds}.
This implies that a more involved picture will be found 
when including backreaction on the spacetime geometry,
with a much larger range for the masses of hairy BH as compared to
the case discussed here. Furthermore, some basic features of the latter solutions should be preserved when replacing the $Q$-balls with $vortons$. These are a special class of scalar solitons made from loops of vortices, which are
balanced against collapse by rotation,
being the  four dimensional field theory analogues of the
higher dimensional black rings of vacuum general relativity \cite{Radu:2008pp}. Some of these solutions were shown to be stable \cite{Garaud:2013iba}.
We expect that all gravitating vortons in \cite{Kunz:2013wka}
to possess BH generalizations with many similar properties to the KBHsSH in this work.
This type of study can moreover be used to set constraints on the properties
of the scalar field,
by comparing astrophysical observations with 
the set of hairy BHs predictions -- see $e.g.$
\cite{AmaroSeoane:2010qx}
for a similar approach in the spherically symmetric
solitonic limit of solutions.

 Yet another possible generalization for future research is to include a matter content more generic than scalar fieds in the theory, and look for the corresponding hairy solutions. One could test the conjecture, put forward in \cite{Herdeiro:2014ima}
that
{\it ``A (hairless) BH which is afflicted by the superradiant instability of a given field must allow
hairy generalizations with that field."}\footnote{Of course, the test field must source a time and azimuthal independent energy-momentum tensor, which, for instance, immediately excludes a single real scalar field.}
In particular, this suggests the existence of Kerr BHs with Proca hair -- see~\cite{Pani:2012vp,Herdeiro:2011uu,Wang:2012tk,Sampaio:2014swa} for superradiant instabilities of Proca fields around BHs. 
The simplest case in this line of extensions, however, are  
 generalizations of the KBHsSH with a $U(1)$ gauged scalar field leading to Kerr-Newman BHs with scalar hair (KNBHsSH).
The condition (\ref{cond}) is replaced in this case with
$w=m\Omega_H+q \Phi_H$, where $q$ is the gauge coupling constant and $\Phi_H$ is the electrostatic potential on the horizon.
Similarly, to the case described here a set of KNBHsSH emerges as  backreacting charged scalar clouds around Kerr-Newman BHs~\cite{Benone:2014ssa,Hod:2014baa}.
An even more complicated picture  should exist when allowing
for scalar multiplets
gauged with respect to a non-Abelian gauge group.
The
simplest set of such solutions has been already discussed
 in a different context in
 \cite{Kleihaus:2007vf},
for a triplet Higgs field 
and an $SU(2)$ gauge field.

Of a more theoretical interest, one may consider generalizations 
of the solutions in this work  to different spacetime dimensions, 
 $D\neq 4$,
 and also possibly with different asymptotics. 
Together with the self-interacting case,
knowledge of these solutions may lead to valuable insights into the more relevant $D=4$ asymptotically flat solutions,
by establishing which properties of hairy BHs are generic,
and also which ingredients are crucial.
Some results in this directions exist already in the literature.
For example,
a family of  asymptotically Anti-de Sitter
rotating BHs with scalar hair and a regular horizon
has been studied in \cite{Dias:2011at} within $D=5$
Einstein's gravity minimally coupled to a complex scalar field doublet;
their asymptotically flat counterparts have been studied in 
\cite{Brihaye:2014nba}.
While the properties of the solutions in \cite{Dias:2011at} 
are rather similar to the case in this work -- and indeed was the first example of this type of solution --
some of the features in \cite{Brihaye:2014nba}
could hardly be anticipated.
For example, the hair of the 
$D=5$ asymptotically flat hairy BH solutions  is intrinsically nonlinear,
since a  Myers-Perry BH  background does not allow for
  scalar clouds of  massive, test scalar fields.

We close with a natural question: can there be a solution of KBHsSH in an analytic closed form? 
In this context, we remark that in more that forty years, 
no closed form expression could be found even for the
simpler case of spherically symmetric BSs -- see however~\cite{Sakamoto:1998hq} for an exception in $D=3$. One hope is to consider solution generating techniques, or in the context of supergravity models, backgrounds preserving some amount of supersymmetry.
In the latter context, however, it is known that the existence of Killing spinors is incompatible with an ergo-region~\cite{Gauntlett:1998fz,Gibbons:1999uv,Herdeiro:2000ap}. 
Thus, even if hairy BHs/BSs could be found in this setup,
 they will hardly be representative of the physical properties of the generic case.

\vspace{0.5cm}

\section*{Acknowledgements}
We would like to thank P. Pani and H. Witek for inviting us to contribute to the focus Issue on ``Black holes
and fundamental fields'' to be published in CQG and which catalysed the writing of this paper. We would also like to thank the organizers of the conferences \textit{New Frontiers in Dynamical Gravity}, Cambridge, \textit{99 Years of Black Holes}, Potsdam and \textit{NEB 16 - recent developments in gravity}, Mykonos, where parts of this work were presented, which led to many useful discussions with the participants of these meetings. Finally, we would also like to thank E. Barausse, C. Benone, E. Berti, Y. Brihaye, V. Cardoso, L. C. Crispino, J. C. Degollado, J. Grover, B. Kleihaus, J. Kunz, L. Rezolla, J. Rosa, H. R\'unarsson,  M. Sampaio, U. Sperhake, M. Volkov and M. Zilh\~ao for comments, discussions and encouragement to further develop this work.  The authors are supported by the FCT Investigator program. This paper has also been supported by the grants PTDC/FIS/116625/2010,  NRHEP--295189-FP7-PEOPLE-2011-IRSES and by the CIDMA strategic funding UID/MAT/04106/2013.

\newpage

\appendix

\section{New coordinates for Kerr}
\label{appendixa}

As we described in the main text we have used the metric ansatz \eqref{ansatz}.
%
The Kerr metric can also be written in this form.
The corresponding expressions of the metric functions read:

 \begin{eqnarray}
\nonumber
&&
e^{2F_1}=(1-\frac{c_t}{r})^2+c_t(c_t-r_H)\frac{\cos^2\theta}{r^2},
\\
\label{Kerr1}
&&
e^{2F_2}=e^{-2F_1}
\left(
          \left(
(1-\frac{c_t}{r})^2+\frac{c_t(c_t-r_H)}{r^2}
          \right)^2
+c_t(r_H-c_t)(1-\frac{r_H}{r})\frac{\sin^2\theta}{r^2}
\right),
\\
\nonumber
&&
F_0=-F_2,~~~~
W=e^{-2(F_1+F_2)}
\sqrt{c_t(c_t-r_H)}(r_H-2c_t)
\frac{(1-\frac{c_t}{r})}{r^3}~~.
\end{eqnarray}

Expressed in this form, the solution contains two constant $r_H$ and $c_t$. While $r_H$ fixes the event horizon radius,
the second constant, $c_t<0$ does not have a transparent meaning; however, it can taken as a measure of
non-staticity, since $c_t=0$ is the Schwarzschild metric.

The expressions for various quantities of interest read (we recall $G=1=c$)
\begin{eqnarray}
\nonumber
&&
M=\frac{1}{2}(r_H-2c_t),
\\
&&
\label{Kerr2}
J=\frac{1}{2}\sqrt{c_t(c_t-r_H)}(r_H-2c_t),
\\
\nonumber
&&
A_H=4\pi (r_H-c_t)(r_H-2c_t),
\\
\nonumber
&&
T_H=\frac{r_H}{4\pi (r_H-c_t)(r_H-2c_t)},
\\
&&
\nonumber
\Omega_H=\frac{\sqrt{c_t(c_t-r_H)}}{(r_H-c_t)(r_H-2c_t)}~.
\end{eqnarray}
Note that the formal limit  $r_H=2 c_t$
is just flat space expressed in an unusual coordinate system.

The relation between the radial coordinate $r$ in (\ref{Kerr1}) and $R$, the radial coordinate
of the Kerr metric in BL form is
\begin{eqnarray}
\nonumber
r=R-\frac{a^2}{R_H},
\end{eqnarray}
where
\begin{eqnarray}
\nonumber
 R_H=M+\sqrt{M^2-a^2}
\end{eqnarray}
is the (outer) horizon radius for the Kerr metric in BL coordinates and $a=J/M$. 
We notice also the simple relation
\begin{eqnarray}
\nonumber
r_H=R_H-\frac{a^2}{R_H}.
\end{eqnarray}
Observe that for extremal Kerr $r_H\rightarrow 0$. The coordinates $\theta,\varphi$ and $t$
are the same for both parametrizations.



\section{ Asymptotic expansions}
\label{coef_asym_exp}

An aproximate form of the solution can easily be constructed on the boundary of the domain of integration.
Of special interest is the expansion as $r\to \infty$,
which is used to evaluate the expression of the
  quadrupole moment.
  
Starting with the scalar field, its asymptotic behaviour is of the form
\begin{equation}
\phi(r,\theta)=f(\theta)  \frac{e^{-\sqrt{\mu^2-w^2}r}}{r}+\dots,
\label{scalar_asy}
\end{equation}
as to describe bound state solutions of the KG equation, which requires $w <\mu $.
The function $f(\theta)$
which enters the asymptotics of the scalar field can be expressed as
\begin{eqnarray}
\label{r-infty-s2}
f(\theta)=\sum_{k=0}^{\infty}f_k P_{m+2k}^{m}(\cos \theta),
 \end{eqnarray} 
 where 
$P_{m+2k}^{m}(\cos \theta)$ are the associated Legendre functions
and $f_k$ are constants fixed by numerics. 

One can also construct an approximate expression for the metric functions 
in inverse powers of $r$  
(note that for the purposes
here the scalar field contribution can be neglected,
 since $\phi$ decays faster than any power of $r$). 
The  leading order terms in such asymptotic expansion read
\begin{eqnarray}
\label{r-infty}
\nonumber
&&
F_0(r,\theta)=\frac{c_t}{r}
+\frac{c_t r_H}{2 r^2}
+\frac{f_{03}(\theta)}{r^3}
+\dots,
\\
&&
\nonumber
F_1(r,\theta)=-\frac{c_t}{r}
+\frac{f_{12}(\theta)}{2 r^2}
+\frac{f_{13}(\theta)}{r^3}
+\dots,
\\
&&
\label{inf-1}
F_2(r,\theta)=-\frac{c_t}{r}
+\frac{a_5-\frac{1}{4}c_t(c_t+r_H)}{r^2}
+\frac{f_{23}(\theta)}{r^3}
+\dots,
\\
&&
\nonumber
W(r,\theta)=\frac{c_\varphi}{r^3}
+\frac{3c_\varphi c_t }{r^4}
+\frac{w_5(\theta)}{r^5}
+\dots,
\end{eqnarray} 
where the expressions for the $\theta$-dependent coefficients are:
\begin{eqnarray}
\label{r-infty-s1}
\nonumber
&&
f_{03}(\theta)=b_1+\frac{c_t r_H^2}{2}-\frac{1}{8}\left(24 b_1+8 a_5 c_t-2c_t^3-4a_5 r_H+3 c_t^2 r_H+3c_t r_H^2 \right)\cos^2 \theta,
\\
&&
\nonumber
f_{12}(\theta)= -\frac{c_t}{4}(c_t+r_H)+a_5\cos 2\theta,
\\
&&
\nonumber
f_{13}(\theta)=  \frac{1}{16}\bigg(
8 b_1+a_5(8c_t-4 r_H)
-c_t(2c_t^2+c_tr_H+r_H^2)
\\
\nonumber
&&
{~~~~~~~~~~~~~~~~~~~~~}+
(
24 b_1+8 a_5c_t-2c_t^3+12 a_5r_H+3c_t^2r_H+3c_tr_H^2
)\cos 2\theta
\bigg) ,
\\
&&
\label{inf-2}
f_{23}(\theta)=  
\frac{1}{16}\bigg(
8 b_1+4a_5(2c_t+3 r_H)
-c_t(2c_t^2+c_tr_H+r_H^2)
\\
\nonumber
&&
{~~~~~~~~~~~~~~~~~~~~~}+
(
24 b_1+8 a_5c_t-2c_t^3-4 a_5r_H+3c_t^2r_H+3c_tr_H^2
)\cos 2\theta
\bigg) ,
\\
&&
\nonumber
w_{5}(\theta)=  
\frac{1}{20}\bigg(
 -36 a_5 c_\varphi +15 c_\varphi c_t(7c_t+r_H)+45 w_t+75 w_t \cos 2\theta
 \bigg) .
\end{eqnarray} 
In these expressions, 
$c_t$,
$c_\varphi$
$a_5$,
$b_1$
and
$w_t$
are arbitrary constants
which are extracted from the numerical output.
Note that for the Kerr BH, the solution
have
\begin{eqnarray}
\label{r-infty-sa1}
\nonumber
&&w_t=\frac{2}{15}c_t\sqrt{c_t(c_t-r_H)}(c_t-r_H)(2c_t-r_H),~~
b_1=\frac{1}{6}c_t^2(2c_t-3r_H),~~
\\
\nonumber
&&
a_5=\frac{1}{4}c_t(c_t-r_H),~~
c_\varphi=\sqrt{c_t(c_t-r_H)}(r_H-2c_t).
\end{eqnarray}

\section{Reference solutions plots}
\label{sec_plots}
In this appendix we provide the plots mentioned in Section~\ref{sec_sample}.

\begin{figure}[h!]
\centering
\includegraphics[height=1.69in]{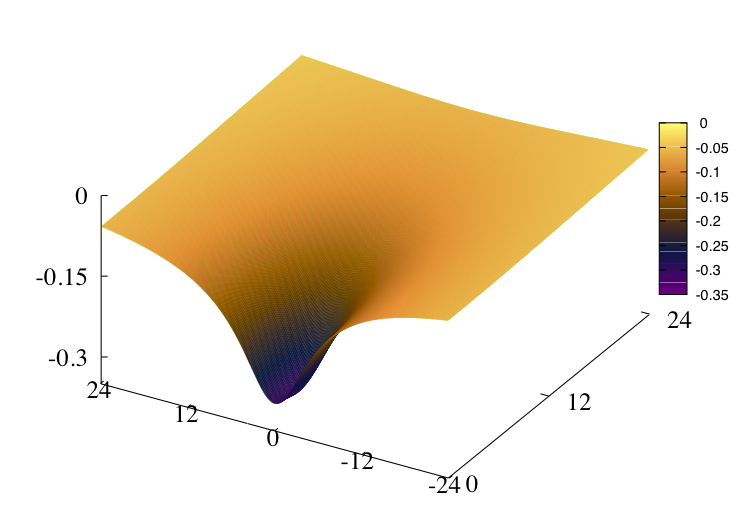}\ \ \ \ \ 
\includegraphics[height=1.69in]{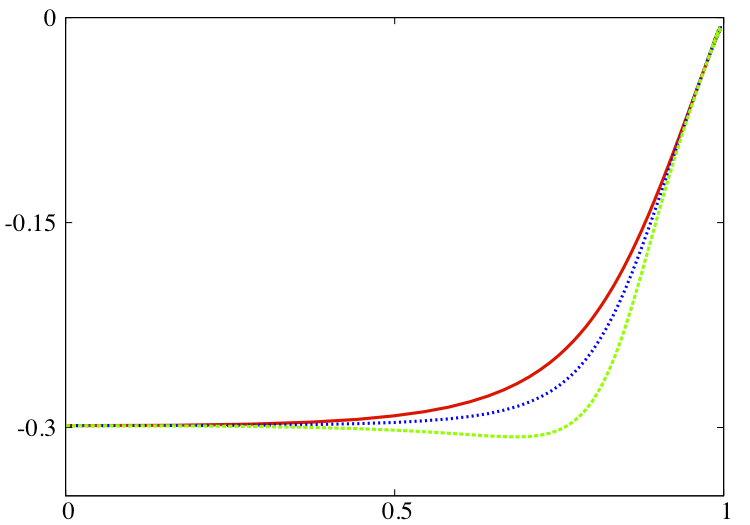}\\
\includegraphics[height=1.69in]{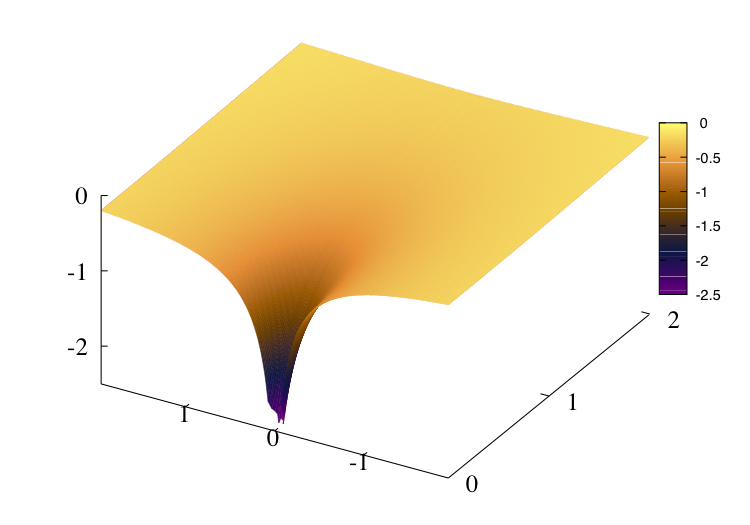}\  \ \ \ \ 
\includegraphics[height=1.69in]{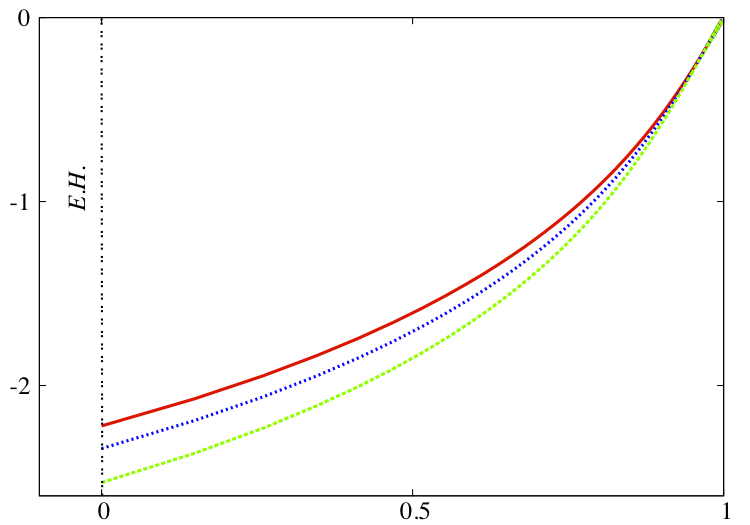}\\
\includegraphics[height=1.69in]{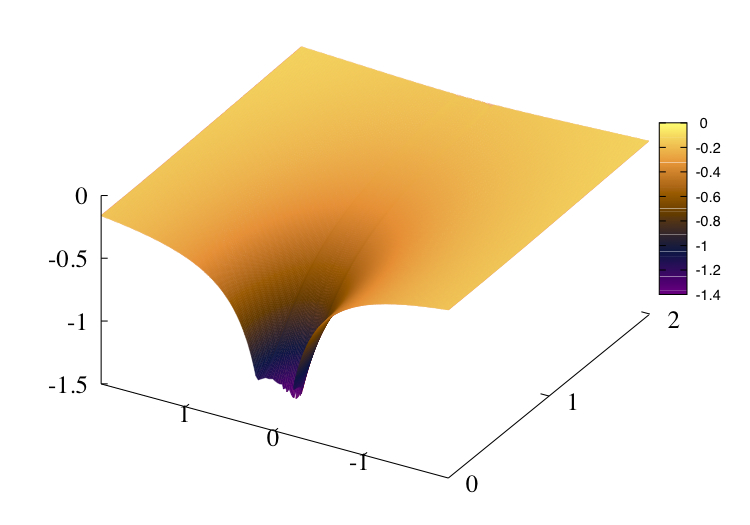}\  \ \ \ \ 
\includegraphics[height=1.69in]{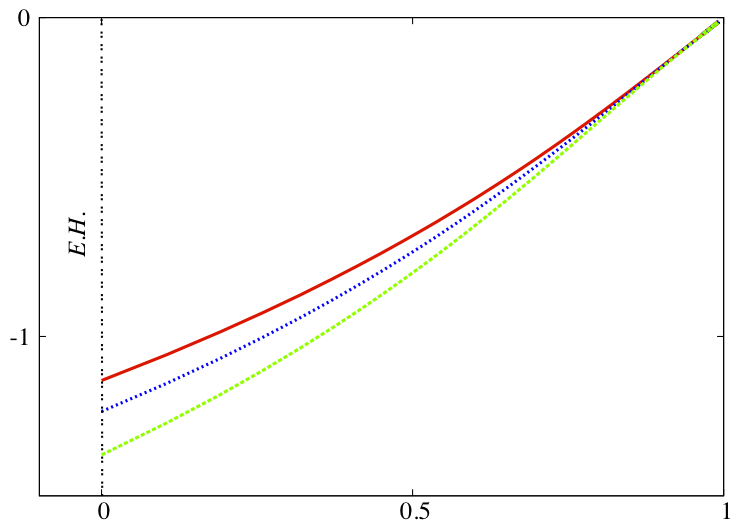}\\
\includegraphics[height=1.69in]{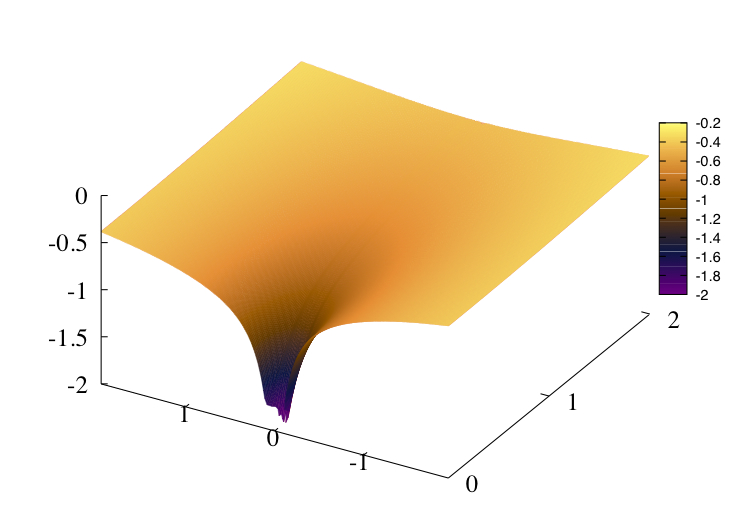}\ \ \ \ \ 
\includegraphics[height=1.69in]{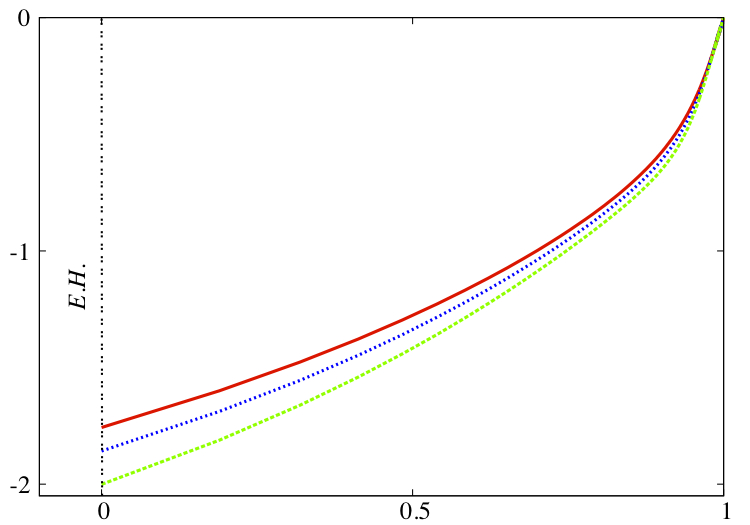}\\
\includegraphics[height=1.69in]{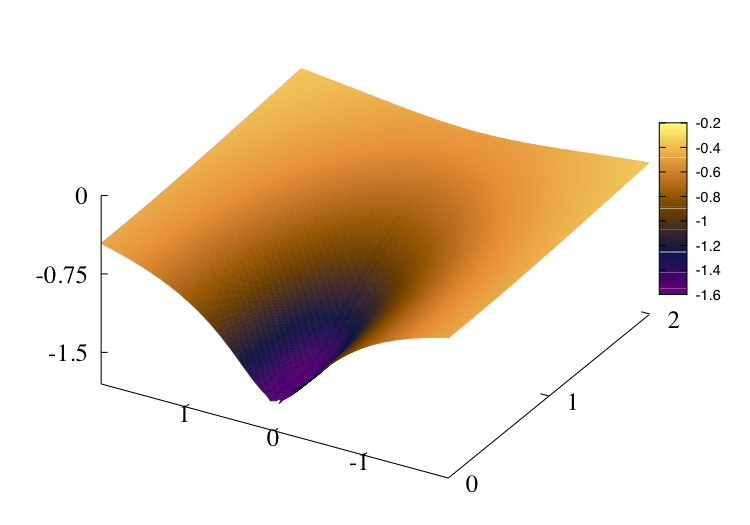}\ \ \ \ \ 
\includegraphics[height=1.69in]{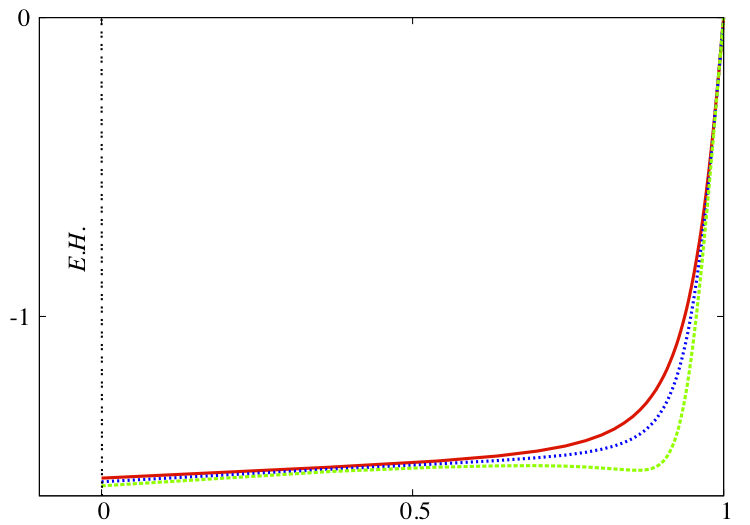}\\
\caption{The metric function $F_0$ for example solutions {\bf I}--{\bf V}, $cf.$ Section~\ref{sec_sample}.} 
\label{F0}
\end{figure}

\newpage

\begin{figure}[h!]
\centering
\includegraphics[height=1.69in]{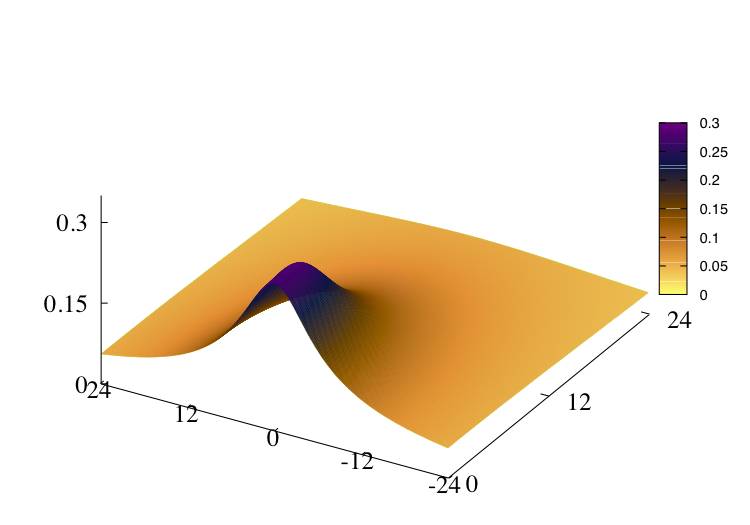}\ \ \ \ \ 
\includegraphics[height=1.69in]{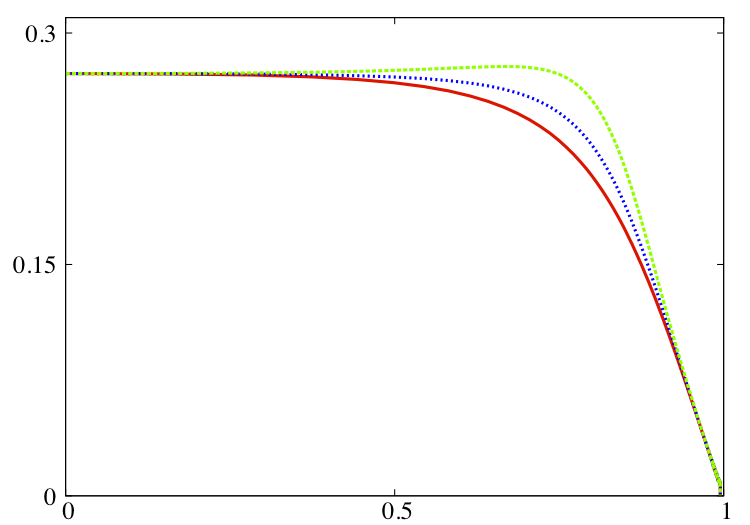}\\
\includegraphics[height=1.69in]{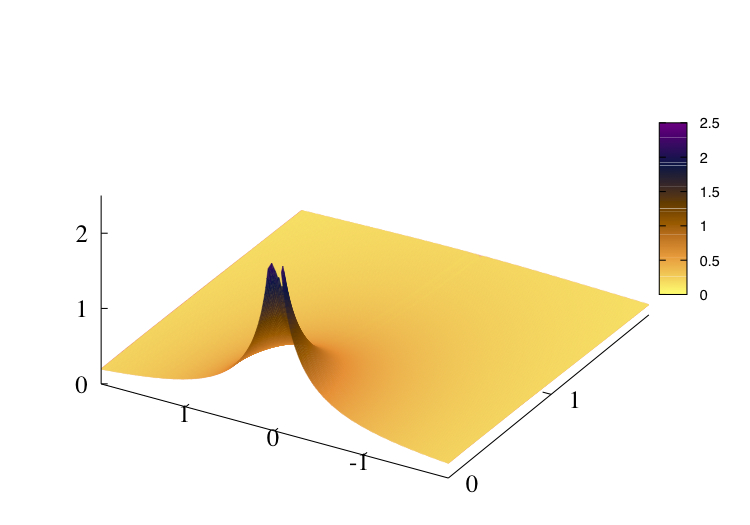}\  \ \ \ \ 
\includegraphics[height=1.69in]{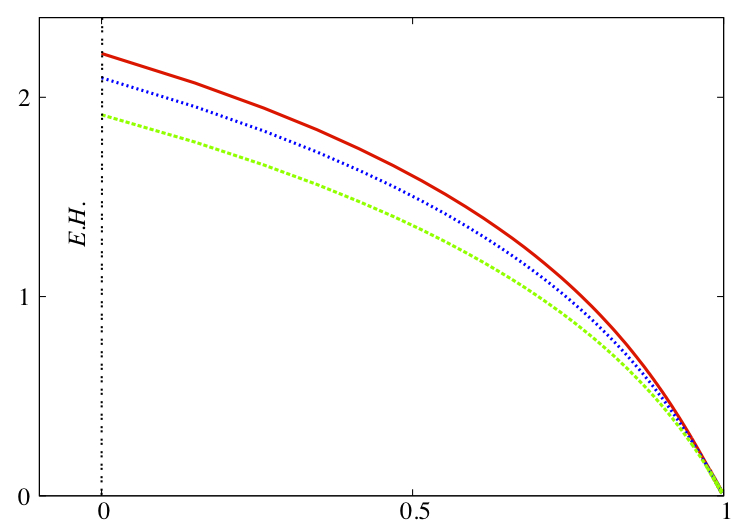}\\
\includegraphics[height=1.69in]{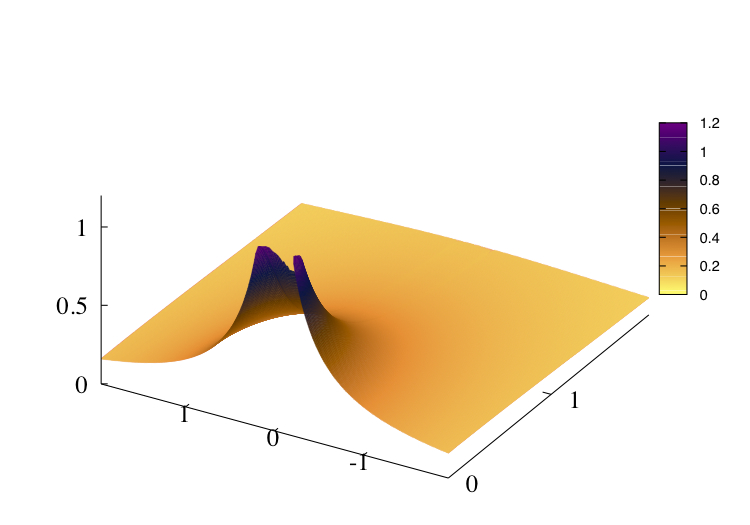}\  \ \ \ \ 
\includegraphics[height=1.69in]{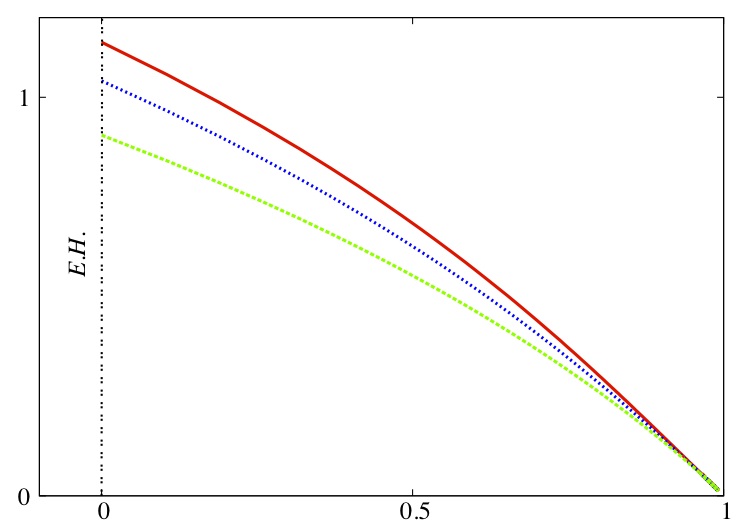}\\
\includegraphics[height=1.69in]{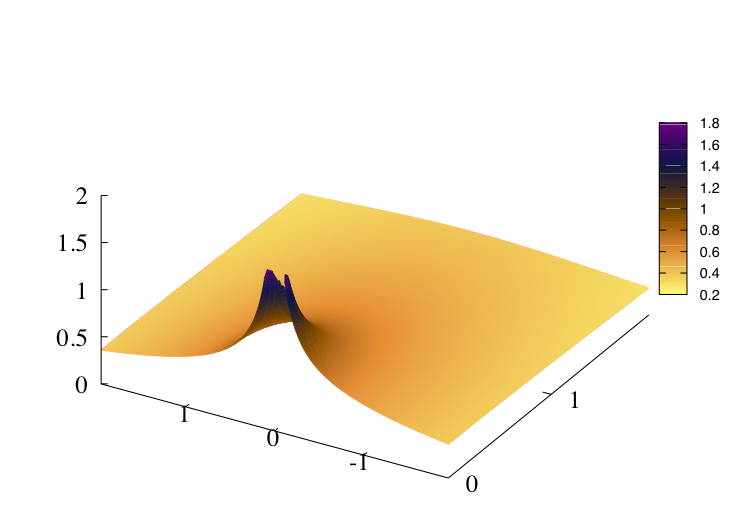}\ \ \ \ \ 
\includegraphics[height=1.69in]{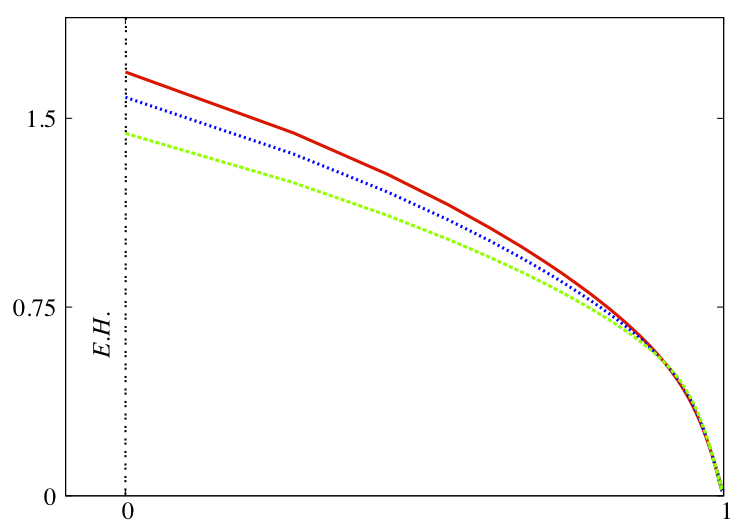}\\
\includegraphics[height=1.69in]{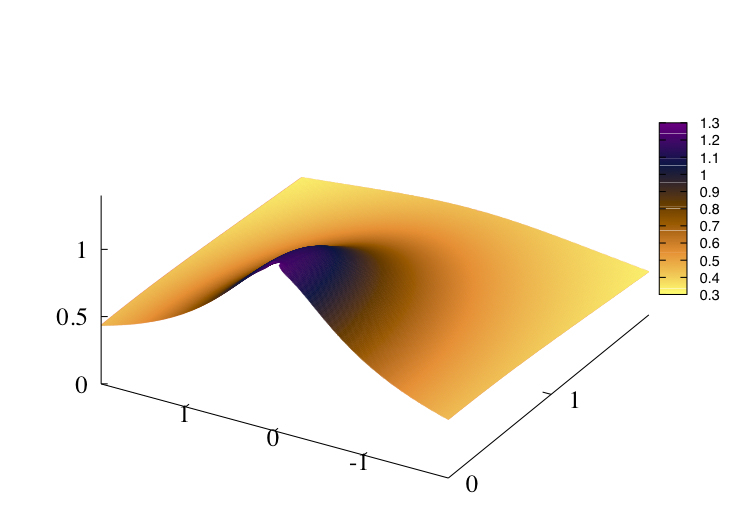}\ \ \ \ \ 
\includegraphics[height=1.69in]{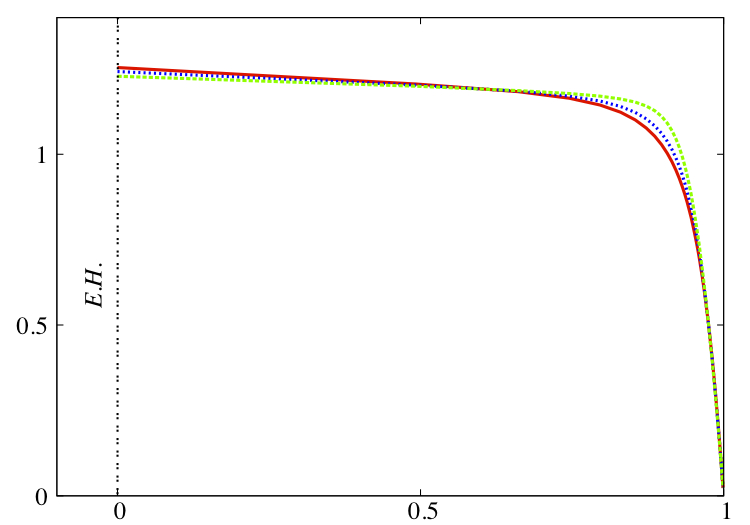}\\
\caption{The metric function $F_1$ for example solutions {\bf I}--{\bf V}, $cf.$ Section~\ref{sec_sample}.
} 
\label{F1}
\end{figure}

\newpage

\begin{figure}[h!]
\centering
\includegraphics[height=1.69in]{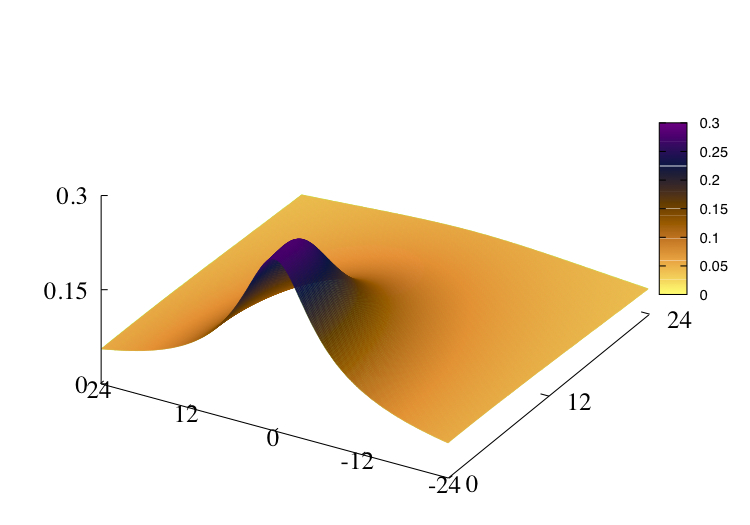}\ \ \ \ \ 
\includegraphics[height=1.69in]{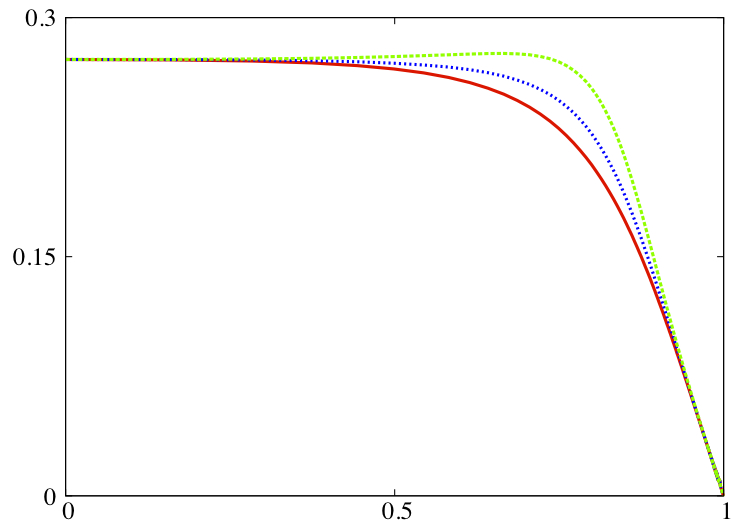}\\
\includegraphics[height=1.69in]{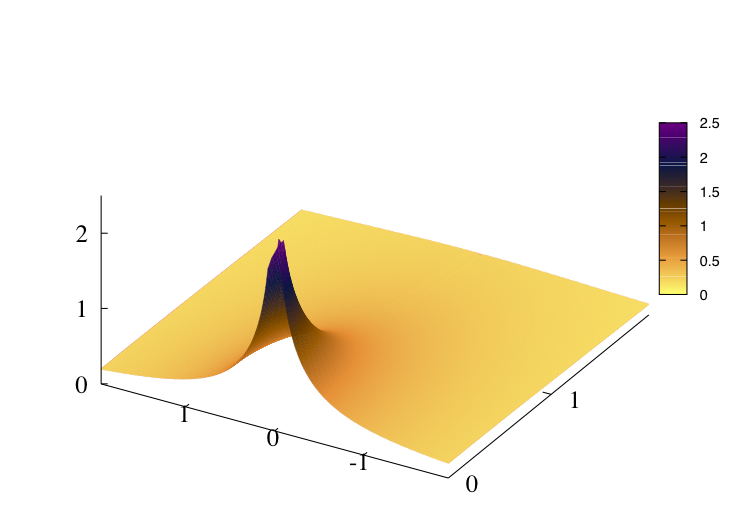}\  \ \ \ \ 
\includegraphics[height=1.69in]{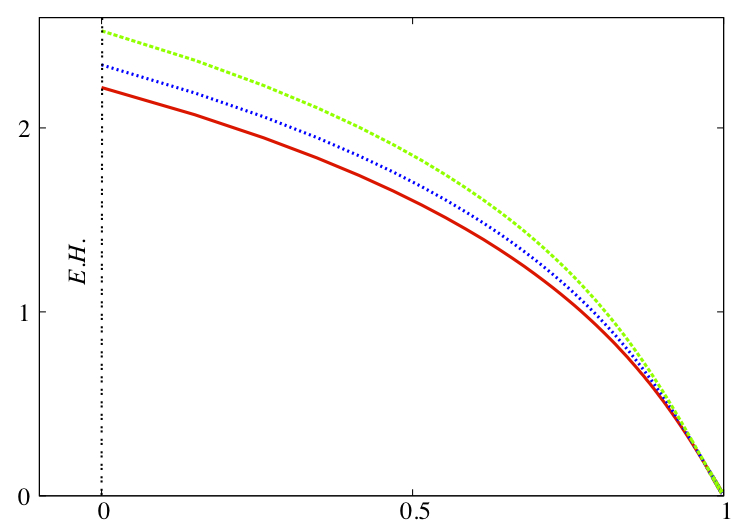}\\
\includegraphics[height=1.69in]{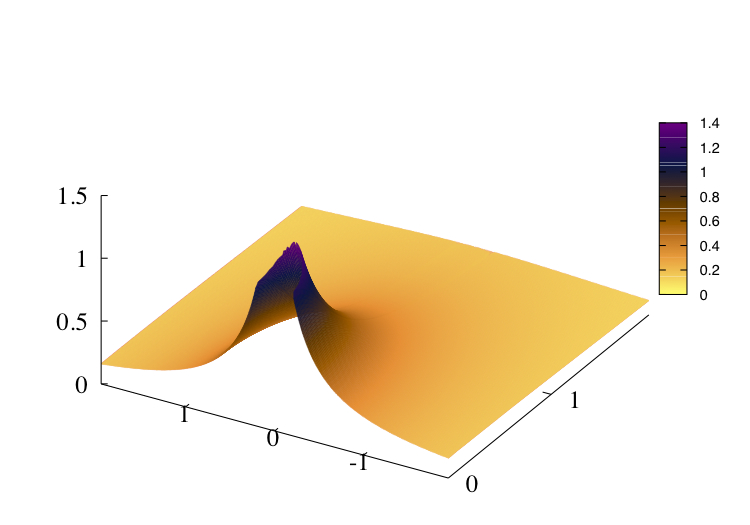}\  \ \ \ \ 
\includegraphics[height=1.69in]{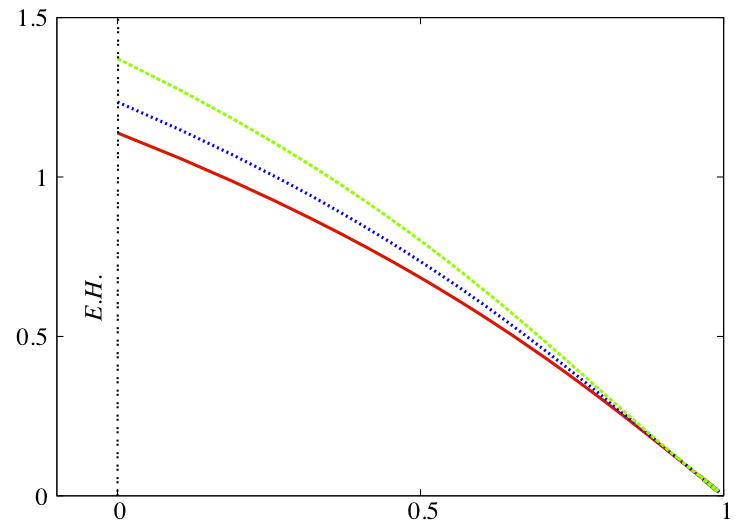}\\
\includegraphics[height=1.69in]{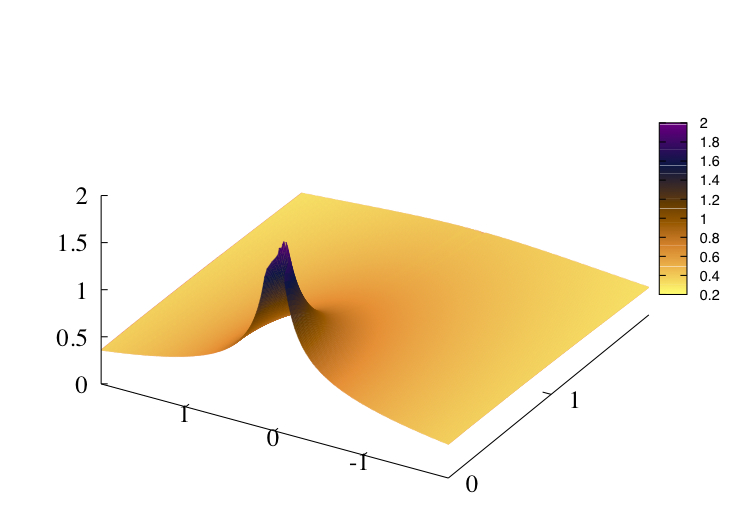}\ \ \ \ \ 
\includegraphics[height=1.69in]{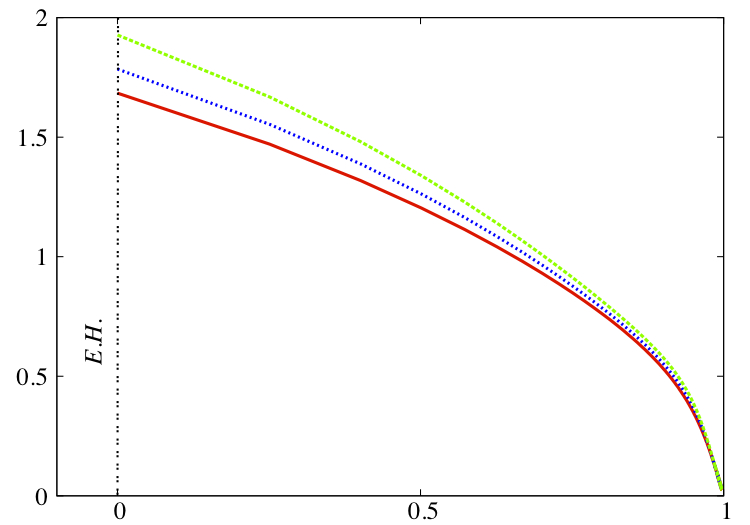}\\
\includegraphics[height=1.69in]{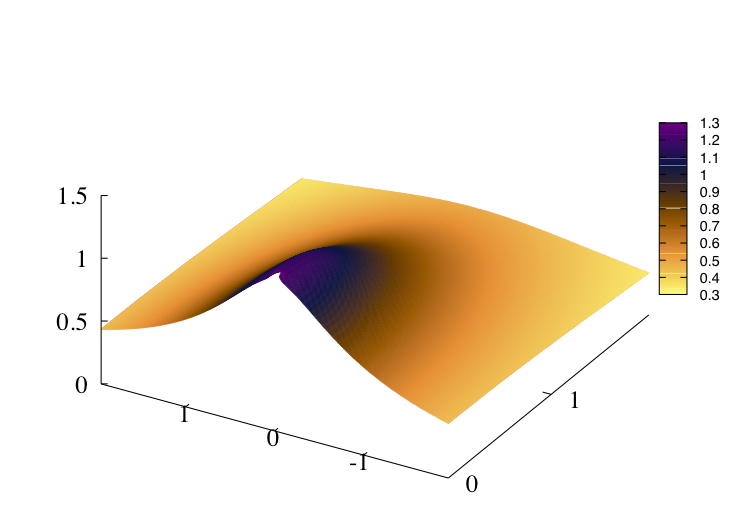}\ \ \ \ \ 
\includegraphics[height=1.69in]{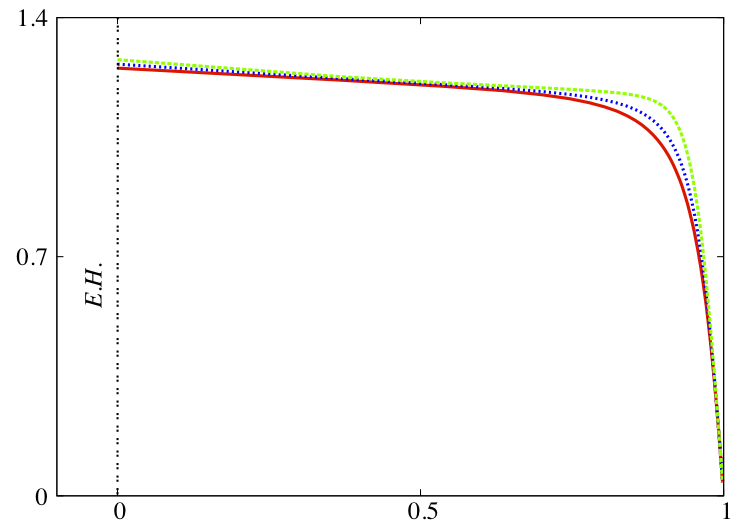}\\
\caption{The metric function $F_2$ for example solutions {\bf I}--{\bf V}, $cf.$ Section~\ref{sec_sample}.
} 
\label{F2}
\end{figure}

\newpage

\begin{figure}[h!]
\centering
\includegraphics[height=1.69in]{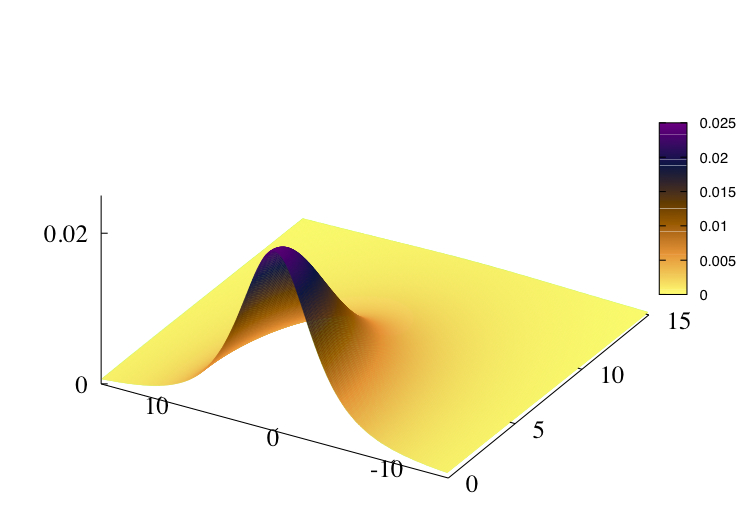}\ \ \ \ \ 
\includegraphics[height=1.69in]{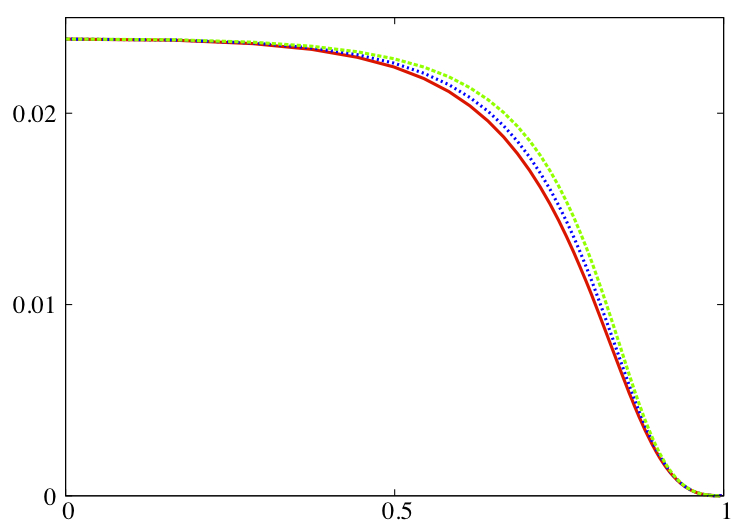}\\
\includegraphics[height=1.69in]{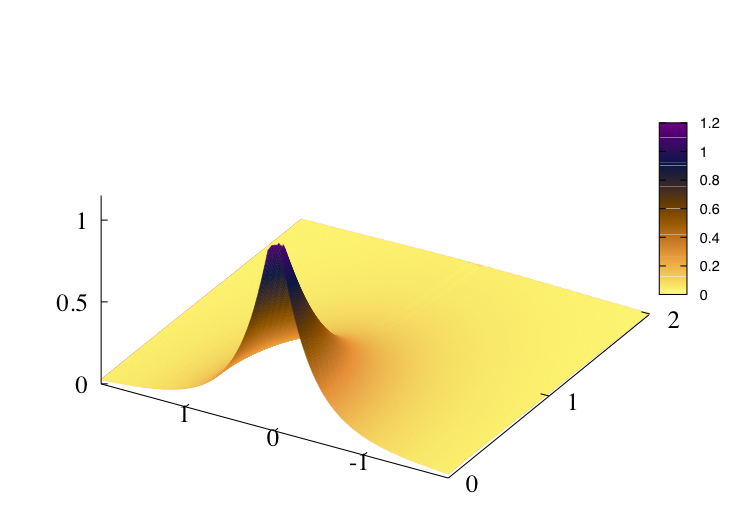}\  \ \ \ \ 
\includegraphics[height=1.69in]{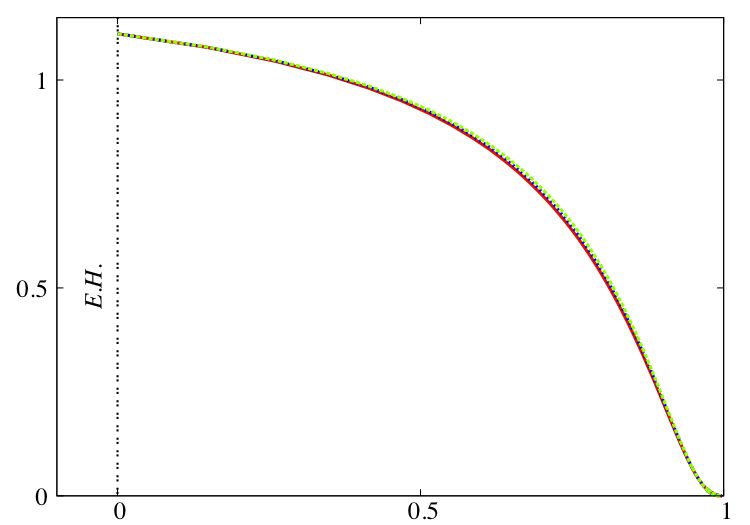}\\
\includegraphics[height=1.69in]{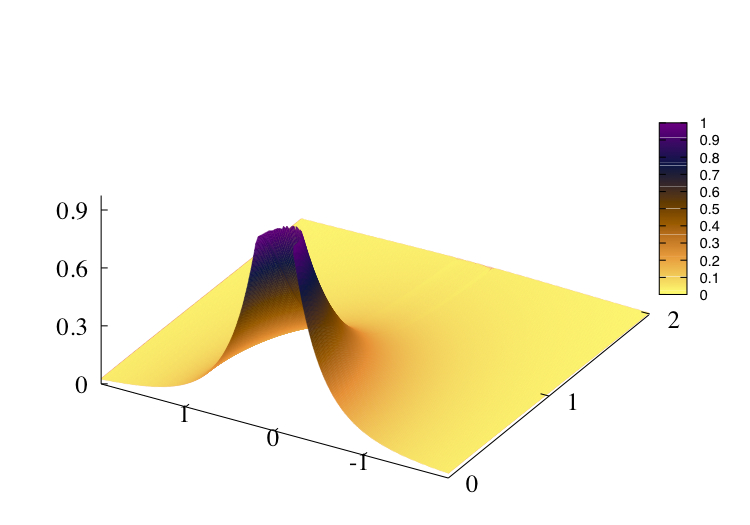}\  \ \ \ \ 
\includegraphics[height=1.69in]{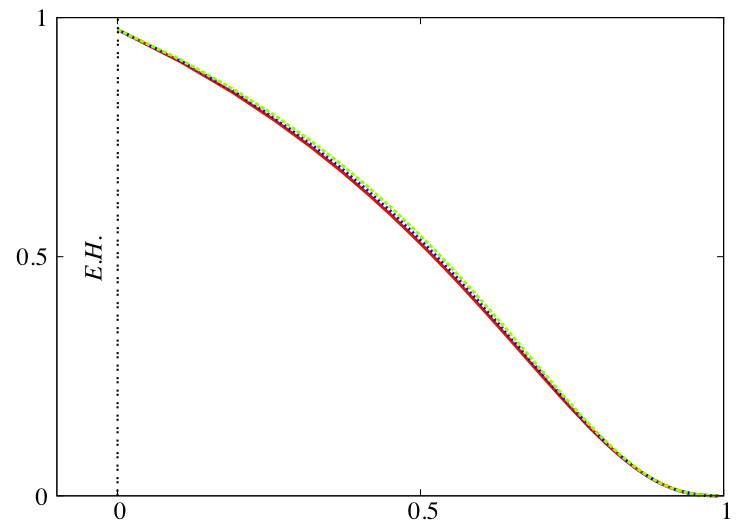}\\
\includegraphics[height=1.69in]{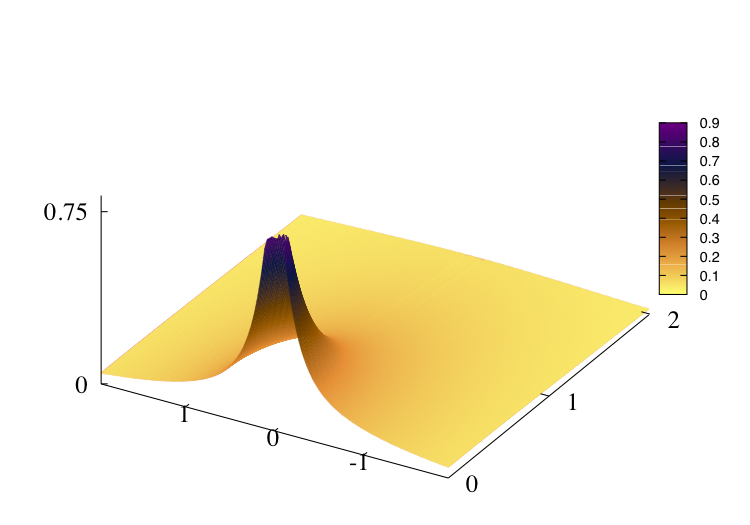}\ \ \ \ \ 
\includegraphics[height=1.69in]{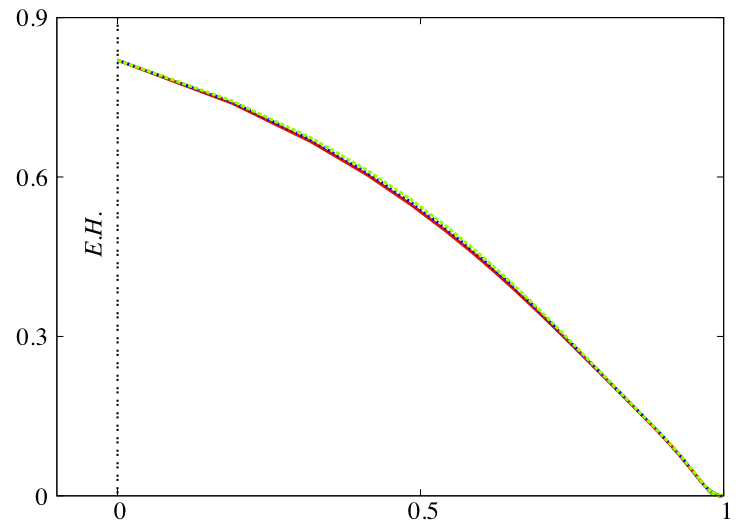}\\
\includegraphics[height=1.69in]{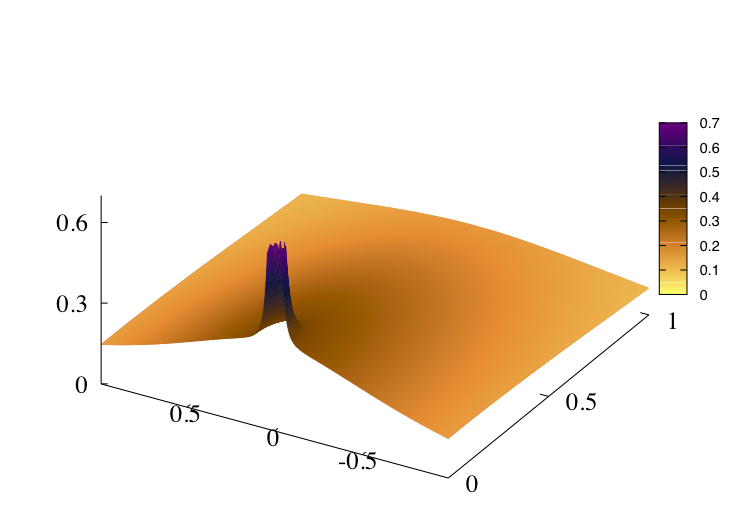}\ \ \ \ \ 
\includegraphics[height=1.69in]{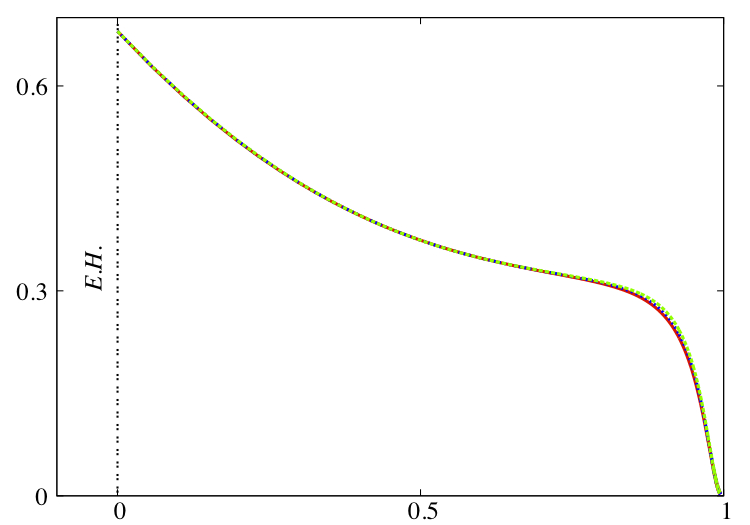}\\
\caption{The metric function $W$ for example solutions {\bf I}--{\bf V}, $cf.$ Section~\ref{sec_sample}.} 
\label{W}
\end{figure}

\newpage

\begin{figure}[h!]
\centering
\includegraphics[height=1.69in]{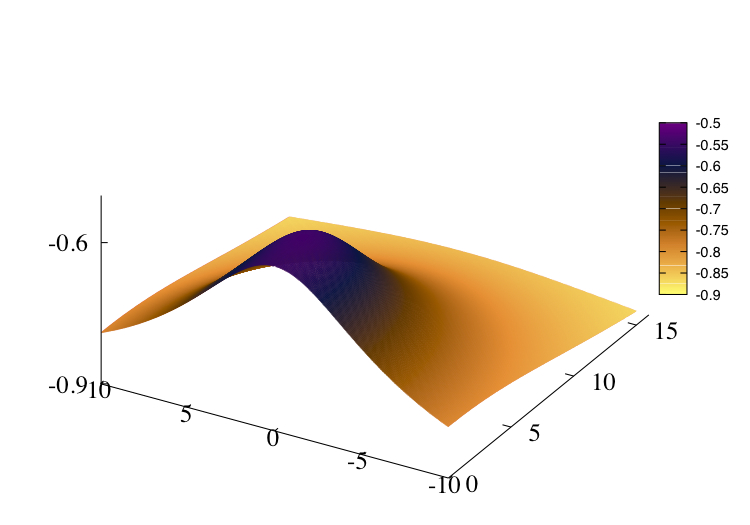}\ \ \ \ \ 
\includegraphics[height=1.69in]{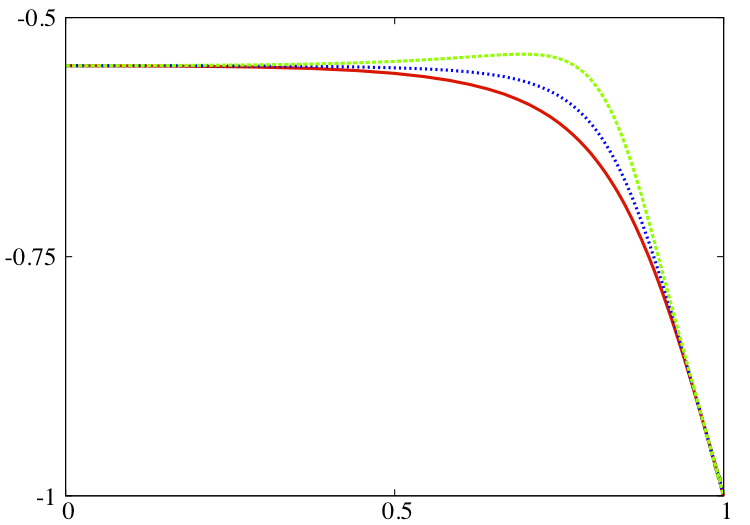}\\
\includegraphics[height=1.69in]{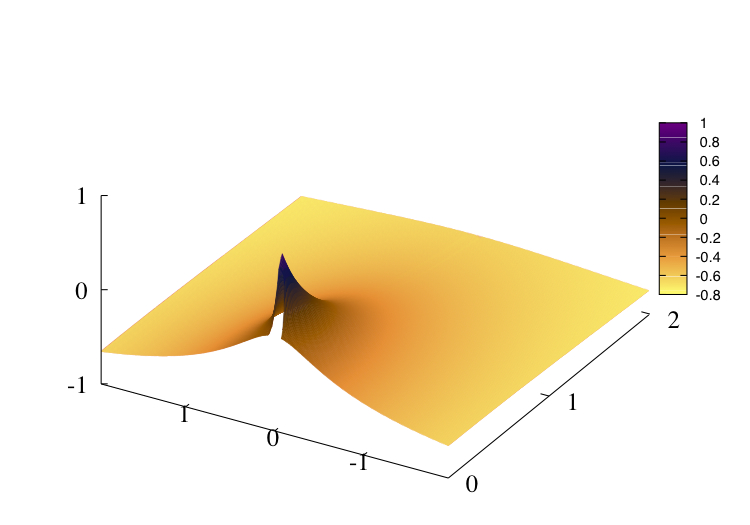}\  \ \ \ \ 
\includegraphics[height=1.69in]{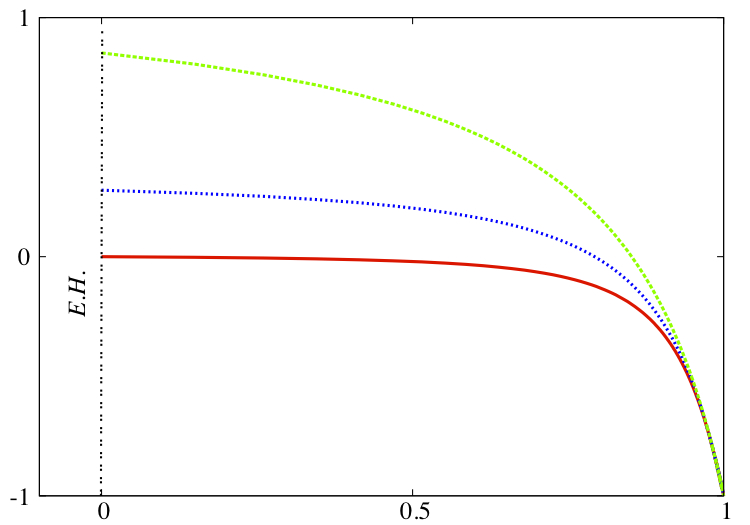}\\
\includegraphics[height=1.69in]{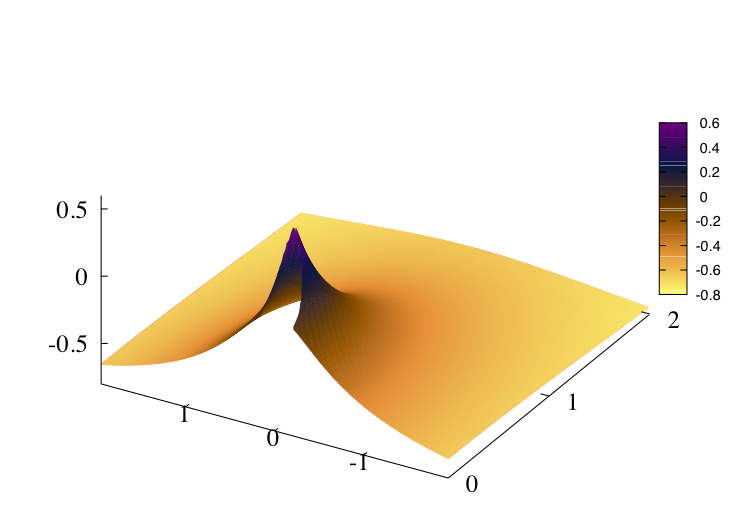}\  \ \ \ \ 
\includegraphics[height=1.69in]{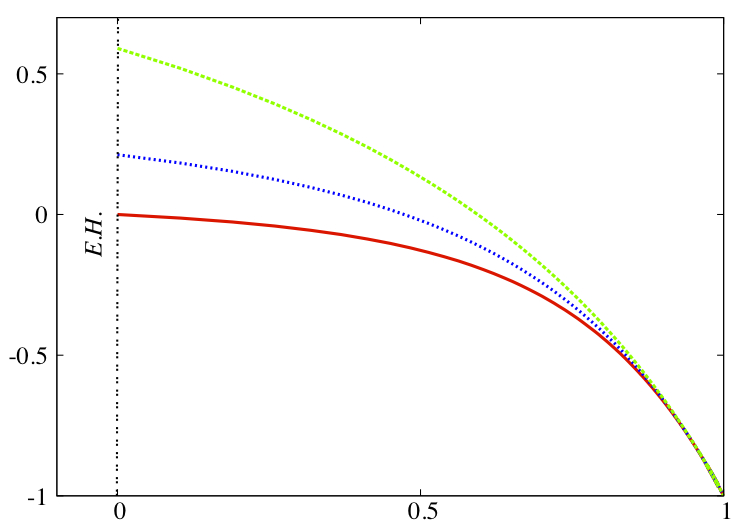}\\
\includegraphics[height=1.69in]{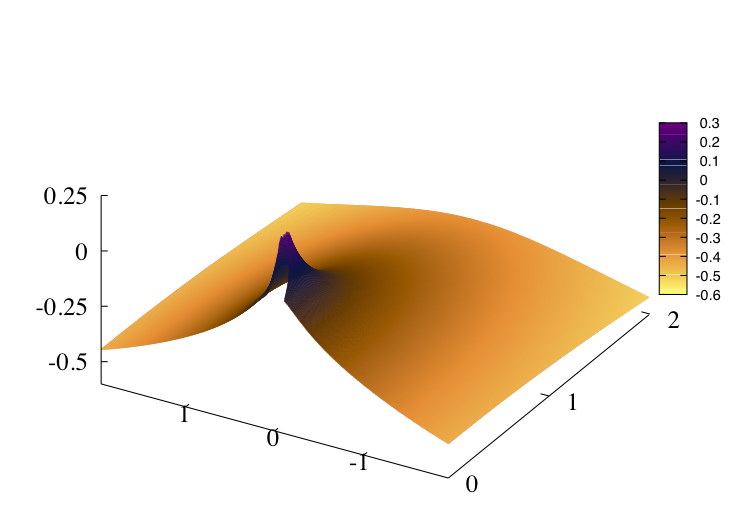}\ \ \ \ \ 
\includegraphics[height=1.69in]{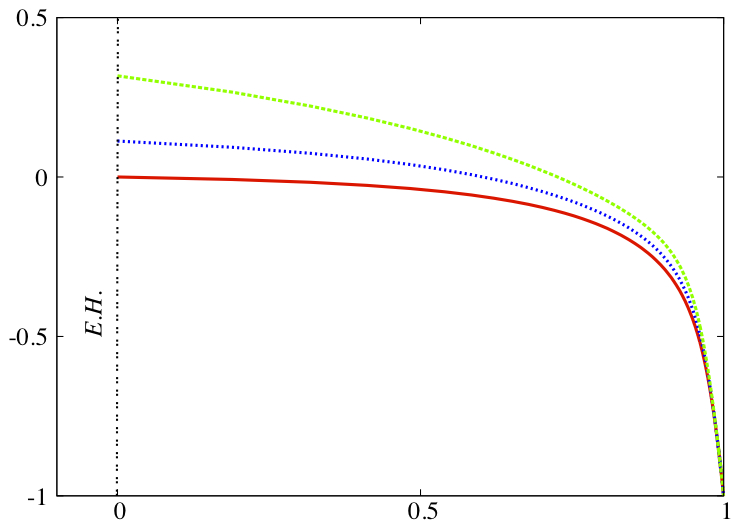}\\
\includegraphics[height=1.69in]{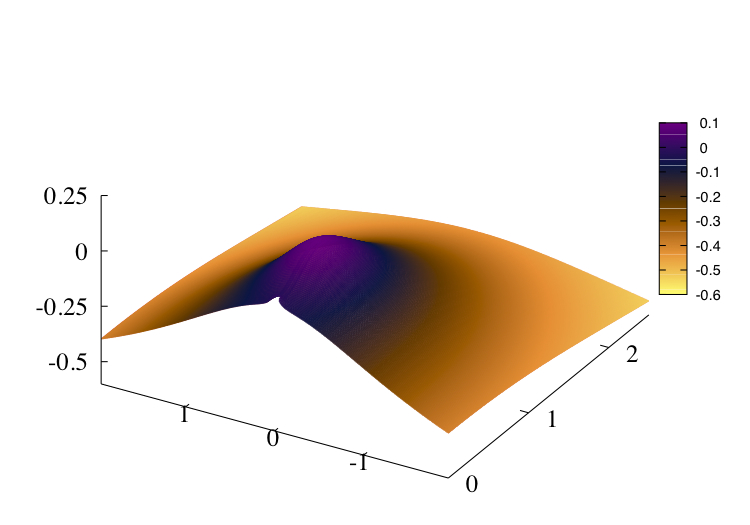}\ \ \ \ \ 
\includegraphics[height=1.69in]{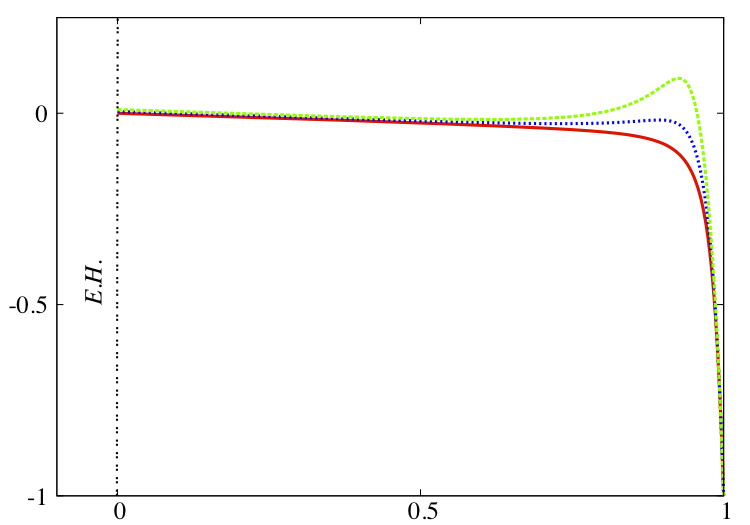}\\
\caption{The metric coefficient $g_{tt}$ for example solutions {\bf I}--{\bf V}, $cf.$ Section~\ref{sec_sample}.} 
\label{gtt}
\end{figure}

\newpage

\begin{figure}[h!]
\centering
\includegraphics[height=1.99in]{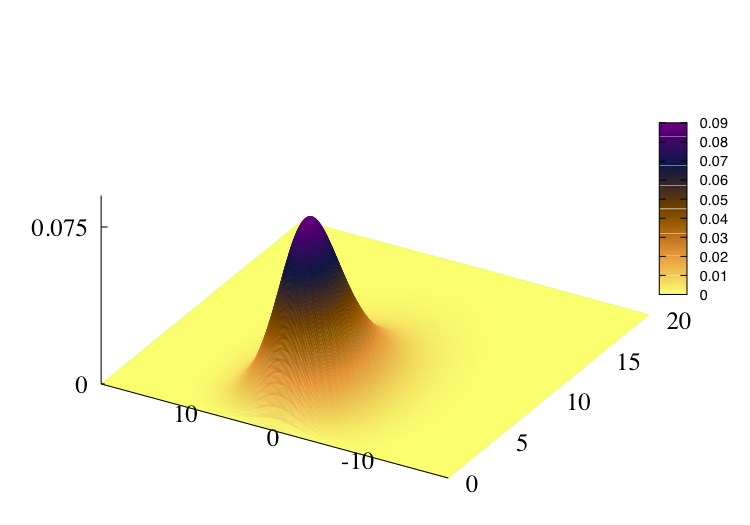}\ \ \ \ \ 
\includegraphics[height=1.69in]{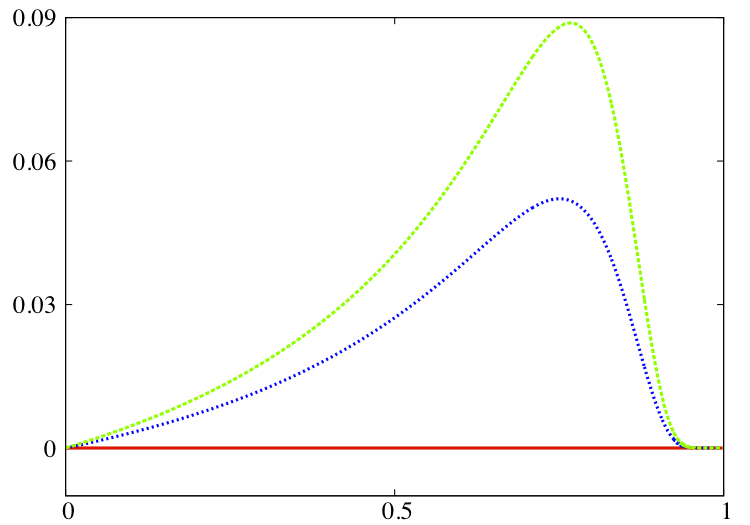}\\
\includegraphics[height=1.99in]{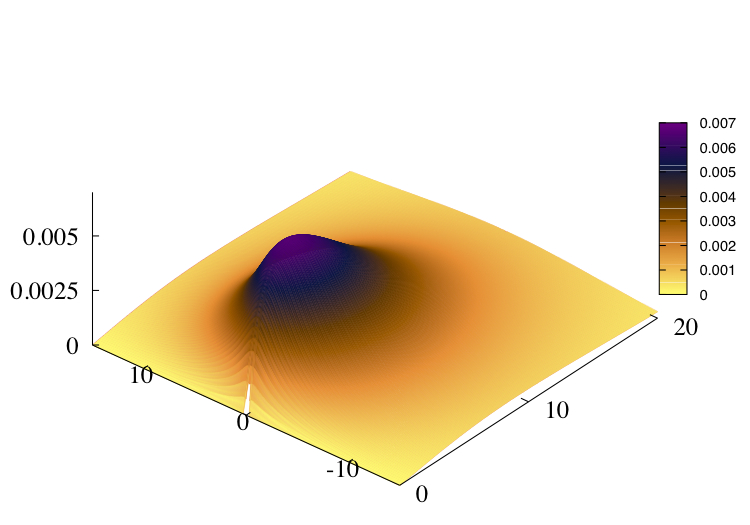}\  \ \ \ \ 
\includegraphics[height=1.69in]{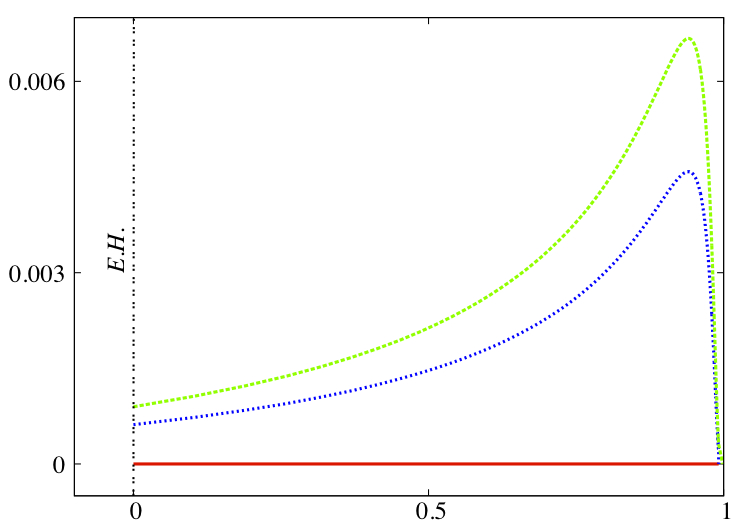}\\
\includegraphics[height=1.99in]{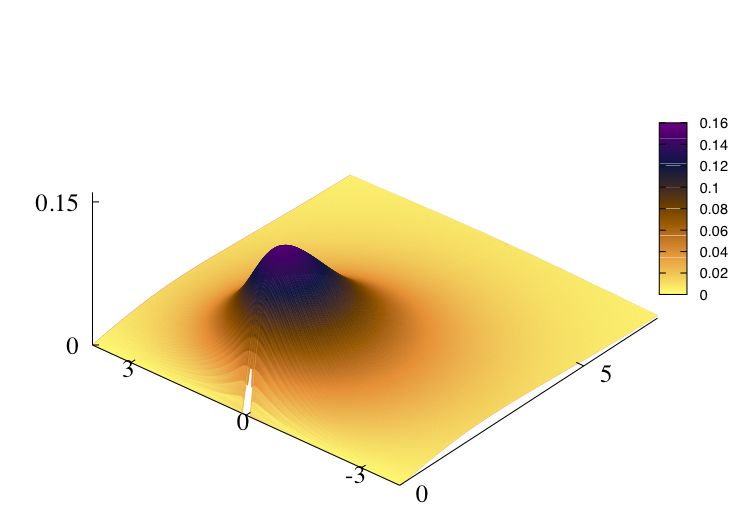}\ \ \ \ \ 
\includegraphics[height=1.69in]{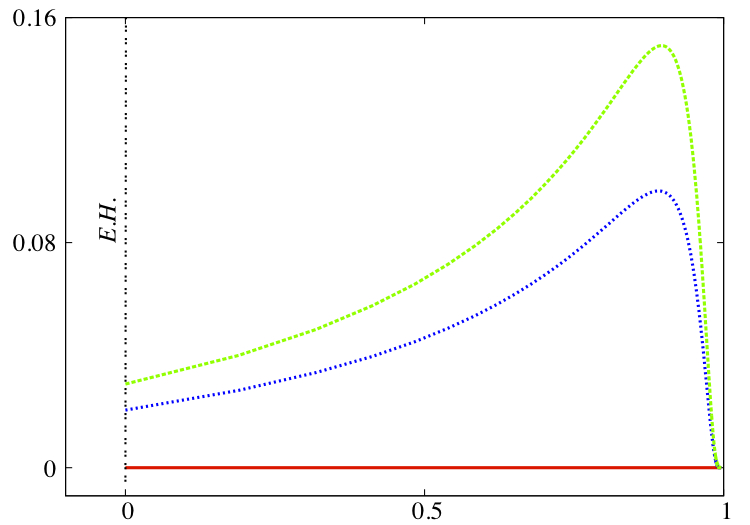}\\
\includegraphics[height=1.99in]{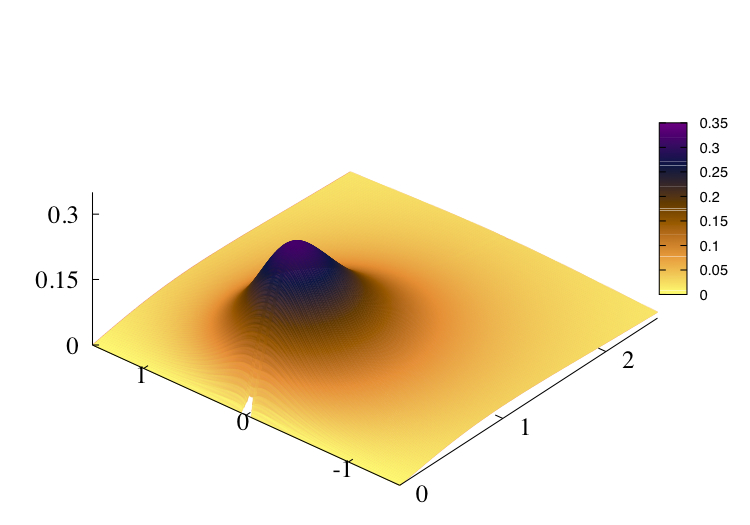}\ \ \ \ \ 
\includegraphics[height=1.69in]{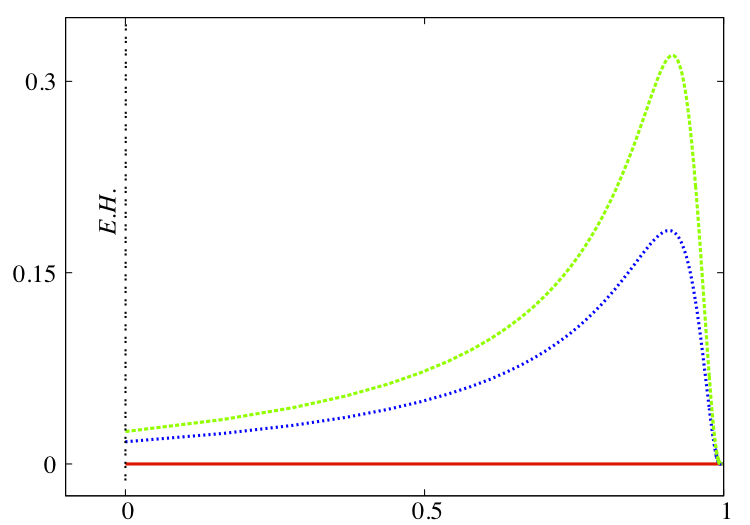}\\
\caption{The scalar field profile function $\phi$ for example solutions {\bf I} and {\bf III}--{\bf V}, $cf.$ Section~\ref{sec_sample}.
} 
\label{Z}
\end{figure}

\newpage

\begin{figure}[h!]
\centering
\includegraphics[height=1.99in]{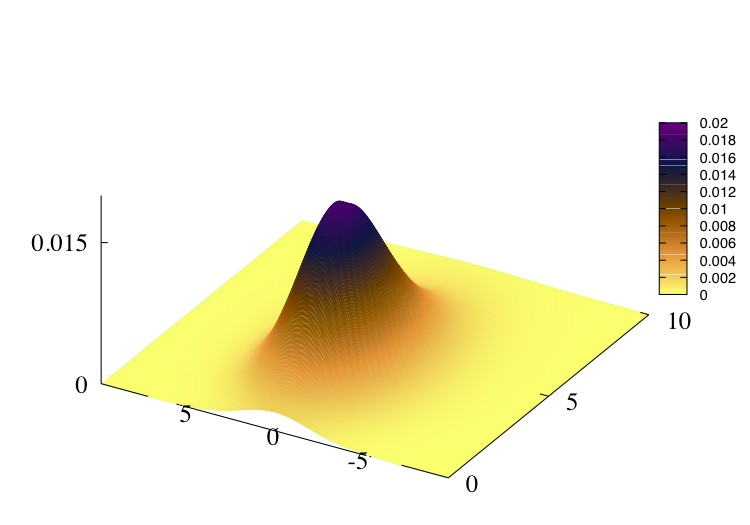}\ \ \ \ \ 
\includegraphics[height=1.69in]{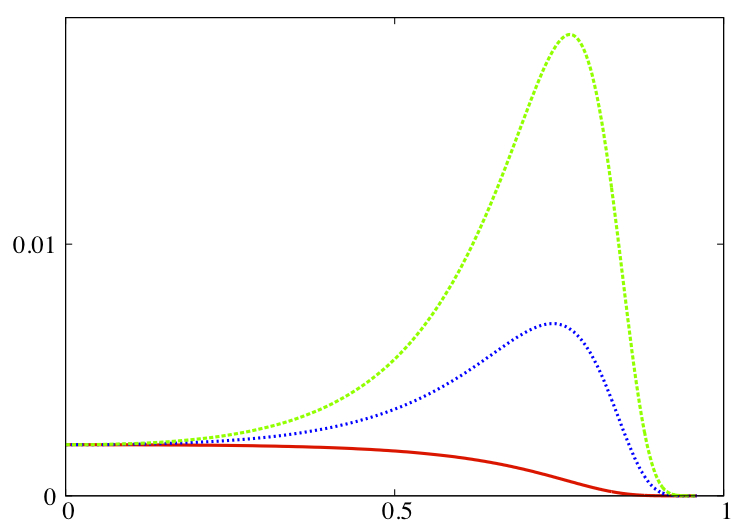}\\
\includegraphics[height=1.99in]{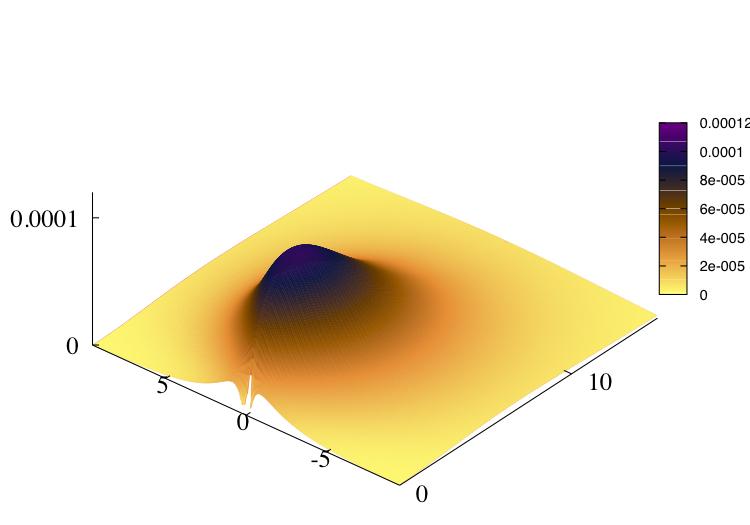}\  \ \ \ \ 
\includegraphics[height=1.69in]{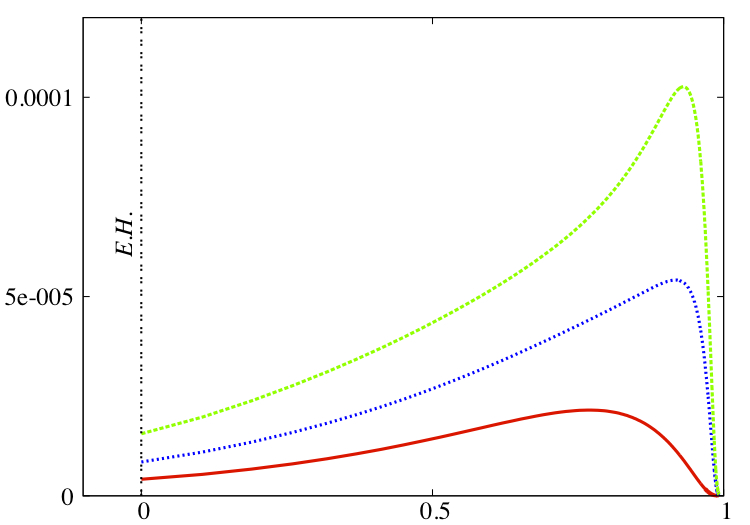}\\
\includegraphics[height=1.99in]{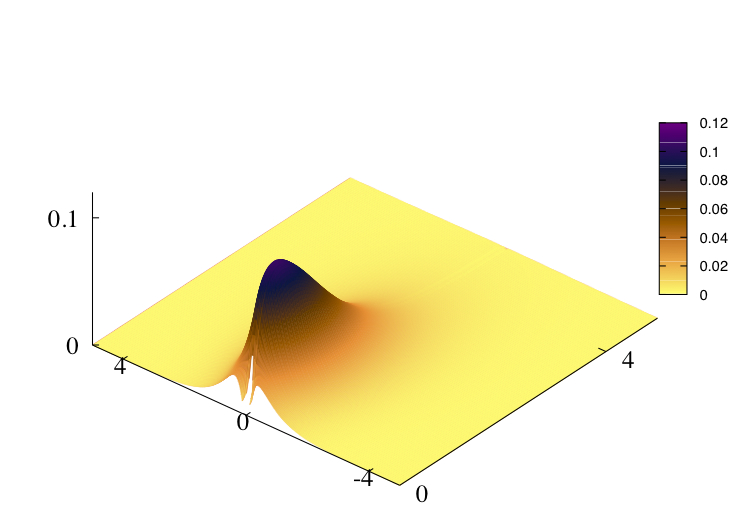}\ \ \ \ \ 
\includegraphics[height=1.69in]{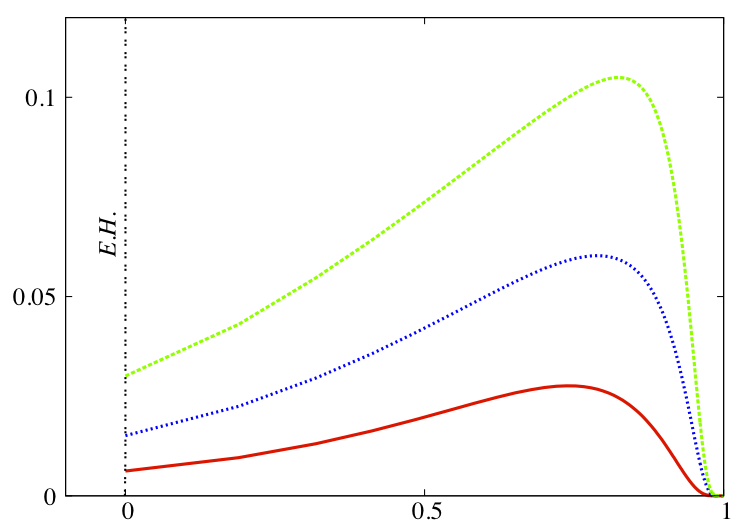}\\
\includegraphics[height=1.99in]{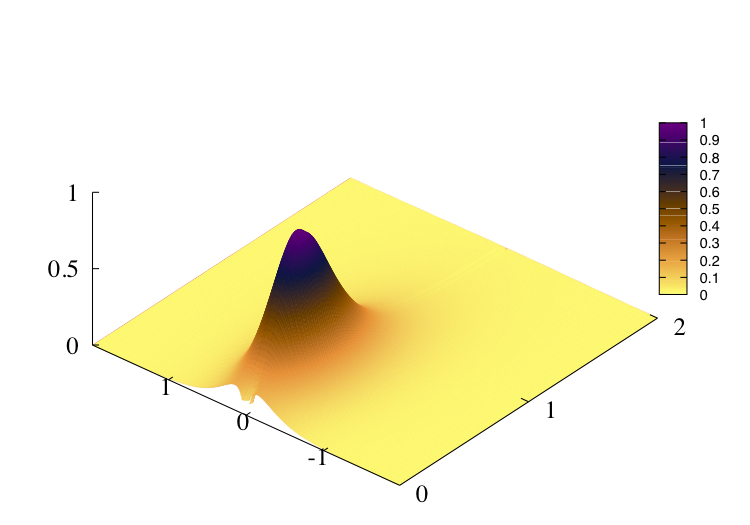}\ \ \ \ \ 
\includegraphics[height=1.69in]{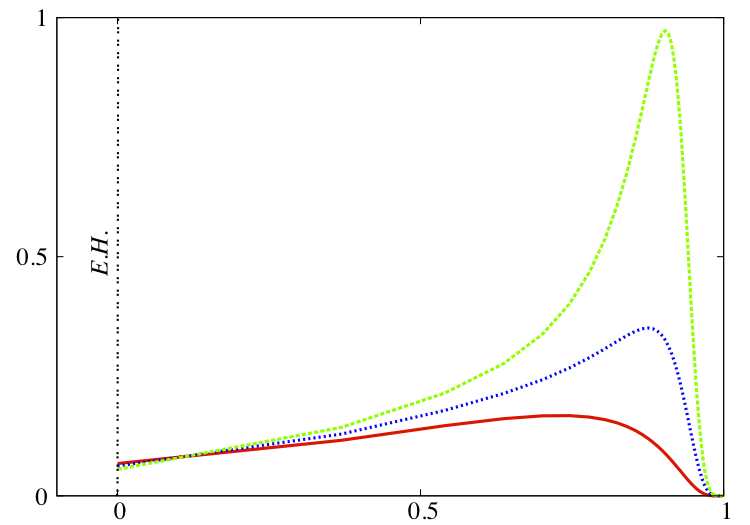}\\
\caption{The scalar ``energy density" for example solutions {\bf I} and {\bf III}--{\bf V}, $cf.$ Section~\ref{sec_sample}.
} 
\label{E}
\end{figure}

\newpage

\begin{figure}[h!]
\centering
\includegraphics[height=1.99in]{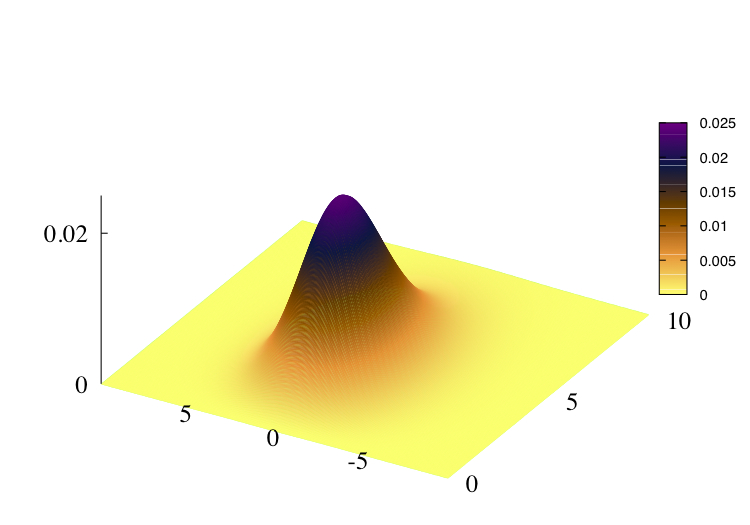}\ \ \ \ \ 
\includegraphics[height=1.69in]{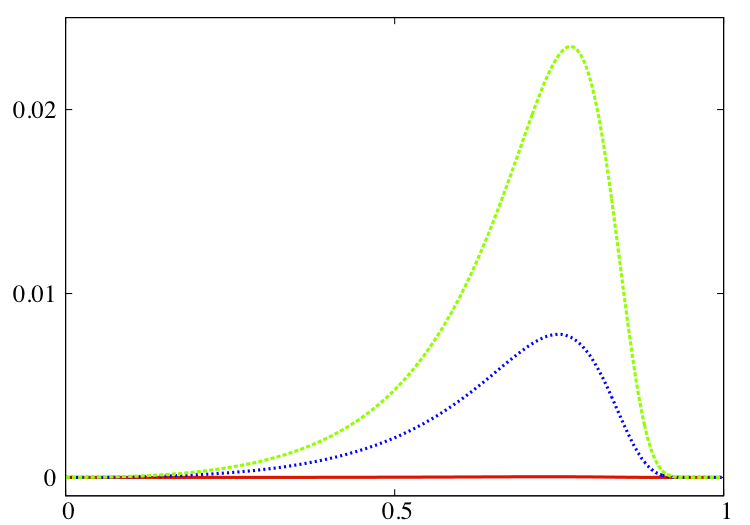}\\
\includegraphics[height=1.99in]{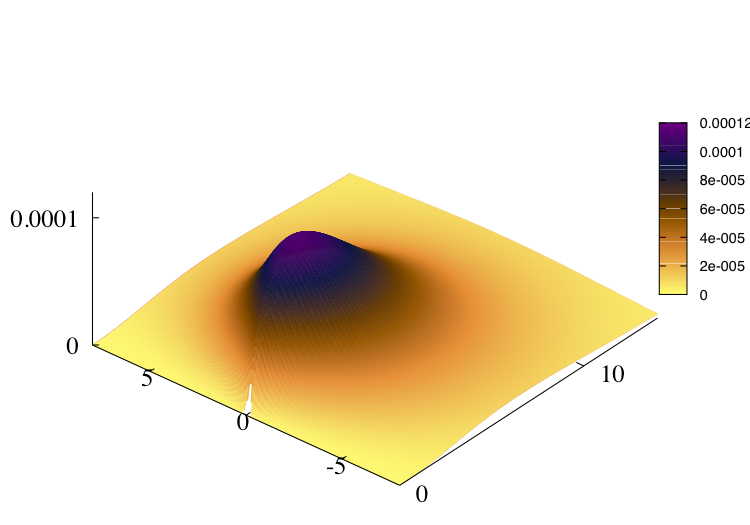}\  \ \ \ \ 
\includegraphics[height=1.69in]{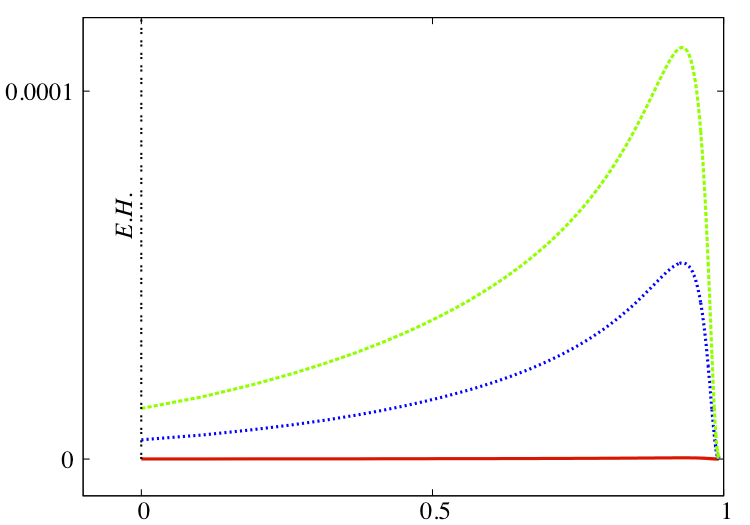}\\
\includegraphics[height=1.99in]{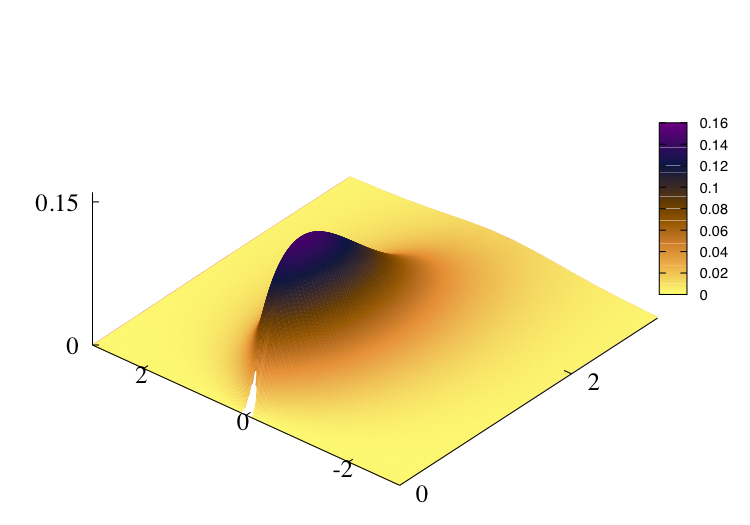}\ \ \ \ \ 
\includegraphics[height=1.69in]{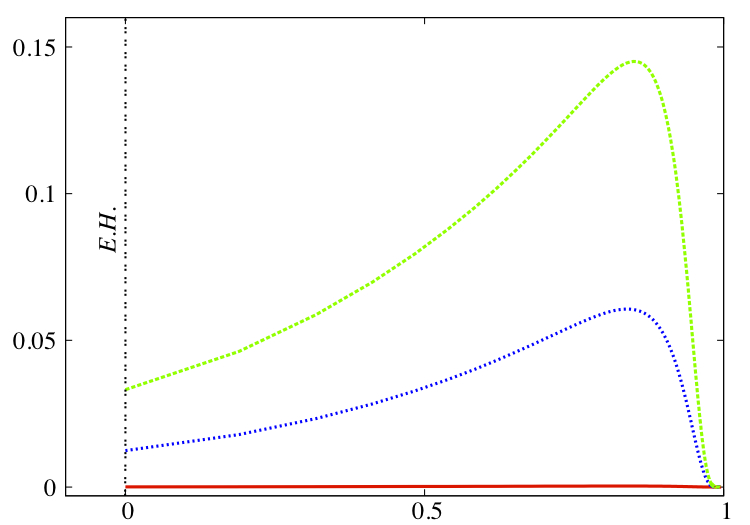}\\
\includegraphics[height=1.99in]{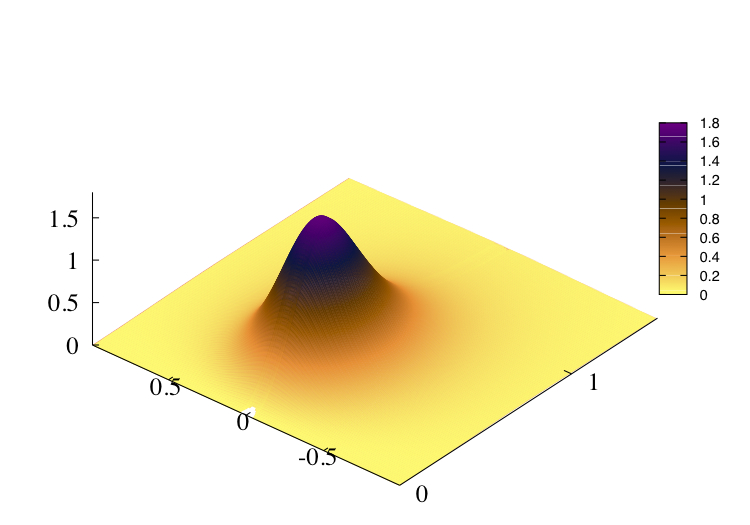}\ \ \ \ \ 
\includegraphics[height=1.69in]{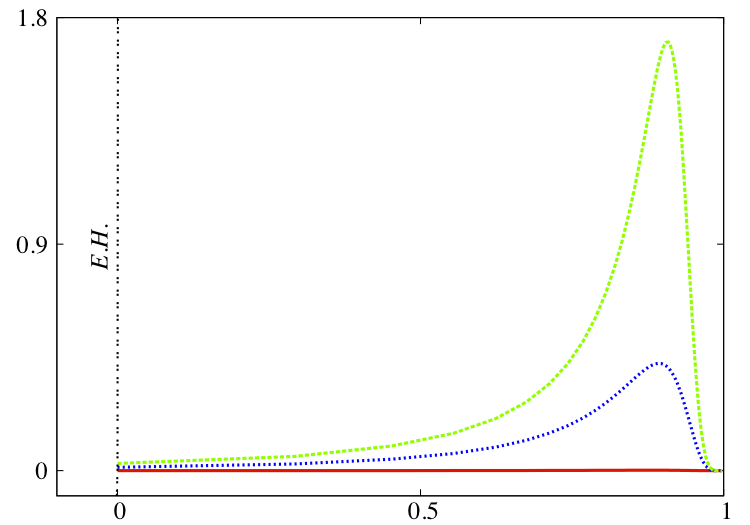}\\
\caption{The scalar ``angular momentum density" for example solutions {\bf I} and {\bf III}--{\bf V}, $cf.$ Section~\ref{sec_sample}.
} 
\label{J}
\end{figure}

\newpage

\begin{figure}[h!]
\centering
\includegraphics[height=1.99in]{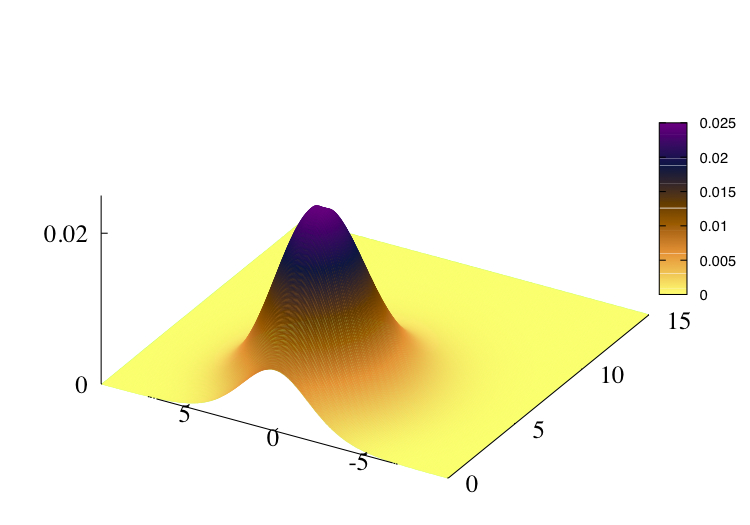}\ \ \ \ \ 
\includegraphics[height=1.69in]{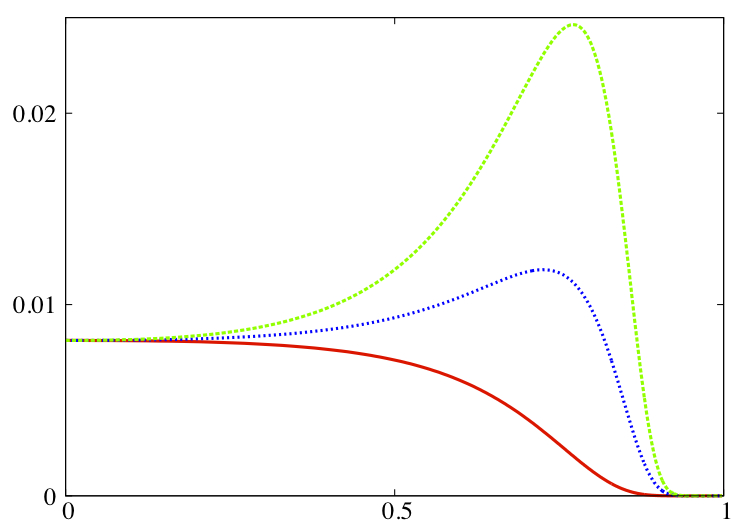}\\
\includegraphics[height=1.99in]{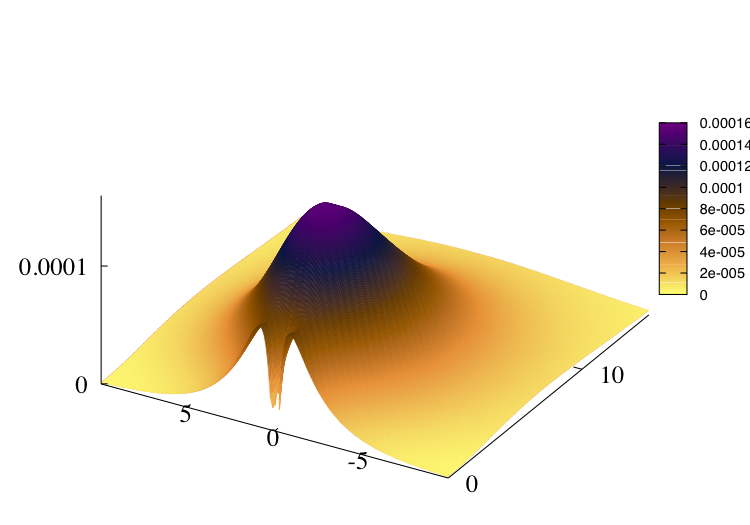}\  \ \ \ \ 
\includegraphics[height=1.69in]{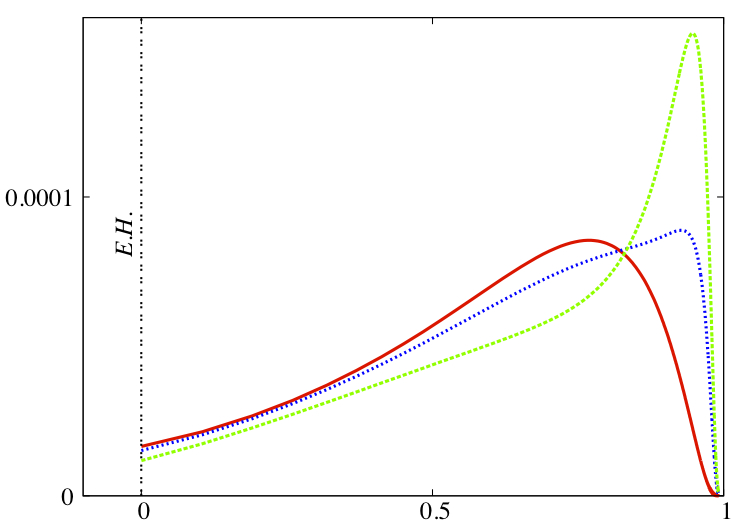}\\
\includegraphics[height=1.99in]{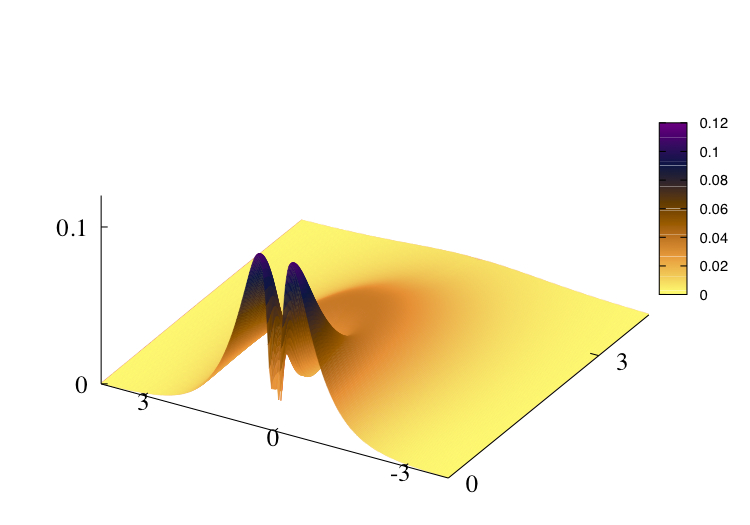}\ \ \ \ \ 
\includegraphics[height=1.69in]{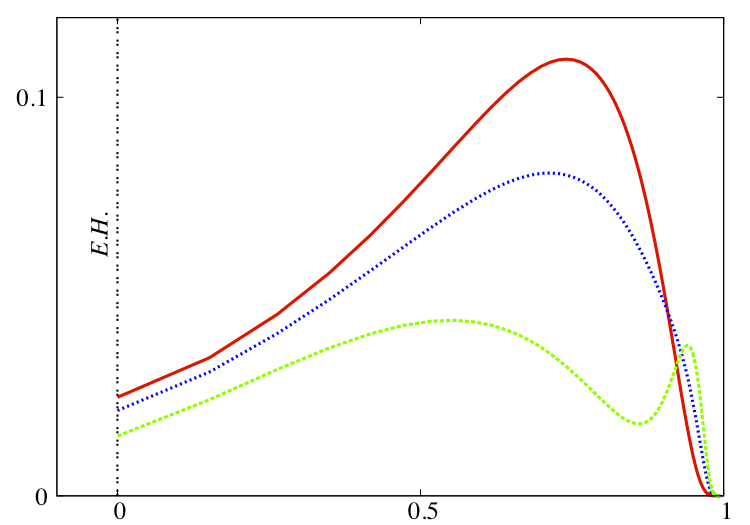}\\
\includegraphics[height=1.99in]{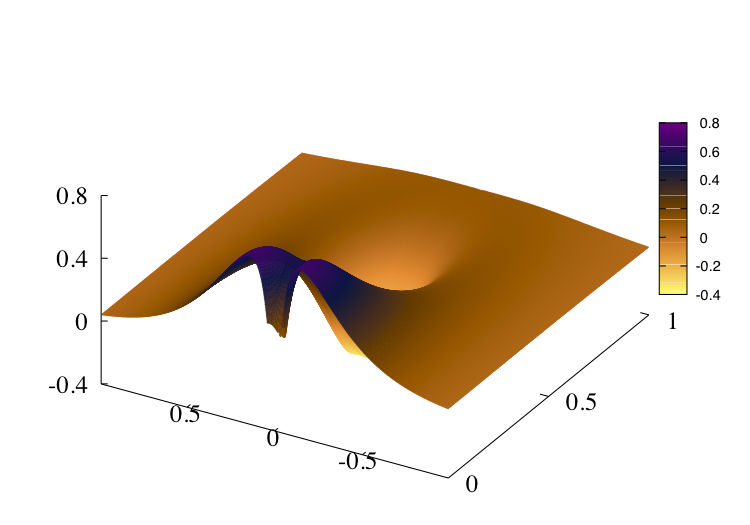}\ \ \ \ \ 
\includegraphics[height=1.69in]{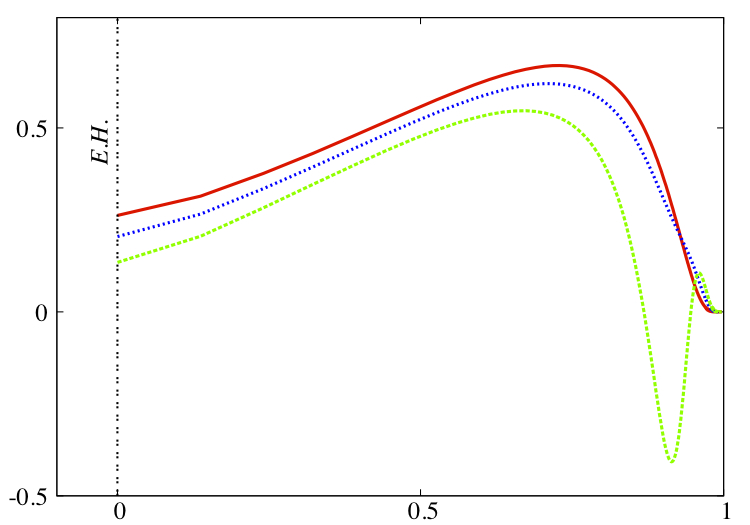}\\
\caption{The Ricci scalar for example solutions {\bf I} and {\bf III}--{\bf V}, $cf.$ Section~\ref{sec_sample}.
} 
\label{R}
\end{figure}

\newpage

\begin{figure}[h!]
\centering
\includegraphics[height=1.69in]{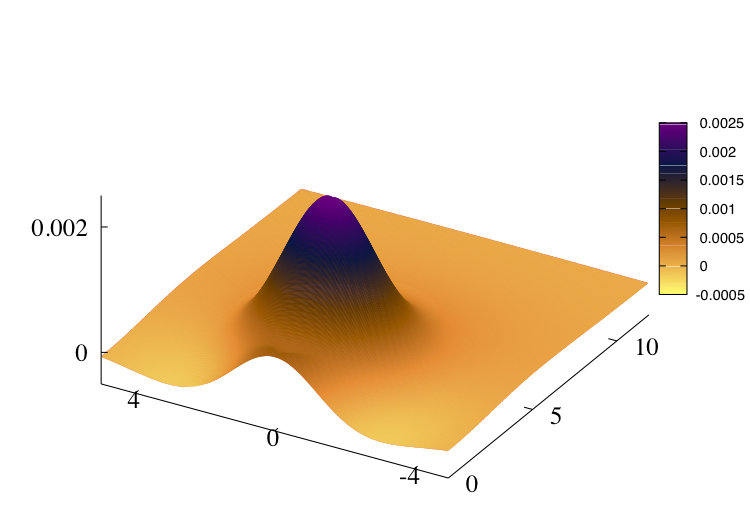}\ \ \ \ \ 
\includegraphics[height=1.69in]{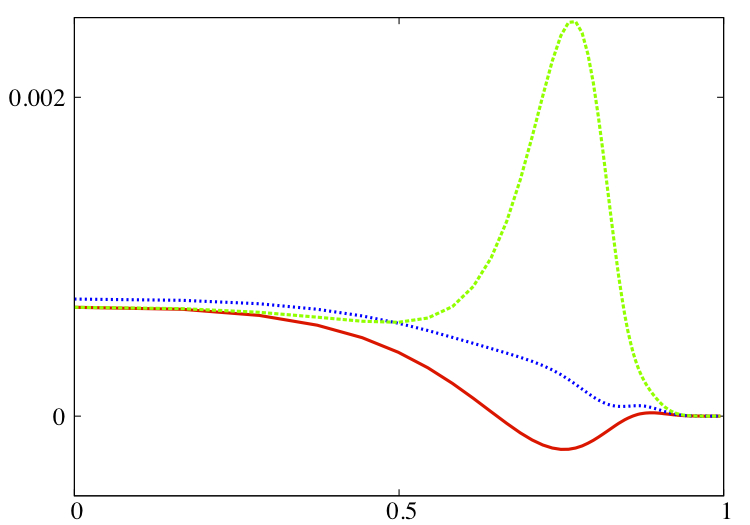}\\
\includegraphics[height=1.69in]{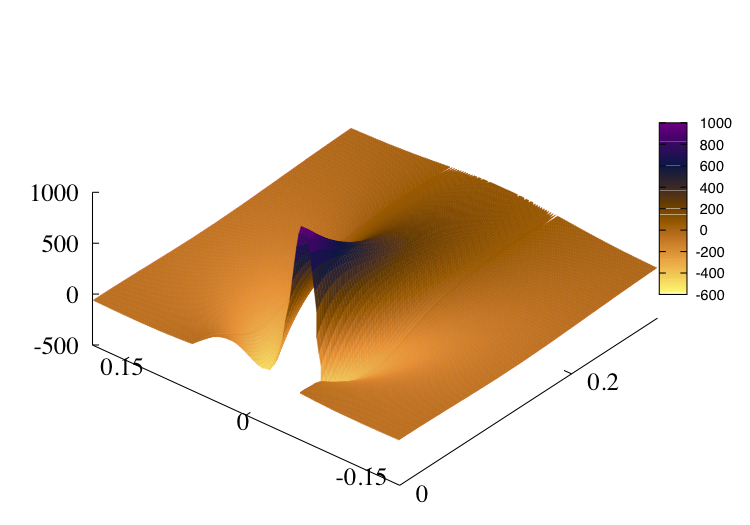}\  \ \ \ \ 
\includegraphics[height=1.69in]{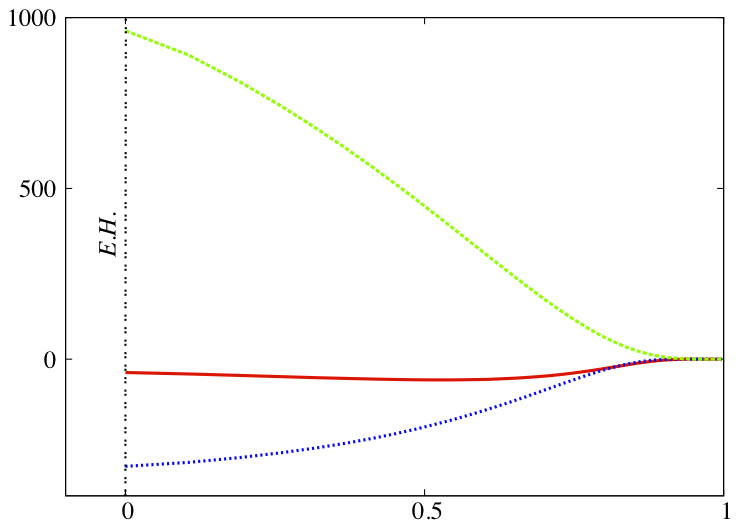}\\
\includegraphics[height=1.69in]{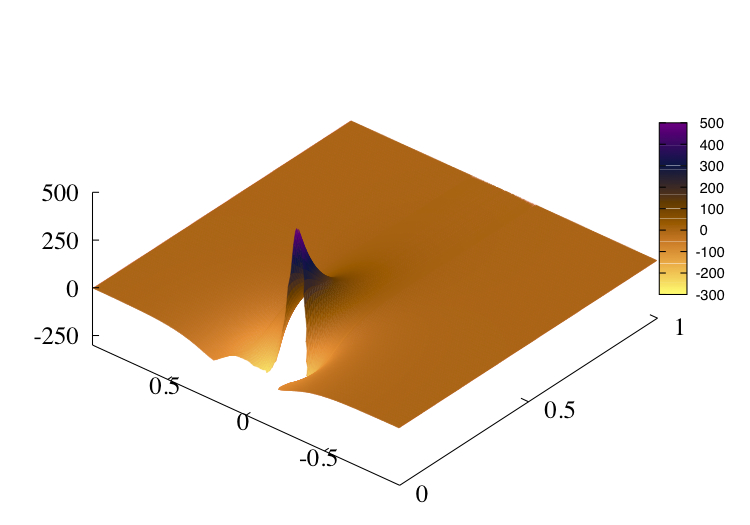}\  \ \ \ \ 
\includegraphics[height=1.69in]{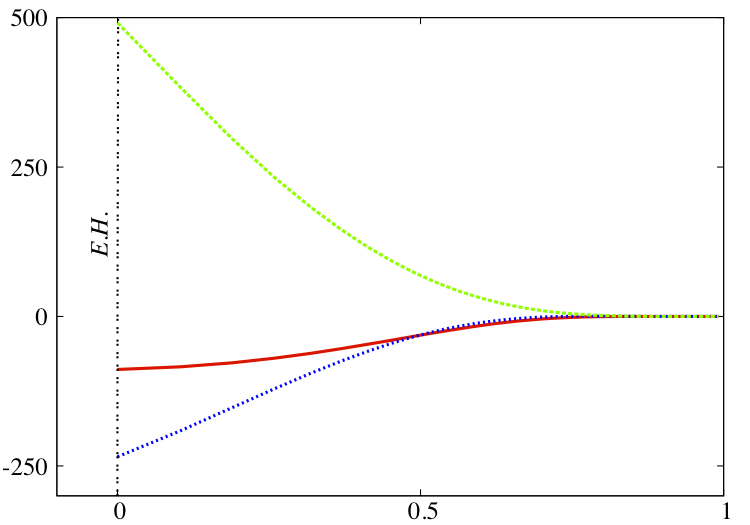}\\
\includegraphics[height=1.69in]{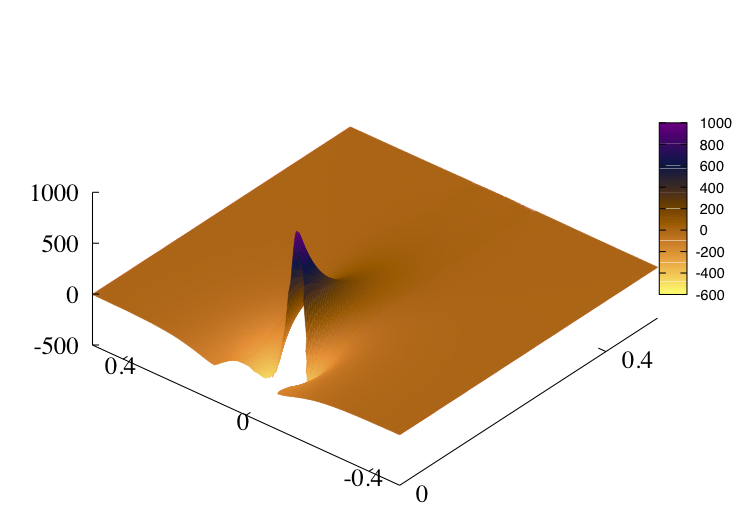}\ \ \ \ \ 
\includegraphics[height=1.69in]{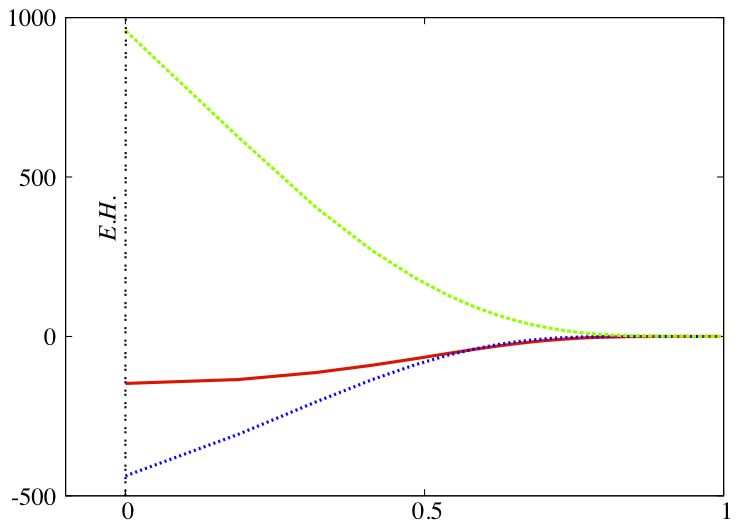}\\
\includegraphics[height=1.69in]{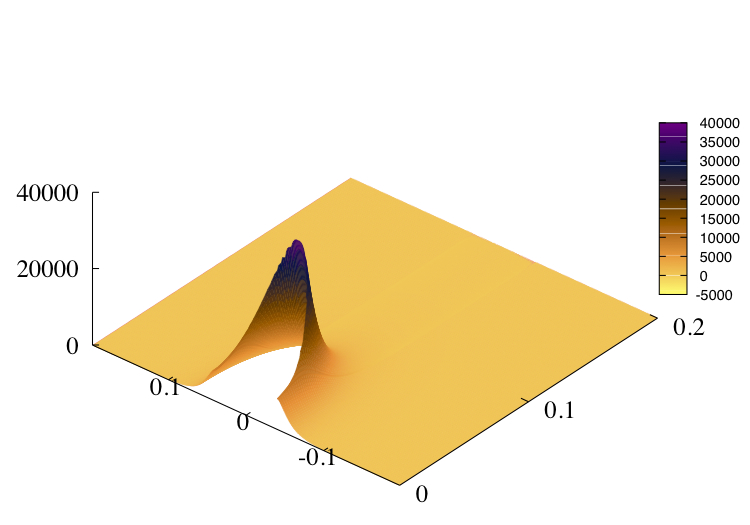}\ \ \ \ \ 
\includegraphics[height=1.69in]{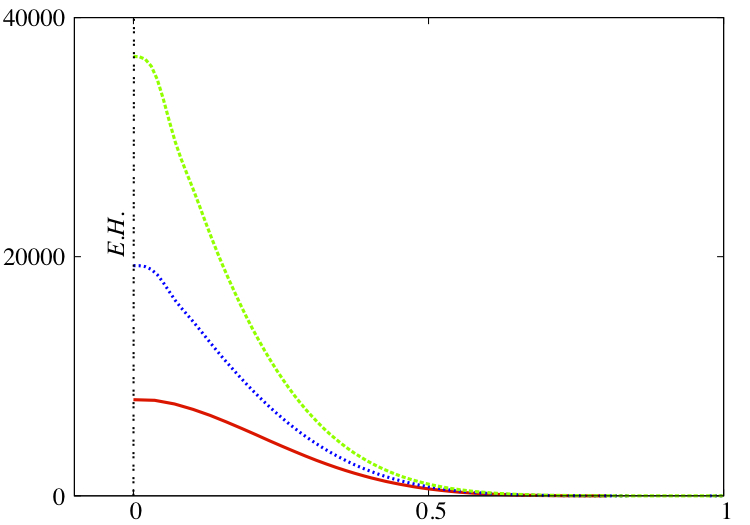}\\
\caption{The Kretschmann scalar for example solutions {\bf I}--{\bf V}, $cf.$ Section~\ref{sec_sample}.} 
\label{K}
\end{figure}

\bibliographystyle{h-physrev4}
\bibliography{long}

\end{document}